\documentclass[a4paper,12pt]{article}

\usepackage{jheppub}
\usepackage[english]{babel}
\pdfoutput=1
\usepackage[utf8]{inputenc}
\usepackage[T1]{fontenc}
\usepackage{amsmath}
\usepackage{float}
\usepackage{graphicx}
\usepackage{comment}
\usepackage[dvipsnames]{xcolor}
\usepackage{color}
\usepackage{epsfig}
\usepackage{dcolumn}
\usepackage{hyperref}
\usepackage{bm}
\usepackage{amssymb}
\usepackage{slashed}
\usepackage[caption=false]{subfig}
\usepackage{epstopdf}
\usepackage{setspace}
\usepackage[normalem]{ulem}	
\usepackage{multicol}
\usepackage{mathrsfs}
\usepackage{cancel}
\usepackage{fancybox}
\pdfoutput=1

\newcommand{\ie}{\textit{i.e. }}

\newcommand{\Deltatpp}{\Delta t_{\rm{pp}}}

\usepackage{tikz,pgfplots}
\usetikzlibrary{arrows,shapes,trees,calc,positioning,patterns}
\newcommand{\vEW}{v_\text{\tiny EW}}
\definecolor{corlinks}{RGB}{0,0,128}
\definecolor{cormenu}{RGB}{0,0,128}
\definecolor{corurl}{RGB}{0,0,128}

\newcommand{\mphi}{\widetilde{m}_\phi}

\definecolor{colRed0}{rgb}{0.85, 0.05, 0.12}
\definecolor{colRed1}{rgb}{0.92, 0.1, 0.05}
\definecolor{colRed2}{rgb}{0.95, 0.35, 0.05}
\definecolor{colYellow1}{rgb}{1., 0.82, 0.}
\definecolor{colBlue1}{rgb}{0.0, 0., 0.4}
\definecolor{colBlue2}{rgb}{0.1, 0.3, 0.9}
\definecolor{colBlue3}{rgb}{0.15, 0.4, 0.75}
\definecolor{colBlue4}{rgb}{0.3, 0.8, 0.93}
\definecolor{colGreen0}{rgb}{0.0, 0.15, 0.05}
\definecolor{colGreen1}{rgb}{0.0, 0.35, 0.1}
\definecolor{colGreen2}{rgb}{0.1, 0.65, 0.2}
\definecolor{colGreen3}{rgb}{0.3, 0.85, 0.5}
\definecolor{colBrown1}{rgb}{0.3, 0.18, 0.12}
\definecolor{colBrown2}{rgb}{0.5, 0.3, 0.20}
\definecolor{colViolet1}{rgb}{0.4, 0.18, 0.42}
\definecolor{colViolet2}{rgb}{0.5, 0.3, 0.70}

\hypersetup{
colorlinks=true,
urlcolor=corlinks,
linkcolor=corlinks,
menucolor=cormenu,
citecolor=corlinks,
pdfborder= 0 0 0
}

\DeclareRobustCommand{\Eq}[1]{Eq.~(\ref{#1})}
\DeclareRobustCommand{\Fig}[1]{Fig.~\ref{#1}}
\DeclareRobustCommand{\Ref}[1]{Ref.~\cite{#1}}

\newcommand{\ew}{\textnormal{\tiny EW}}

\newcommand{\MPl}{M_\text{Pl}}

\newcommand{\phidotSR}{\dot\phi_\mathrm{SR}}
\newcommand{\GeV}{\text{ GeV}}
\newcommand{\MeV}{\text{ MeV}}
\newcommand{\keV}{\text{ keV}}
\newcommand{\TeV}{\text{ TeV}}

\newcommand{\fcos}{f}
\newcommand{\fZZ}{F}
\newcommand{\fFF}{F_\gamma}

\title{Relaxion Fluctuations  (Self-stopping Relaxion) and Overview of Relaxion Stopping Mechanisms}

\date{\today}

\author[a,b]{Nayara Fonseca,}
\author[a,c]{Enrico Morgante,}
\author[a]{Ryosuke Sato,}
\author[a,d]{G\'eraldine Servant}

\affiliation[a\,]{DESY, Notkestrasse 85, D-22607 Hamburg, Germany}
\affiliation[b\,]{Abdus Salam International Centre for Theoretical Physics,
Strada Costiera 11, 34151, Trieste, Italy}
\affiliation[c\,]{PRISMA$^+$  Cluster  of  Excellence  and  Mainz  Institute  for  Theoretical  Physics, Johannes  Gutenberg-Universit\"at  Mainz,  D-55099  Mainz,  Germany}
\affiliation[d\,]{II. Institute of Theoretical Physics, Univ. Hamburg, D-22761 Hamburg, Germany}

\emailAdd{nfonseca@ictp.it}
\emailAdd{emorgant@uni-mainz.de}
\emailAdd{ryosuke.sato@desy.de}
\emailAdd{geraldine.servant@desy.de}

\abstract{
In implementations of the electroweak  scale cosmological relaxation mechanism proposed so far, the effect of the quantum fluctuations of the homogeneous relaxion field has been ignored.
We show that they can grow during the classical cosmological evolution of the relaxion field passing through its many potential barriers.
The resulting production of relaxion particles can act as an efficient stopping mechanism for the relaxion. 
We revisit the original relaxion proposal and determine under which conditions inflation may no longer be needed as a source of friction.
We review alternative stopping mechanisms and determine in detail the allowed parameter space for each of them (whether happening before, during and after inflation), also considering and severely constraining the case of friction from electroweak gauge boson production in models with large and Higgs-independent barriers.
}

\begin{document}

\begin{flushright} 
DESY 19-203
MITP/19-080
\end{flushright}

\maketitle

\section{Introduction}

As data from the LHC have magnified the Higgs naturalness problem
by enlarging the gap between the electroweak  (EW) scale and the scale of new physics, a new approach has been pursued in the last few years to address the origin of the EW scale 
as the result of a cosmological process.
The relaxation mechanism by
Graham--Kaplan--Rajendran \cite{Graham:2015cka}
is a new mechanism to solve the Higgs naturalness problem through the cosmological evolution of an axion-like field coupled to the Higgs, called the {\it relaxion}. 
While classically rolling down its potential, the relaxion field $\phi$ scans down the Higgs mass parameter, starting from a value of the order of the cutoff scale, until a stopping mechanism comes into play precisely when the Higgs mass parameter approaches zero.
Such proposal implies a radically new strategy to tackle experimentally the hierarchy problem. While the original model  \cite{Graham:2015cka} still requires the addition of new fermions at the weak scale, it was shown in Ref.~\cite{Espinosa:2015eda} that  the relaxion mechanism can be implemented without requiring any new physics at the weak scale and the only signature of the weak scale stabilisation mechanism is the existence of very light and weakly coupled axion-like particles with no effect at future colliders. Similarly, the alternative implementation in Ref.~\cite{Hook:2016mqo,Fonseca:2018xzp} is also very challenging as far as detection is concerned, as the relaxion is the only predicted new physics manifestation. 

While this proposal has been followed-up by a large literature addressing a variety of questions 
(see e.g.~\cite{Fonseca:2018xzp} for an almost complete list of references), there are still many open interrogations and directions for future exploration of this scenario.
Strangely enough, none of the previous works considered the effect of relaxion quantum fluctuations. In fact, only the evolution of the homogeneous zero-mode has been considered.
The purpose of this work is to work out the implications of relaxion particle production on the relaxion mechanism.
Our conclusions are based on the results of a companion paper \cite{Fonseca:2019ypl} that is general and applies to any axion-like particle in the early universe subject to rolling along a potential which also features a large number of wiggles, as ubiquitous in string axion monodromy models for instance.
If the field fast-rolls and overpasses many barriers, it can produce an exponentially  large number of its quanta which backreact on the homogeneous field. We call this effect axion fragmentation.
An important question is whether  this can be efficient enough to stop the relaxion so that inflation is no longer needed as a source of friction.

In Section \ref{sec:summaryfragmentation}, we summarise the main results of \cite{Fonseca:2019ypl}.
Concretely, the only inputs from \cite{Fonseca:2019ypl} which we use in this analysis are the expression for the time scale of efficient relaxion fragmentation $\Delta t_{\rm frag}$, the field excursion during this time $\Delta \phi_{\rm frag}$,  and the conditions on the Hubble expansion rate and on the potential slope which allow efficient relaxion fragmentation.
In Section \ref{sec:relaxion growing barriers}, we take the original relaxion proposal where the barriers of the relaxion periodic potential are generated by a new (non-QCD) strongly interacting sector and are Higgs-dependent. We review all conditions  to be imposed on the mechanism and work out in detail the open region of parameter space. We treat three cases:
\begin{enumerate}
\item[$\bullet$] Section \ref{sec:during inflation}:  The relaxation mechanism occurs during inflation, like in the original proposal ~\cite{Graham:2015cka}, and the relaxion is a subdominant component of the total energy density of the universe. We outline three different stopping mechanisms in this case. 
\begin{enumerate}
\item[(i)] through Hubble friction like in ~\cite{Graham:2015cka}
\item[(ii)] because of large barriers (and low Hubble friction) 
\item[(iii)] through axion fragmentation (and low Hubble friction) 
\end{enumerate}
\item[$\bullet$] Section \ref{sec:withoutinflation}: The relaxation mechanism occurs without inflation (self-stopping relaxion), while the relaxion can dominate the energy density of the universe and is fast-rolling. The relaxion stops because of relaxion particle production.
\item[$\bullet$] Section \ref{sec:relaxioninflating}: The relaxation mechanism occurs during an inflation era driven by the relaxion itself. The relaxion stops because of relaxion particle production.
\end{enumerate}
In Section \ref{sec:HMT}, we examine the consequences of relaxion fragmentation on a second interesting class of relaxion model~\cite{Hook:2016mqo,Fonseca:2018xzp}, where the stopping mechanism comes from the production of EW gauge bosons.  In this case, barriers of the relaxion potential are large and Higgs-independent while the relaxion has a significant coupling to EW gauge bosons. The universe starts in the broken EW phase where the Higgs vacuum expectation value is large and of the order of the cutoff scale. When their mass becomes small, EW gauge bosons can be abundantly produced and, at the same time, they take away the relaxion kinetic energy.
In this context, we show that relaxion particle production is very efficient and very severely constrains this model. We summarise our findings in Section \ref{secsummary}. In   Appendix \ref{sec:explicit model},
we recap the origin of the Higgs-dependent barrier  in the models discussed in Section \ref{sec:relaxion growing barriers}. In Appendix \ref{sec:GKR rolling}, we clarify the relaxion stopping condition used in \cite{Graham:2015cka}. All the detailed constraints are presented in Appendix \ref{sec:detailedplots}.  Note that the case of Higgs particle production during relaxation was studied earlier and it was shown that it cannot be used as a stopping mechanism for the relaxion \cite{Fonseca:2018xzp}.
Alternative sources of friction were considered elsewhere in the literature. Finite temperature effects are used in~\cite{Hardy:2015laa}. In~\cite{Ibe:2019udh} the necessary friction is provided by parametric resonance of the Higgs zero mode. In~\cite{Wang:2018ddr} a potential instability is introduced to stop the field. Finally, friction from the production of dark fermions is discussed in~\cite{Kadota:2019wyz}.

\section{Relaxion fragmentation}
\label{sec:summaryfragmentation}

The Higgs--relaxion potential is  of the form:
\begin{equation}\label{eq:relaxion potential}
V (\phi,h)= \Lambda^4 -g\Lambda^3\phi + \frac{1}{2}(\Lambda^2 - g'\Lambda\phi)h^2 + \frac{\lambda}{4}h^4 + \Lambda_b^4  \cos\frac{\phi}{f} \,,
\end{equation}
where $\Lambda$ is the cutoff of the effective theory, and $g,g'\ll1$ should be thought as spurions parametrizing the breaking of the relaxion shift-symmetry, $\lambda$ is the Higgs quartic coupling, and $\Lambda_b$ is the scale  related to the non-perturbative dynamics which generates the periodic potential.
For the moment, we will neglect the dependence of the barriers  amplitude on the Higgs VEV.
Electroweak symmetry breaking happens for $\phi\approx\Lambda/g'$. One of the key points of our discussion will be to understand the mechanism responsible for stopping the evolution of the relaxion field and, consequently, of the Higgs field. Our goal is to work out the implications of relaxion particle production on the relaxion mechanism.
This effect is studied in detail in Ref.~\cite{Fonseca:2019ypl} and is relevant in the 
regime where the relaxion initial velocity is large enough to overpass the potential barriers.
 We summarise the main results in this section.
Ref.~\cite{Fonseca:2019ypl} shows that (in general for any rolling axion-like field) the dynamics of axion fluctuations accompanying the evolution of the axion zero-mode rolling down its potential while passing through a large number of wiggles can be described by the Mathieu equation, at least in the limit of constant velocity. 
Solutions of this equation have instability bands where quantum fluctuations grow exponentially. 
The homogeneous mode gradually looses kinetic energy as more energy goes into fluctuations. We found that this back reaction effect can be large enough as to stop the relaxion.

Let us start by decomposing the relaxion field into a homogeneous mode plus small fluctuations:
\begin{equation}
\phi(x,t) = \phi(t) + \delta\phi(x,t) = \phi(t) + \left( \int \frac{d^3k}{(2\pi)^3}a_k u_k(t) e^{ikx} + h.c. \right)
\end{equation}
where $a_k$ are the usual annihilation operators with $[a_k,a_{k'}^\dagger] = (2\pi)^3\delta^3(k-k')$. The initial conditions for the mode functions $u_k$ at $t\rightarrow -\infty$ are
\begin{equation}
u_k(t) = \frac{e^{-i(k/a)t}}{a\sqrt{2k}} \,.
\end{equation}
The equations of motion for the zero mode $\phi(t)$ and for the mode functions $u_k$ are given by
\begin{align}
\ddot{\phi} + 3H\dot{\phi} + V'(\phi) + \frac{1}{2}V'''(\phi) \int\frac{d^3 k}{(2\pi)^3} |u_k|^2 &=0, \label{eq:zeromode}\\
\ddot{u_k} + 3H\dot{u_k} + \left[ \frac{k^2}{a^2} + V''(\phi)\right] u_k &= 0. \label{eq:fluctuation}
\end{align}
The dependence of \Eq{eq:zeromode} on the mode functions is such that, when $u_k$ grow, the evolution of the zero mode $\phi$ is slowed down, consistently with energy conservation.
For the moment we will work in Minkowski space, neglecting cosmic expansion. Equation~\ref{eq:fluctuation} reads
\begin{equation}\label{eq:fluc eom}
\ddot u_k + \left(k^2 - \frac{\Lambda_b^4}{f^2}\cos\frac{\phi}{f} \right) u_k = 0 \,.
\end{equation}
In the limit of constant velocity $\dot\phi$, \Eq{eq:fluc eom} can be read as a Mathieu equation~\cite{mclachlan}, which is well known \textit{e.g.} in the context of preheating~\cite{Kofman:1997yn} (the difference with preheating though, is that in our case the frequency changes rather than the amplitude of the cosine).
Depending on its parameters, solutions of the Mathieu equation can be unstable and grow exponentially. In the above notation, this happens if $k$ falls in some specific bands around $n \dot\phi /(2f)$, for integer $n\geq 1$. For $n\geq2$ the instability bands have small width and the exponential growth of the corresponding solutions is slow. Modes that fall in the $n=1$ instability band instead grow faster, and the width of the band is larger, thus they are the principal source of friction that decelerate the relaxion.

For $\dot\phi^2/2\gg\Lambda_b^4$ the first instability band can be written as $|k-k_\mathrm{cr}|<\delta k_\mathrm{cr}$, with
\begin{equation} \label{eq:kcr and delta kcr}
k_\mathrm{cr} = \frac{\dot\phi}{2f} \, ,
\qquad
\delta k_\mathrm{cr} = \frac{\Lambda_b^4}{2f \dot\phi}\,.
\end{equation}
The asymptotic behaviour of $u_k$ at large $t$ is
\begin{equation}
%u_k \propto \exp\left( \sqrt{ (\delta k_{\rm cr} )^2 - \left( k- k_{\rm cr}  \right)^2 } t \right) \sin \left(k_{\rm cr} t - \arctan\sqrt{ \frac{\delta k_{\rm cr} - (k - k_{\rm cr} ) }{ \delta k_{\rm cr} + (k - k_{\rm cr} )  } } \right). \label{eq:asymptotic uk}
u_k \sim k_{\rm cr}^{-1/2} \exp\left( \sqrt{ (\delta k_{\rm cr} )^2 - \left( k- k_{\rm cr}  \right)^2 } t \right) \sin k_{\rm cr} t . \label{eq:asymptotic uk}
\end{equation}
The number of exponentially growing modes per unit volume is $\sim k_{\rm cr}^2 \delta k_{\rm cr}$.
As long as $\dot\phi$ is constant, the energy density of the fluctuations within the instability band grows as
\begin{align} \label{eq:rho fluc}
\rho_{\rm fluc}(t) \sim
k_{\rm cr}^2 \delta k_{\rm cr} \times |\dot u_{k_\mathrm{cr}}|^2 \sim
k_{\rm cr}^3 \delta k_{\rm cr} \exp(2 \delta k_{\rm cr} t ).
\end{align}
Energy conservation implies that the same amount is subtracted from the kinetic energy of the zero mode, which thus slows down. Because of this back reaction, the instability band moves towards smaller $k$'s. The growth of the modes around $k_\mathrm{cr}$ stops when this goes out of the instability band.
This happens when the critical mode has decreased by the amount $\delta k_\mathrm{cr}$, and consequently, the kinetic energy of the zero-mode is reduced by

\begin{equation} \label{eq:delta K}
\delta K \approx \frac{1}{2}4 f^2 [(k_{\rm cr}-\delta k_{\rm cr})^2 - (k_{\rm cr})^2] \approx - 4 f^2 k_{\rm cr} \delta k_{\rm cr} = - \dot\phi^2 \frac{\delta k_{\rm cr}}{k_{\rm cr}}\, ,
\end{equation}
where we used the definition of $k_{\rm cr}$ as in \Eq{eq:kcr and delta kcr}. From Eqs.\,(\ref{eq:rho fluc}) and (\ref{eq:delta K}), we can estimate the timescale $\delta t_{\rm amp}$ a given mode $k_{\rm cr}$ spends inside the instability band:
\begin{equation}\label{eq:delta t amp}
\rho_{\rm fluc} \approx - \delta K \approx \dot\phi^2 \times \frac{\delta k_{\rm cr}}{k_{\rm cr}}~~ \Rightarrow ~~ \delta t_{\rm amp} = \frac{1}{2\,\delta k_{\rm cr}} \log \frac{\dot\phi^2}{(k_{\rm cr})^4}\,.
\end{equation}
Therefore, the evolution of the kinetic is approximately  given by  $ \delta K / \delta t_{\rm amp}$ :
\begin{align}\label{eq:K evolution}
\frac{d}{dt} \left(\frac{1}{2}\dot\phi^2\right)
\approx \frac{\delta K}{\delta t_\mathrm{amp}}
\sim -\frac{\Lambda_b^8}{f \dot\phi} \left( \log\frac{16f^4}{\dot\phi^2} \right)^{-1} \,,  
\end{align}
which can be integrated exactly to estimate the time it takes $\Delta t_{\rm frag}$ and the field excursion $\Delta\phi_{\rm frag}$ from the beginning of particle production until the relaxion  stops.
Notice that in general $\Delta t_{\rm frag}$ can be much larger than $\delta t_{\rm amp}$: the latter is the time spent by a single mode inside the instability band, while the former is the total integrated time of the fragmentation process, from the initial condition until the field is trapped in the potential barriers.

A more rigorous calculation involves a WKB approximation and is performed in~\cite{Fonseca:2019ypl}. The time scale of relaxion particle production in terms of the parameters of the relaxion potential is
\begin{equation}
\label{eq:mastertimescale}
\boxed{
\Delta t_{\rm frag} \simeq \frac{2 f \dot{\phi}_0^3}{3 \pi \Lambda_b^8}\log\frac{32 \pi^2 f^4}{\dot{\phi}_0^2}}
\end{equation}
and the corresponding field excursion
\begin{equation}
\label{eq:masterfieldexcursion}
\boxed{
\Delta \phi_{\rm frag} \simeq \frac{ f \dot{\phi}_0^4}{2 \pi \Lambda_b^8}\log\frac{32 \pi^2 f^4}{\dot{\phi}_0^2}}
\end{equation}
where $\dot{\phi}_0$ is the initial velocity of the relaxion, before fragmentation starts. This result agrees, up to $\mathcal{O}(1)$ factors, with the one obtained by integrating Eq.~(\ref{eq:K evolution}).

Equations~(\ref{eq:mastertimescale}) and (\ref{eq:masterfieldexcursion}) were derived neglecting Hubble friction and the slope $-g\Lambda^3 \phi$ of the potential in \Eq{eq:relaxion potential}. The effect of Hubble friction on the growth of perturbation is double. The dominant effect is that the friction term $3 H \dot u_k$ in \Eq{eq:fluctuation} suppresses the growth of $u_k$. Secondarily, the physical wave number $k/a$ corresponding to a given comoving mode $k$ decreases as the scale factor $a(t)$ grows. While the zero mode $\phi$ decelerates, the instability band moves towards smaller physical modes too, thus cosmic expansion prolongs the time that each given mode spends inside the instability band. As for the  slope term  $-g\Lambda^3 \phi$, it enters in \Eq{eq:fluctuation} instead only through the evolution of the zero mode: if the slope is large, the field accelerates and the instability band moves to the UV.%
\footnote{An interesting case is the one in which the acceleration due to the slope and the friction from fragmentation exactly cancel each other, leading to a slow roll regime even in the presence of a steep potential~\cite{Fonseca:2019ypl}, similarly to what happens in axion inflation with a coupling to dark photons~\cite{Anber:2009ua} and in trapped inflation~\cite{Green:2009ds}. We leave this scenario for future study.} As a result, both a large Hubble friction and a large slope suppress the growth of the relaxion fluctuation, thus making fragmentation ineffective.
The actual conditions under which these terms can be neglected depend on the initial velocity of the zero-mode. 
The following  equation, which   relates the slope $g \Lambda^3$, the Hubble friction $H$ and the height of the wiggles $\Lambda_b$, must be satisfied in order  for fragmentation to be efficient \cite{Fonseca:2019ypl}:

\begin{equation} \label{eq:master eq}
g \Lambda^3 < 
2H\dot\phi_0 + \displaystyle\frac{\pi\Lambda_b^8}{2f\dot\phi_0^2} \left( W_0\left( \displaystyle\frac{32\pi^2f^4}{e\dot\phi_0^2} \right) \right)^{-1}\,,
\end{equation}
where $W_0$ is the 0-th branch of the product logarithm function (also known as the Lambert $W$ function).
Intuitively, this equation express the fact that the slope can not be too large, otherwise its acceleration would be stronger than the slow down due to fragmentation.
Here we will be interested in two special cases of condition (\ref{eq:master eq}):

\begin{enumerate}
\item
During inflation, which corresponds to the case where the initial velocity is the slow-roll velocity,  $\dot\phi_0 = \phidotSR = g\Lambda^3/(3H)$, fragmentation is effective for
\begin{equation}
\boxed{
H \phidotSR^3 < \frac{\pi \Lambda_b^8}{2f} \left( W_0\left( \frac{32\pi^2 f^4}{e \phidotSR^2} \right) \right)^{-1}.}
\label{eq:condition 3}
\end{equation}
\item
When $\dot\phi_0 < \phidotSR $ instead, which includes the case of negligible Hubble expansion, fragmentation can stop the field for
\begin{equation}
\boxed{g\Lambda^3 < \frac{\pi\Lambda_b^8}{2f\dot\phi_0^2} \left( W_0\left( \frac{32\pi^2 f^4}{e\dot\phi^2_0} \right) \right)^{-1}.} \label{eq:slope bound (H=0)}
\end{equation}
\end{enumerate}
The last two inequalities tell us that for effective fragmentation,  the field velocity should not be too large, otherwise each mode does not spend enough time in the instability band. As seen in Eqs.~(\ref{eq:delta t amp}) and~(\ref{eq:kcr and delta kcr}), this time is controlled by the width of the instability band, which goes like $1/\dot\phi$.

In summary, the results of~\cite{Fonseca:2019ypl} show that the evolution of the relaxion can be stopped by the growth of its own perturbations, once certain conditions are satisfied. In the following, we will use this result, and in particular Eqs.~(\ref{eq:mastertimescale}), (\ref{eq:masterfieldexcursion}), (\ref{eq:condition 3}) and (\ref{eq:slope bound (H=0)}) to determine the stopping point of the relaxion evolution. 
Depending on the model, fragmentation can offer an alternative stopping mechanism for the relaxion, thus offering new possibilities such as realizing the relaxion mechanism after inflation. On the other hand, fragmentation is a serious concern in models in which it is active at all times, which correspond to the scenarios where constant and sizable wiggles are present  during the scanning of the Higgs mass parameter~\cite{Hook:2016mqo, Fonseca:2018kqf, Fonseca:2018xzp}. This feature makes these  scenarios more constrained  than the ones in which the barriers only appear at the critical point where the EW symmetry is broken. These two options will be analysed in Sec.~\ref{sec:relaxion growing barriers} and ~\ref{sec:HMT}, respectively.

A last comment should be spent on the validity of the perturbative expansion that leads to Eqs.~(\ref{eq:zeromode}) and~(\ref{eq:fluctuation}). The issue of perturbativity arises when $\delta\phi\sim f$, which is the case in the last phases of fragmentation. This is not a problem, though, because we are only interested here in the time scale for stopping, and not on the precise dynamics of the relaxion fluctuations. During their growth, the fluctuations are, for most of the time, perturbative, and only at the end of the evolution they become of order $\delta\phi\sim f$, without growing more than this (at least not as long as the relaxion keeps rolling with $\dot\phi>\Lambda_b^2$). Thus, we expect non-perturbative dynamics not to change the picture dramatically, entering Eqs.~(\ref{eq:mastertimescale}) and (\ref{eq:masterfieldexcursion}) only as an $\mathcal{O}(1)$ normalization factor. It would be valuable to check this statement with a proper numerical simulation, which we postpone for future work~\cite{lattice}.

\subsection{Comparison of relaxion stopping mechanisms}

\begin{figure}[!h]
\centering
\includegraphics[width=.72\textwidth]{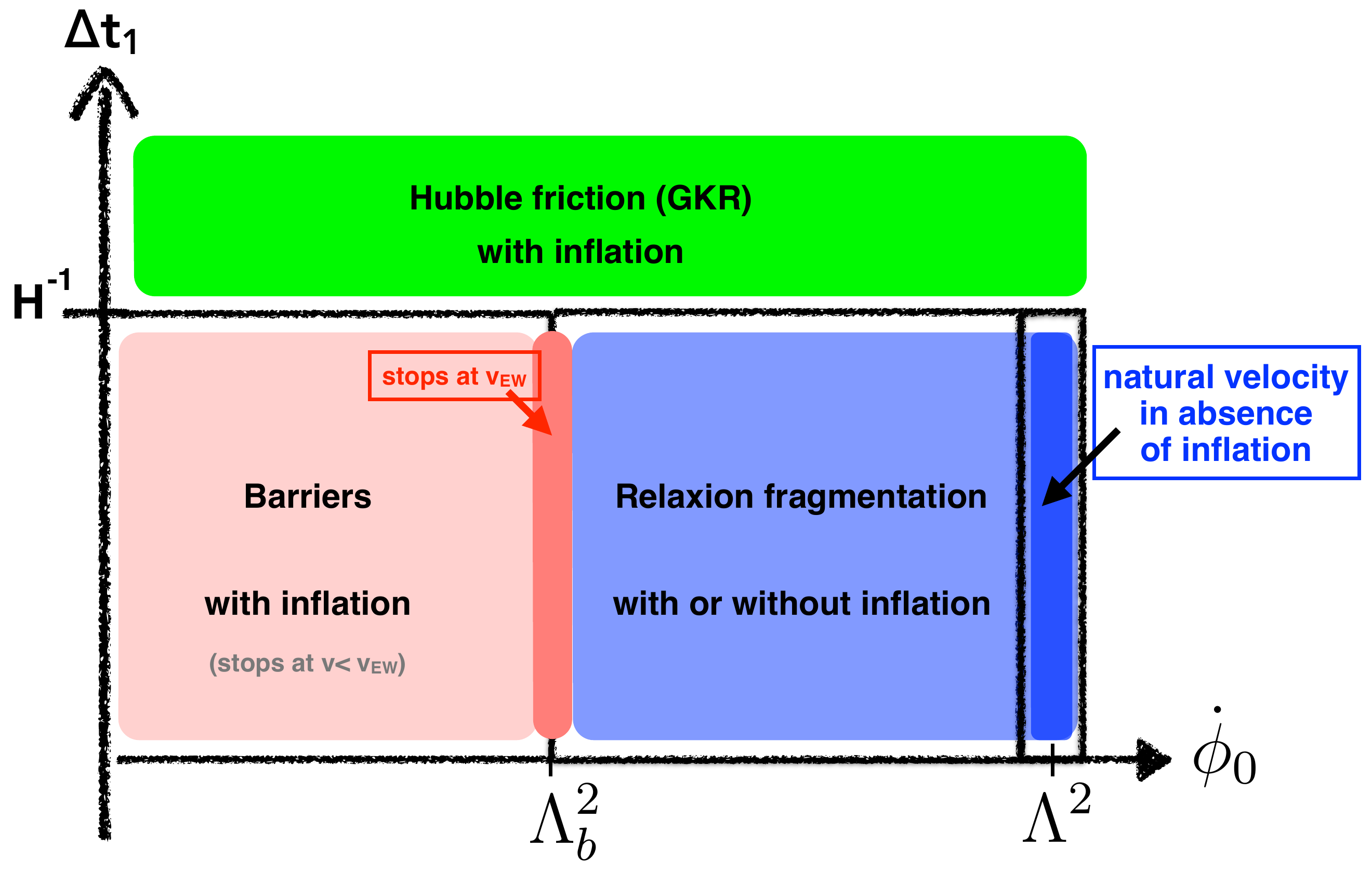}
\caption{\label{fig:sketch} Sketch of relaxion stopping mechanisms. The initial velocity $\dot{\phi}_0$ is either the relaxion velocity before the barriers appear in models with Higgs-dependent barriers, or the velocity before particle production starts in models where relaxion stops due to particle production (either relaxion particles or  EW  gauge bosons). $\Delta t_1$ is the time scale it takes for the relaxion to roll over a distance $2 \pi f$,
% and is of order $\Delta t_1 \sim 6 \pi f H/g \Lambda^3$.
to be compared to the Hubble time $H^{-1}$. The final size of the barriers is $\Lambda_b$ while the cutoff scale is $\Lambda$.
In our notation (see Eq.~(\ref{eq:barrierscaling})), $\Lambda_b$ is the size of the barriers when the relaxion stops at $\langle h\rangle=v_\ew$. In the red region, corresponding to the `large barriers' case of Sec.~\ref{sec:during inflation}, the relaxion stops when the barriers amplitude equals the initial kinetic energy. Thus, for $\dot\phi_0^2/2 < \Lambda_b^4$, the final value of the Higgs vev would be lower than the measured one (light red region).}
\end{figure}

In the rest of the paper, we will analyse the implications of relaxion fragmentation. 
As we will extensively discuss and as summarised in Fig.~\ref{fig:sketch}, there are two key quantities controlling the effective stopping mechanism. First is the time scale for the relaxion to roll over a distance $2 \pi f$, denoted $\Delta t_1$,  and second is the initial relaxion velocity $\dot{\phi}_0$. Depending on these two values, the relaxion may stop either because of Hubble friction, of large barriers, or of particle production.  There are two main types of relaxion models, those with Higgs-dependent  and Higgs-independent barriers respectively. We will examine those two cases in Sections \ref{sec:relaxion growing barriers} and \ref{sec:HMT} respectively.

\section{Consequences I: Relaxation with Higgs-dependent barriers}
\label{sec:relaxion growing barriers}
%\label{sec:GKR}

We consider the original relaxion model, which was first introduced in~\cite{Graham:2015cka} and later studied in a large literature.
In the non-QCD model in \cite{Graham:2015cka}, the relaxion potential features Higgs-dependent barriers that scale as%
\footnote{This notation does not coincide with the one of~\cite{Graham:2015cka}, where the barriers are denoted by $\Lambda^4\cos\phi/f$ with $\Lambda\propto \langle h\rangle^n$, nor with~\cite{Espinosa:2015eda}, which writes $\epsilon\Lambda_c^{4-n}\langle h\rangle^n$. The notation (\ref{eq:barrierscaling}) makes it clear that the barriers are proportional to $\langle h\rangle^2$ and that $\Lambda_b$ is the size of the barrier once the Higgs has reached its stopping point with $\langle h\rangle = v_\ew$. Thus, $\Lambda_b$ is not the confinement scale nor a parameter of the Lagrangian. It is determined by the dynamics of the stopping mechanism, and it depends on the initial relaxion velocity and on the measured value of $v_\ew$. $\Lambda_b$ is one of the parameters we are scanning over in our various contour plots.}
\begin{equation}
\label{eq:barrierscaling}
V (\phi,h)= \Lambda^4 -g\Lambda^3\phi + \frac{1}{2}(\Lambda^2 - g'\Lambda\phi)h^2 + \frac{\lambda}{4}h^4 + \Lambda_b^4 \frac{\langle h\rangle^2}{v_\ew^2} \cos\frac{\phi}{f} \,,
\end{equation}
and the initial conditions are such that the EW symmetry is initially not broken. 
For the stability of the potential (\ref{eq:barrierscaling}), the spurions should satisfy $ g \gtrsim g'/(4 \pi)$  since the term $\sim g' \Lambda^3 \phi$ is generated by closing the Higgs loop in the third term in Eq\,.(\ref{eq:barrierscaling}).
The initial condition must be such that $\mu_h^2 = (\Lambda^2 - g'\Lambda\phi)\approx\Lambda^2$, and we assume $\dot \phi_0>0$. Electroweak symmetry breaking happens for $\phi\approx\Lambda/g'$. After this point, the Higgs VEV $\langle h\rangle$ grows up to its final value $v_\ew$. 

Loop effects generate a Higgs-independent amplitude for the cosine, such that there are small constant wiggles during the whole field excursion (see for details App.\,\ref{sec:explicit model}). In this paper we work in the regime in which the potential has local minima. We postpone the study of fragmentation from wiggles that do not generate local minima to future investigation.

\subsection*{List of conditions}
%\label{sec:listofconditions}

There are a number of conditions that we will need to assume for a successful relaxation mechanism. We start by listing the ones that do not depend on the embedding of the mechanism in the cosmological history, which we will discuss later.
\begin{itemize}
\item \textbf{Initial conditions and total field excursion:}
First of all, to avoid  fine-tuning in the initial conditions, the total field excursion of the relaxion must be larger than $\Lambda/g'$, so that the Higgs mass can scan the range from the cut-off down to the EW scale. For definiteness, we assume that initially $\phi=0$, so that
\begin{equation}\label{eq:field range}
\Delta\phi = \frac{\Lambda}{g'}\,.
\end{equation}
\item \textbf{Precision of the mass scanning:}
In order not to overshoot its measured value $m_h^2$, the scanning of the Higgs mass should happen with enough precision. Thus we impose
\begin{equation}\label{eq:enoughprecision}
g' \Lambda (2\pi f) < \frac{m_h^2}{2} \,.
\end{equation}
\item \textbf{Large barriers:}
After the Higgs has grown to $v_\ew$, the barriers should be large enough to prevent the field from further rolling down, despite the slope $-g\Lambda^3$. Imposing that $V'>0$ for some values of $\phi>\Lambda/g'$ we get
\begin{equation}\label{eq:visiblewiggles}
\frac{\Lambda_b^4}{f} \geq g\Lambda^3\,.
\end{equation}
\item \textbf{Symmetry breaking pattern:}
In the Lagrangian of \Eq{eq:relaxion potential}, the scale $f$ should be thought as the scale of spontaneous breaking of a global symmetry, whose Goldstone boson is the relaxion. The spurious $g$ and $g'$ control the explicit breaking of the residual shift symmetry, as well as the Higgs mass parameter. For the consistency of this picture, we impose
\begin{equation} \label{eq:eftvalidity}
f > \Lambda.
\end{equation}
\item \textbf{Microscopic origin of the barriers:}
The last term in \Eq{eq:relaxion potential} must originate from the interaction of some field charged under the Standard Model gauge group and under the relaxion global symmetry. Explicit examples of such a kind were proposed in~\cite{Graham:2015cka} and~\cite{Flacke:2016szy}. A general feature of these constructions is that the term $\Lambda_b^4\langle h^2\rangle/v_\ew^2 \cos\phi/f$ is accompanied by the similar term $\Lambda_b^4 h^2/v_\ew^2 \cos\phi/f$ by which the Higgs interacts with $\phi$. Closing a Higgs loop, a constant term is generated, which must be subdominant compared to the previous one. The actual size of this term is model dependent, and we will here assume that the model discussed in Appendix~\ref{sec:explicit model} is realized. Thus we impose
\begin{equation}\label{eq:stability}
\Lambda_b < \sqrt{4\pi} v_\ew \,.
\end{equation}
Notice that this condition will turn out to be important in determining the upper bound on the cut-off of new physics, and thus weakening it will result in a larger allowed parameter space. Nonetheless, in the simplest explicit UV constructions, \Eq{eq:stability} has the correct numerical coefficient up to $\mathcal{O}(1)$ factors.
\item \textbf{Higgs field tracking the minimum of its potential:}
After EW symmetry breaking, the evolution of the Higgs field should follow closely the minimum of the potential, otherwise, after the relaxion stops, the Higgs would continue growing. If we denote by $v$ the minimum of the potential during the relaxion's evolution, we want $\langle h\rangle = v$. This happens if the evolution of $v$ is adiabatic, \ie if
\begin{equation}\label{eq:Higgs tracking}
\left| \frac{\dot v}{v^2} \right |_{v = v_\ew} < 1 \Longleftrightarrow \frac{g' \Lambda \dot\phi}{2 \lambda v_\ew^3} <1
\end{equation}
Equation~(\ref{eq:Higgs tracking}) guarantees that, as the Higgs potential changes due to the relaxion's evolution, the Higgs field has enough time to adapt and follow closely the minimum. A more detailed analysis can be performed by expanding the Higgs field $h = v + x$, solving the equation of motion of $x$ and imposing that the solution does not grow~\cite{Hook:2016mqo}. The resulting condition differs from the above for just a factor $\mathcal{O}(1)$ (see also the discussion in Ref.~\cite{Fonseca:2018xzp} and Fig.~5 therein).
\item \textbf{Sub-Planckian decay constant:}
We assume that the relaxion $\mathrm{U}(1)$ symmetry is broken below the Planck scale,
\begin{equation}\label{eq:fsubpl}
f<\MPl.
\end{equation}
Since the relaxion is  a pseudo-Nambu-Goldstone boson, the  linear terms in $\phi$ in the potential Eq.(\ref{eq:relaxion potential}) should arise from  the expansion of a second oscillatory potential, besides the $\Lambda_b^4 \cos{(\phi/f)}$, with a  much larger period, implying that $F \gg f$. The effective decay constant $F$ needs to be at least as large as the relaxion field excursion,  $F >  \Lambda/g'$, which in turn can have trans-Planckian field values. There are different frameworks to generate hierarchical decay constants such as multi-axion alignment 
 and mixing in the axion moduli space (see e.g. \cite{Kim:2004rp, Choi:2014rja}).  Therefore, the condition in Eq.\,(\ref{eq:fsubpl}) can be seen as a choice of simplicity, to avoid additional model building which would be necessary in order to embed a scenario with multiple effective super-Planckian  decay constants.

\item \textbf{No shift-symmetry restoration after reheating:}
After reheating, if the temperature is larger than the confinement scale at which the barrier term in the potential is generated, we expect that this term is erased by thermal fluctuations, and the relaxion starts rolling again. To avoid this possibility, we impose a lower bound on this scale assuming that this is not smaller than the minimal reheating temperature required for a successful Big Bang nucleosynthesis, $\sim 10\MeV$. We expect the scale $\Lambda_b$ to be related to the confinement scale by some small couplings. Thus, we impose
\begin{equation}
\Lambda_b > 1\MeV\,.
\label{eq:BBN}
\end{equation}
We should anyway note three points. First, the correct bound of \Eq{eq:BBN} is model dependent, and can be easily made weaker assuming small Yukawa couplings. Second, as we will see in the following, \Eq{eq:BBN} does not lead to important constraints on the parameter space, so that determining the actual limit has no relevant consequences on our study.
Third, if this condition is violated and the shift symmetry is restored after reheating, a very interesting scenario opens up, in which the relaxion starts rolling again and is stopped a second time when the Universe cools down and the barriers appear again. We will not discuss this scenario here for simplicity, but we refer the reader to Refs.~\cite{Choi:2016kke, Banerjee:2018xmn, Abel:2018fqg} in which this scenario is analysed and many consequences are discussed.

\end{itemize}

\subsection*{Parameter space}
%\label{sec:parameterspace}

The mechanism can be described in terms of 7 free quantities: 
\begin{equation}
g, g', \Lambda, \Lambda_b, f, H, \dot\phi_0 \, .
\label{eq:ourparameterspace}
\end{equation}
In addition, we define the quantity $\mphi$:
\begin{equation} \label{eq:massappr}
\mphi \equiv \frac{\Lambda_b^2}{f} \,,
\end{equation}
which is related to the relaxion mass in a way that depends on the actual realization of the mechanism, as we will detail in the next section. 
To simplify the problem, we will assume a fixed ratio $g/g'$, which we will take equal 1 unless otherwise specified. Moreover, we will relate $f$ and $\dot\phi_0$ to the other parameters using the fact that the final Higgs VEV should match the observed value, and choosing a sensible value for the field velocity. Thus, the parameter space has dimension 4, and can be characterized by $g'$, $\Lambda$, $\Lambda_b$ and one among $H$ or $f$. To constrain the parameter space we adopt the following logic. We will combine all the constraints in order to eliminate the variables $\Lambda_b$, $H$ or $f$, and derive all the equations that constrain the variables $g'$, $\Lambda$ only. In other words, this is equivalent to projecting the 4 dimensional hypersurface to the $g'$, $\Lambda$ plane. Then, we will present contours in this plane for the other free quantities, as well as the constraints  on the other variables for a few selected benchmarks.

%%%%%%%%%%%%%%%%%%%%%%%%%%%%%%%%%%%%%%%%%%%%%%%%%%%%%%%%%%%%%%%%%%%%%%%%%%%%

\subsection{Relaxation during inflation}
\label{sec:during inflation}

We first consider the case in which relaxation happens during inflation. This is the scenario proposed in~\cite{Graham:2015cka}, and the most studied in the literature (see Fig.\,\ref{fig:sketchinflation} in App.\,\ref{sec:cosmologicalhistory} for a sketch of the energy density of the universe during relaxation). 
We define the slow-roll velocity
\begin{equation}
\label{ref:slowrollvelocity}
\phidotSR \equiv \frac{g\Lambda^3}{3H},
\end{equation}
During inflation, the relaxion slow-rolls thanks to the large inflationary Hubble friction, which is dominated by some sector other than the relaxion. We thus assume
\begin{equation}\label{eq:i4}
H > \frac{\Lambda^2}{\sqrt{3} \MPl } \,.
\end{equation}
Moreover, the symmetry breaking pattern that leads to the Lagrangian \Eq{eq:relaxion potential} requires
\begin{equation}\label{eq:i2}
H < f,
\end{equation}
and
\begin{equation}\label{eq:i3}
H < \Lambda \,.
\end{equation}
Finally, the evolution should be dominated by the classical rolling of the relaxion field and not by the quantum fluctuations:
\begin{equation}
\label{eq:i1}
\frac{\phidotSR}{H} > \frac{H}{2\pi} \,.
\end{equation}

\begin{figure}
\centering
\begin{minipage}{.45\textwidth}
\includegraphics[width=\textwidth]{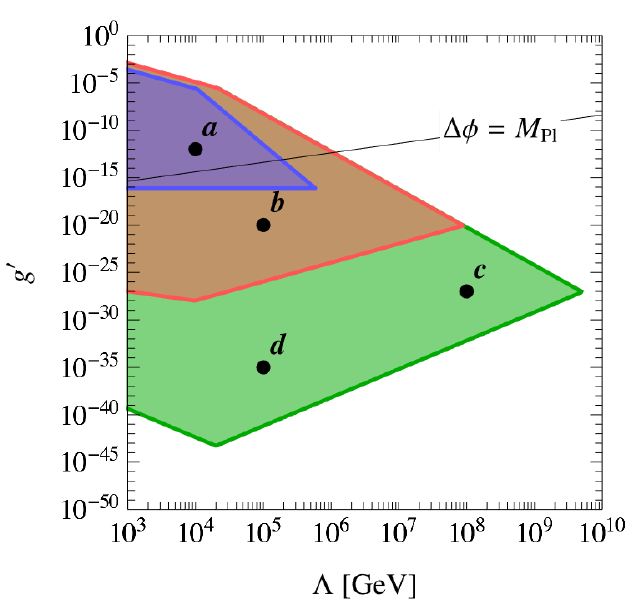}
\end{minipage}
\begin{minipage}{.45\textwidth}
\begin{itemize}
\item[{\tikz\fill[rounded corners = .5mm, blue](0,0)rectangle(0.3,0.2);}] Fragmentation
\item[{\tikz\fill[rounded corners = .5mm, red](0,0)rectangle(0.3,0.2);}] Large barriers
\item[{\tikz\fill[rounded corners = .5mm, Green](0,0)rectangle(0.3,0.2);}] Hubble friction (GKR)
\end{itemize}
\end{minipage}
\caption{\label{fig:summaryINFL}
Parameter space projected in the $\Lambda, g'$ plane of the relaxion mechanism 
 taking place with Higgs-dependent barriers, during inflation,  
where the  relaxion is a subdominant component of the energy density of the universe (non-QCD case of \cite{Graham:2015cka}), as described in Section \ref{sec:during inflation}.
Shown are 
 three distinct stopping mechanisms. `Hubble friction' corresponds to the mechanism discussed in \cite{Graham:2015cka}. Each of these regions is associated with {\it distinct} values of the inflationary scale and the relaxion mass, as illustrated in Fig.~\ref{fig:comparisonINFL}.  
 In other words, none of these regions overlap in the full parameter space.
 Benchmark point {\bf a} can be reached by all three stopping mechanisms. Benchmark point {\bf b} cannot be reached by fragmentation. Benchmark points {\bf c} and {\bf d}  can only occur through `Hubble friction'. In the region below the  line $\Delta \phi = \Lambda/g'= \MPl$, the field excursion  is super-Planckian.}
\end{figure}

\begin{figure}
\centering
\includegraphics[width=.32\textwidth]{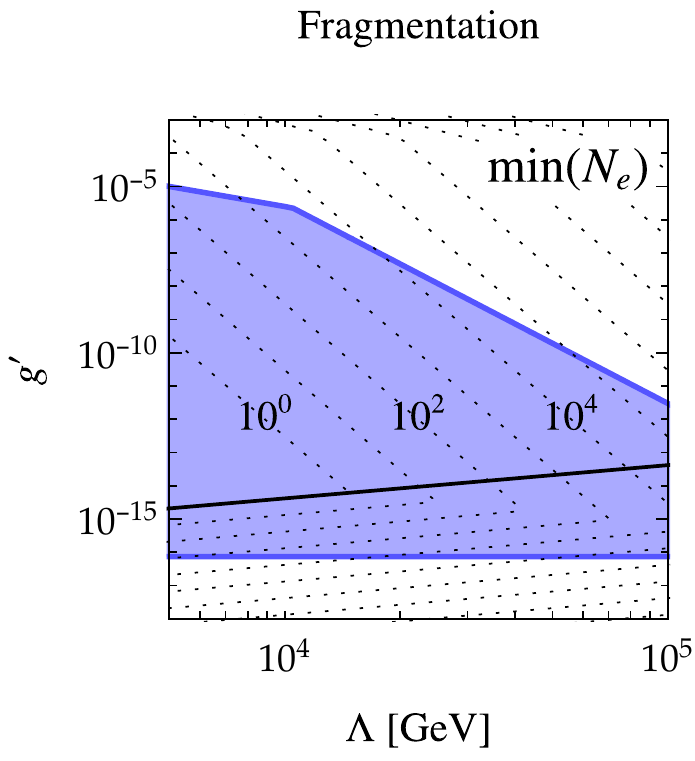}
\includegraphics[width=.32\textwidth]{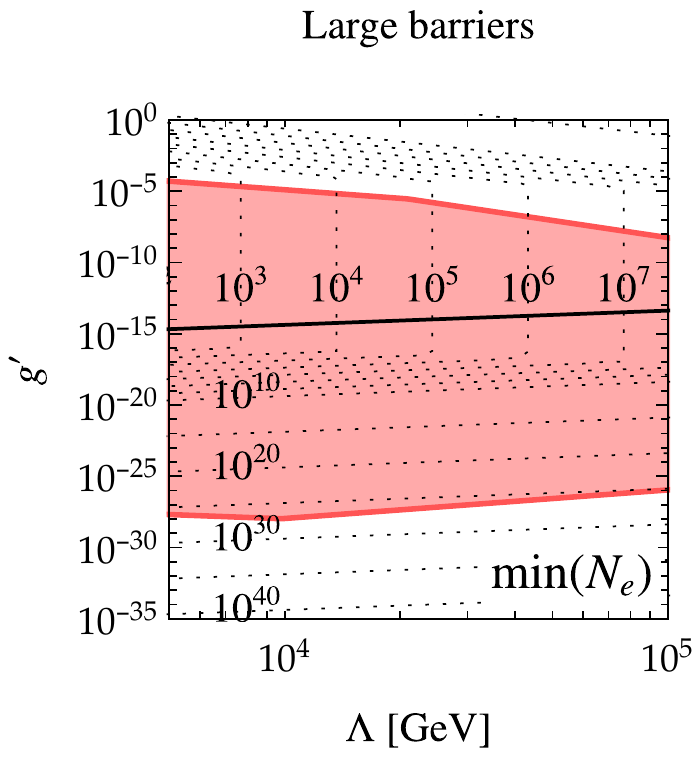}
\includegraphics[width=.32\textwidth]{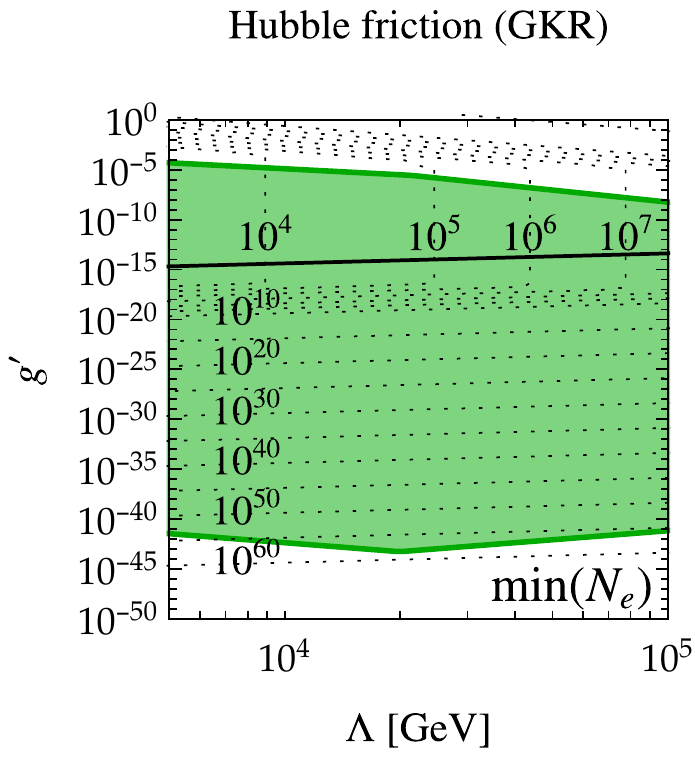}
\caption{\label{fig:efoldsduringinflation} Minimal number of efolds in the scenario where relaxation takes place during an inflation era for the three stopping mechanisms discussed in Sec.~\ref{sec:during inflation}. Same color code as in Fig.~\ref{fig:summaryINFL}. Below the solid line the field excursion $\Lambda/g'$ is super-Planckian. The contours of $\min(N_e)_\mathrm{frag}/\min(N_e)_\mathrm{GKR}$ are shown in Fig.~\ref{fig:efoldsduringinflationGKRfrag} in App.~\ref{sec:GKRcondition}.}
\end{figure}

\begin{figure}[!h]
\includegraphics[height=.32\textwidth]{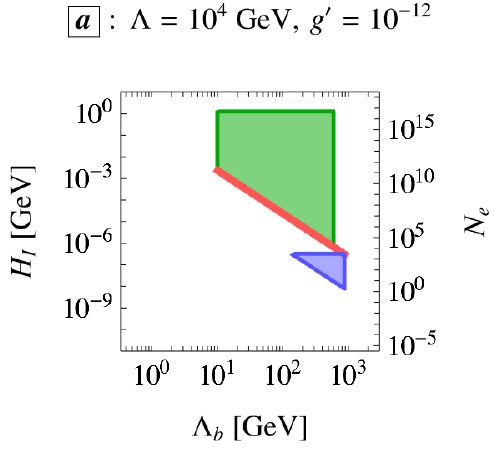}
\includegraphics[height=.32\textwidth]{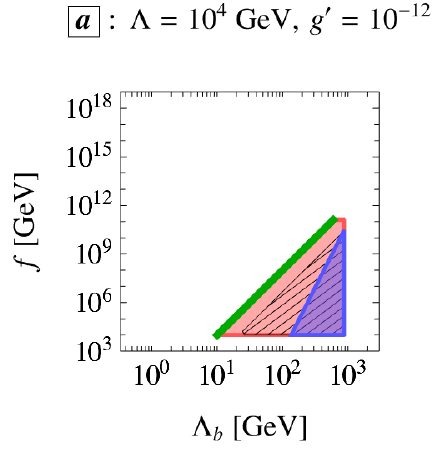}
\includegraphics[height=.32\textwidth]{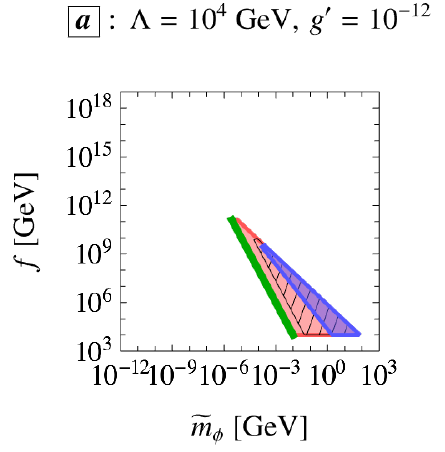}
\includegraphics[height=.32\textwidth]{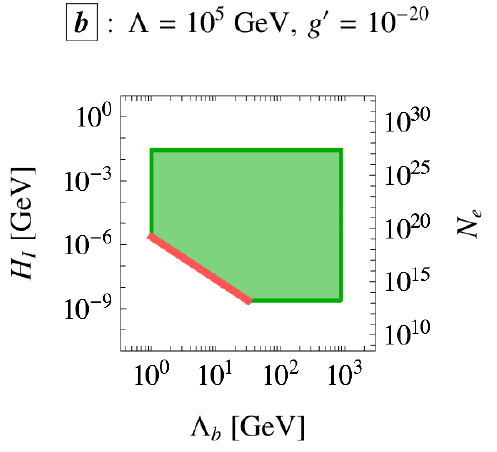}
\includegraphics[height=.32\textwidth]{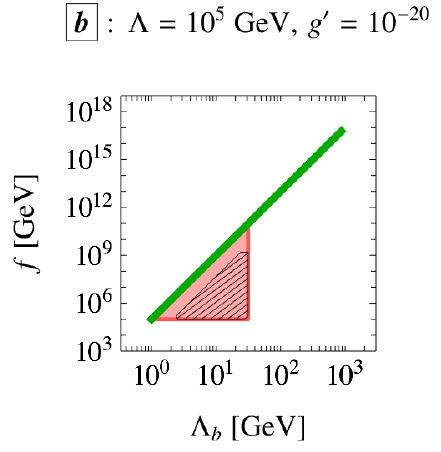}
\includegraphics[height=.32\textwidth]{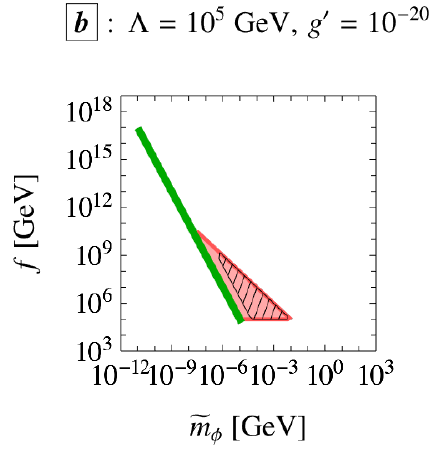}
\includegraphics[height=.32\textwidth]{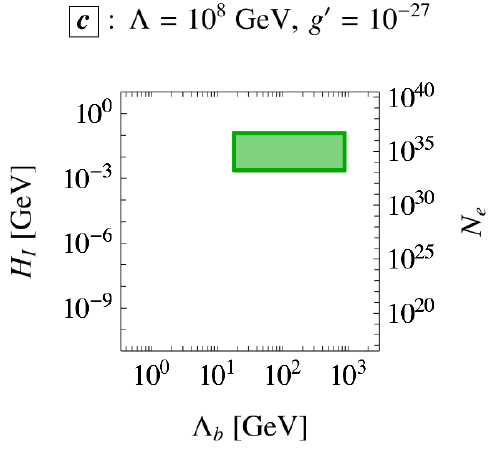}
\includegraphics[height=.32\textwidth]{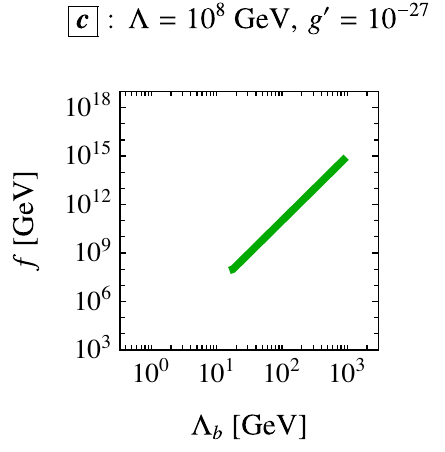}
\includegraphics[height=.32\textwidth]{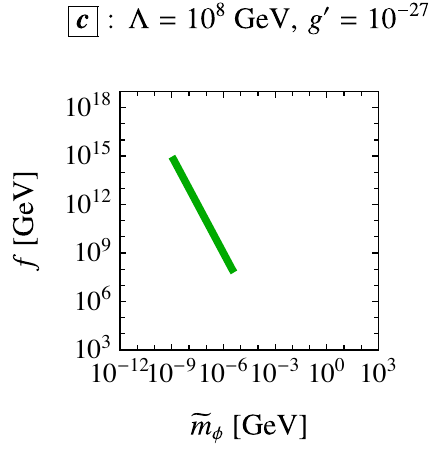}
\includegraphics[height=.32\textwidth]{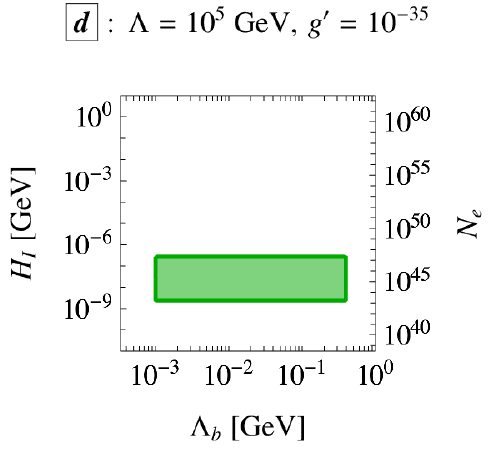}
\includegraphics[height=.32\textwidth]{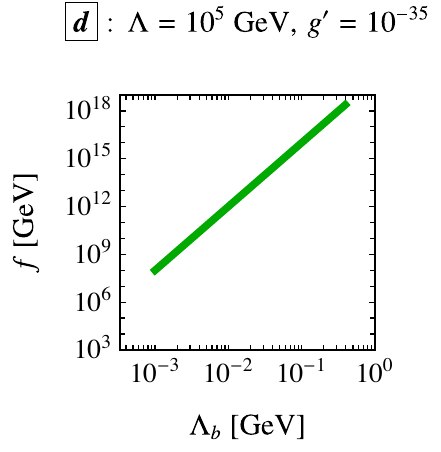}
\includegraphics[height=.32\textwidth]{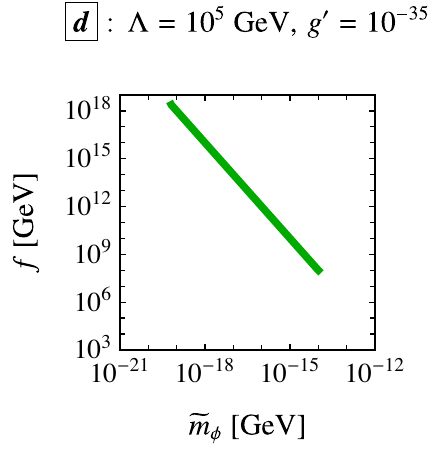}
\caption{\label{fig:comparisonINFL}
Parameter space  of the relaxion mechanism taking place with Higgs-dependent barriers, during inflation,   
where the  relaxion is a subdominant component of the energy density of the universe (non-QCD case of \cite{Graham:2015cka}), as described in Section \ref{sec:during inflation}. Same color code as in Fig.~\ref{fig:summaryINFL}.
The hashed part of the red region corresponds to the case in which, in the `large barriers' case, fragmentation is active and makes the relaxion stop in a shorter amount of time. Accurate predictions in this region are made harder by the action of two friction sources (see discussion below Eq.~(\ref{eq:Lambdab = velocity})).
For each benchmark point of Fig.~\ref{fig:summaryINFL} associated with the three stopping mechanisms defined in Section \ref{sec:during inflation}, we show the corresponding distinct regions in the ($\Lambda_b$, $H_I$ or $N_e$), ($\Lambda_b$, $f$), and ($\widetilde m_\phi$, $f$) planes. A similar plot in the plane ($\Lambda_b$, $M_I/\Lambda$) is displayed in Fig.~\ref{fig:comparisonINFL Lb Mi/L}.}
\end{figure}

After EW symmetry breaking, wiggles turn on in the relaxion potential and the relaxion stops as soon as the relaxion's kinetic energy is smaller than the potential barriers. 
Under the slow-roll assumption, one neglects the first term $\ddot{\phi} $ in \Eq{eq:zeromode}, and therefore, if the effect of quantum fluctuations is small, the relaxion stops as soon as $V'=0$, which requires sufficiently large barriers after EW symmetry breaking. This is the stopping condition used in ~\cite{Graham:2015cka}.
There is an underlying assumption behind this reasoning, which is that the time scale to roll over one wiggle  is much larger than a Hubble time.
As we discuss here, there are actually more stopping possibilities.
Depending on the strength of Hubble friction and on the velocity of the relaxion field, the relaxion can stop at three different times corresponding to three different stopping conditions and  three separate regions in parameter space, whose projection in the ($\Lambda,g'$) plane is shown in Fig.~\ref{fig:summaryINFL}. For each benchmark point $a$, $b$, $c$, $d$, the constraints in the ($\Lambda_b,H_I$), ($\Lambda_b,f$) and ($\mphi,f$) planes are shown in Fig.~\ref{fig:comparisonINFL}. 
We define these regions below:
\begin{enumerate}
\item \textit{Hubble friction (GKR):} If Hubble friction is strong, and in particular if 
 it takes longer than about a Hubble time for the relaxion to roll a distance  between two consecutive maxima of the potential, i.e $\Delta t_1 \gg H^{-1}$, where $\Delta t_1$ is
the time to cross one wiggle, then the slow-roll approximation is always valid. However, the velocity is not well approximated by the  \emph{average} slow-roll velocity  defined in 
(\ref{ref:slowrollvelocity})
but by the \emph{instantaneous} slow-roll velocity $$-V'/(3H).$$ Physically, this happens because Hubble friction has enough time to modify the field velocity as the slope of the potential varies. In this case, an arbitrarily-small barrier is sufficient to stop the relaxion (except for quantum fluctuations), in other words, the field stops as soon as  $V'=0$, \ie for $g \Lambda^3 \approx \Lambda_b^4/f$ (see Appendix~\ref{sec:GKR rolling} for more details).%
We label this case as \textit{Hubble friction} or  \textit{GKR} as this is the case discussed in~\cite{Graham:2015cka}.
It is characterized by
\begin{align}
\Delta t_1 & = \frac{2\pi f}{\phidotSR} > \frac{1}{H}\,, \label{eq:1wiggleslow} \\
f & = \frac{\Lambda_b^4}{g \Lambda^3}\,.
\end{align}
The parameter space consistent with this scenario  appears in green in Figs.~\ref{fig:summaryINFL}, ~\ref{fig:efoldsduringinflation} and~\ref{fig:comparisonINFL}. The precise origin of the limits of each region is depicted in Figs.~\ref{fig:gpLambdaINFLGKR} and \ref{fig:HILambdabINFLGKR} in App.\,\ref{sec:detailedplots}.
In the following, we will only be interested in the constraints that arise from a successful implementation of the relaxion mechanism itself. Additional constraints may be derived from colliders and other particle physics experiments, as well as cosmology and astrophysics. We refer the reader to Refs.~\cite{Kobayashi:2016bue, Choi:2016luu, Flacke:2016szy, Beauchesne:2017ukw, Frugiuele:2018coc}  for an overview of these constraints.

As noted in Ref.~\cite{Graham:2015cka}, quantum effects should be taken into account to define the stopping point of the relaxion. We can distinguish two regimes. Firstly, before the classical stopping point, the slope is small and the condition in Eq.~\ref{eq:i1} breaks down. Quantum fluctuations with a typical size $H/(2\pi)$ dominate at this point over the classical rolling of the zero mode. Secondly, after the classical stopping point, the potential has classically stable minima, which can be overcome by quantum tunnelling. This process is only stopped when the barriers grow large enough to make the lifetime of the relaxion's minima larger that the age of the Universe. These two regimes are studied, respectively, in Appendices~\ref{sec:quantum fluctuations} and \ref{sec:bounce}. We find that the region in which the classical description is not applicable is very small, thus the classical estimate of the stopping condition is reliable.
\item \textit{Large barriers:} If, on the other hand, friction is weak, \ie for $\Delta t_1 < H^{-1}$, the field no longer tracks the slow-roll velocity. After traversing a large number of wiggles, the field is stopped by friction when the size of the wiggles $\Lambda_b^2$ is larger than the average velocity $\phidotSR = g\Lambda^3/(3H)$, and wiggles dominate the potential slope, $\Lambda_b^4/f > g \Lambda^3$ (see Appendix~\ref{sec:GKR rolling}).
In this case we impose
\begin{align}
\Delta t_1 & = \frac{2\pi f}{\phidotSR} < \frac{1}{H}\,, \label{eq:1wigglefast} \\
\Lambda_b^4 & = \frac{1}{2}\phidotSR^2\,. \label{eq:Lambdab = velocity}
\end{align}
The choice $\dot\phi^2/2=\Lambda_b^4$ in the `large barrier' case is not a tuning in the initial conditions, but rather a consequence of our definition of $\Lambda_b$ as the amplitude of the barriers once the relaxion has stopped and the Higgs vev has reached its measured value.

The corresponding parameter space is shown in red in Figs.~\ref{fig:summaryINFL}, \ref{fig:efoldsduringinflation} and \ref{fig:comparisonINFL}. Constraints on this case are shown in Figs.~\ref{fig:gpLambdaINFLBARR} and~\ref{fig:fLambdabINFLBARR} in App.\,\ref{sec:detailedplots}.

Before proceeding to the last case, in which the relaxion is stopped purely due to the effect of fragmentation, we want to stress the role played by fragmentation in the `large barriers' scenario. If Eq.~(\ref{eq:condition 3}) is satisfied, fragmentation will also act as a secondary source of friction. In this case, the estimate $\dot\phi^2/2=\Lambda_b^4$ will be less reliable depending on the efficiency of fragmentation, even though it is clear that the relaxion evolution will be stopped. We highlight the corresponding region with black lines in Fig.~\ref{fig:comparisonINFL}. These regions interpolate between the `large barriers' scenario discussed here and the `relaxion fragmentation' one discussed below. The exact boundary is difficult to compute due to the presence of two effects. On the contrary, in the solid red regions of Fig.~\ref{fig:comparisonINFL}, Eq.~(\ref{eq:condition 3}) is violated, fragmentation is absent and Eq.~(\ref{eq:Lambdab = velocity}) gives the correct stopping condition.
\item \textit{Relaxion fragmentation:} The last case is the one in which we are most interested here. Again, we assume $\Lambda_b^4/f > g \Lambda^3$. If the velocity is large such that 
\begin{align}
\Delta t_1 & < H^{-1} \\
\frac{\phidotSR^2}{2} & > \Lambda_b^4 \,, \label{eq:large velocity}
\end{align}
the relaxion evolution can still be stopped by transferring its kinetic energy into the fluctuations.
For $\dot\phi_0 = \phidotSR$, fragmentation is efficient when \Eq{eq:condition 3} is satisfied. The field range needed to slow down the relaxion evolution is given by \Eq{eq:masterfieldexcursion}.
This must correspond to a final value of the Higgs VEV equal to $v_\ew$. Thus we can use \Eq{eq:masterfieldexcursion} to relate the scale $f$ to the other parameters $\{\Lambda, g, \Lambda_b, \dot\phi_0 \}$:
\begin{equation}\label{eq:fixingf}
\Delta v^2 = \frac{1}{\lambda}g'\Lambda \, \Delta\phi_{\rm frag} \Longrightarrow
f = \frac{2 \pi \lambda\, \Lambda_b^8\, \vEW^2}{ g' \Lambda \,\dot\phi_0^4} \left(\log{\left( \frac{32 \pi^2 f^4}{\dot{\phi}_0^2}\right)}\right)^{-1}\,.
\end{equation}
Notice that Eq.~(\ref{eq:masterfieldexcursion}) was obtained assuming a constant barriers' amplitude. The effect of growing barriers can not be included in our analytical treatment, but we expect it not to modify the time needed to stop the field by more than an $\mathcal{O}(1)$ factor, because the full efficiency of fragmentation is recovered as soon as the barriers grow close to their final value, thus rapidly stopping the field.
In the following, we will drop the logarithmic dependence and approximate $\log{( 32 \pi^2 f^4/\dot{\phi}_0^2}) \approx 50$.
The allowed parameter space for this case is shown blue in the ($\Lambda,g'$) plane in Fig.~\ref{fig:summaryINFL}, \ref{fig:efoldsduringinflation} and in other planes are shown in Fig.~\ref{fig:comparisonINFL} and in Figs.~\ref{fig:gpLambdafraginfl}-\ref{fig:gpLambdaNefoldsfraginfl} in App.\,\ref{sec:detailedplots}.

\end{enumerate}

Note that the  relaxion mass  can significantly differ from the value  naively obtained  when only considering the potential barrier, given by $\mphi$ defined in \Eq{eq:massappr}.  This is clear if we examine the `Hubble friction' stopping mechanism described previously, as in this scenario   $V'' \approx \Lambda_b^4/f^2 \sqrt{1 - (g \Lambda^3f/\Lambda_b^4)^2}$ with $\sqrt{1 - (g \Lambda^3f/\Lambda_b^4)^2} \ll1$.%
\footnote{We thank Hyungjin Kim for pointing  this out.}
On the other hand, for the other stopping mechanisms with large barrier's height, the mass is only  slightly different from $\mphi$, such that this is not a relevant modification for our discussion.   In the following, we simply consider the quantity $\mphi$ having in mind that \Eq{eq:massappr} is a good approximation of the relaxion mass for the `Large barriers' and `Relaxion Fragmentation' stopping mechanisms.

In Fig.~\ref{fig:efoldsduringinflation} and in first plot of Fig.~\ref{fig:comparisonINFL}, we see that stopping relaxation with fragmentation during inflation requires a much smaller number of efolds  $N_e$ than in the original proposal as well as a much smaller inflation scale. This means that relaxation could therefore happen during a second short  (i.e. $N_e\sim {\cal O}(10)$) late stage of inflation in our cosmological history, that is not necessarily responsible for the cosmological perturbations, but induced by some late supercooled phase transition for instance.
This brings a new theoretical framework for relaxation that is free from usual  inflation constraints and the associated model building. It is thus possible to solve the little hierarchy problem up to a cutoff scale ${\cal O}(20)$ TeV, from a minimal relaxion model,  during a short innocuous phase of inflation, as generic in strongly first-order confining phase transitions \cite{Konstandin:2011dr}. The $N_e$ contours in the full relevant parameter space in the case where the relaxion stops because of axion fragmentation are shown in Fig.~\ref{fig:gpLambdaNefoldsfraginfl}.

\newpage

\vspace{-1cm}

\clearpage

%%%%%%%%%%%%%%%%%%%%%%%%%%%%%%%%%%%%%%%%%%%%%%%%%%%%%%%%%%%%%%%%%%%%%%%%%%%%%%%

\subsection{Relaxation without inflation: self-stopping relaxion}
\label{sec:withoutinflation}

Depending on its efficiency, the growth of fluctuations can stop the evolution of the relaxion even if its initial velocity is very large. This opens up an interesting possibility: the relaxion can be stopped even without assuming slow-roll, and in fact without an inflationary background at all. This has important consequences, as the large number of e-folds required for relaxation during inflation is one of the main criticisms raised against this mechanism.

\begin{figure}
\centering
\begin{minipage}{.49\textwidth}
\includegraphics[width=\textwidth]{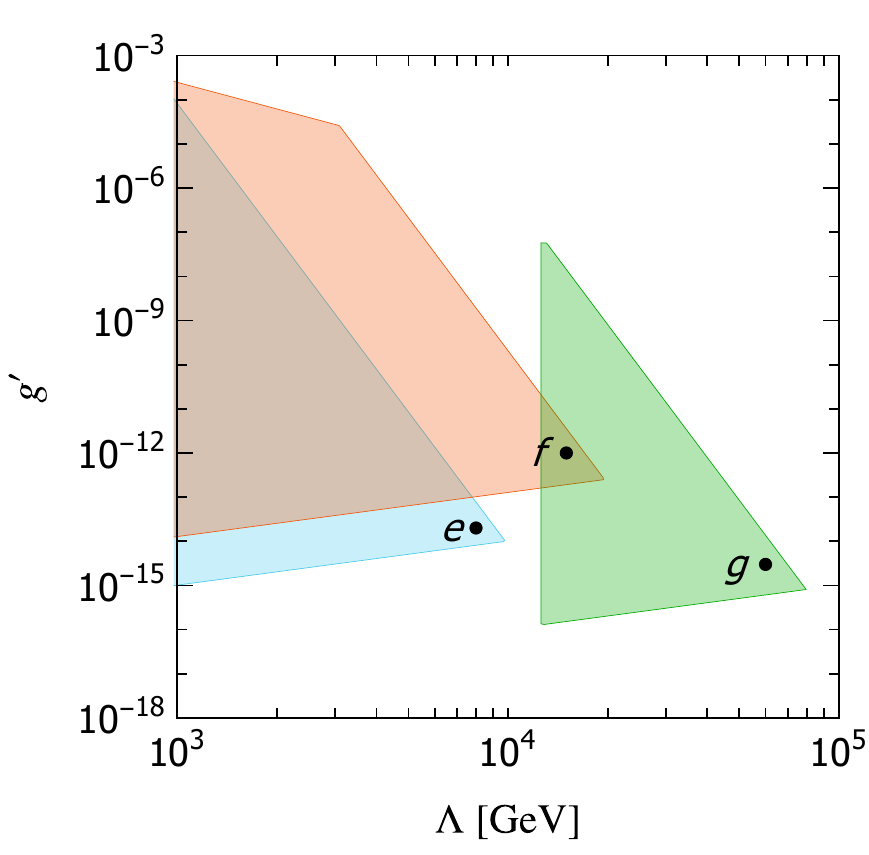}
\end{minipage}
\begin{minipage}{.49\textwidth}
\begin{itemize}
\item[{\tikz\fill[rounded corners = .5mm, colBlue4](0,0)rectangle(0.3,0.2);}] $\dot{\phi}= \sqrt{2\, g/g'}\, \Lambda^2, \,g/g'=1$
\item[{\tikz\fill[rounded corners = .5mm, colRed2](0,0)rectangle(0.3,0.2);}] $\dot{\phi}= \sqrt{2\, g/g'}\, \Lambda^2, \,g/g'=1/(4 \pi)^2$
\item[{\tikz\fill[rounded corners = .5mm, Green](0,0)rectangle(0.3,0.2);}] $\dot{\phi}= 10^{-2}\sqrt{2}\, \Lambda^2$
\end{itemize}
\end{minipage}
\caption{\label{fig:gpLambdaNOINFLATION} Parameter space in the plane $g', \Lambda$ for the scenarios in which relaxation happens \emph{without} inflation.}
\end{figure}

In this subsection, we will discuss how to realize the relaxion mechanism after or before inflation. The relaxion may or may not dominate the energy density of the Universe. The Hubble rate $H$ can be eliminated from the equations, the only condition is
\begin{equation}\label{eq:relaxion hubble}
H \geq \frac{\Lambda^2}{\sqrt{3} \MPl}
\end{equation}
All our results presented in Fig.\,\ref{fig:gpLambdaNOINFLATION} and Figs.\,\ref{fig:gpLambNOInflationApp}, \ref{fig:gpLambContoursApp}, and \ref{fig:contoursMassApp} in App.\,\ref{sec:detailedplots},  do not depend on whether the relaxion dominates or not the energy density of the universe at the time of relaxation. However,  as we will discuss in subsection \ref{subsub:cosmohisto}, the relic abundance of the relaxion  crucially depends on this assumption.
Moreover, we assume that the relaxion does not slow-roll. In this case, it is reasonable to expect that the velocity is of order of the cut-off of the theory. Indeed, if the field is free to roll down a potential with slope $-g\Lambda^3 \phi$ for a field range $\Delta\phi = \Lambda/g'$, starting from rest its final velocity will be
\begin{equation}\label{eq:relaxion velocity}
\dot\phi_0 = \sqrt{\frac{2 \, g}{g'}}\Lambda^2.
\end{equation}
Such an estimate is reliable as long as this value is smaller than the slow roll velocity, otherwise the field would be slowed down by Hubble friction. Therefore, we also impose that
\begin{equation}\label{eq:no slow-roll}
\dot\phi_0 < \phidotSR = \frac{g\Lambda^3}{3 H} \,.
\end{equation}
Provided $\dot H\ll H^2$, the case in which $\dot\phi_0 = \phidotSR$ with the relaxion dominating the energy density would lead to an inflationary period driven by the relaxion, which we will discuss in Sec.~(\ref{sec:relaxioninflating}) below.
As we discussed in Sec.\,\ref{sec:summaryfragmentation}, in the case $\dot\phi_0<\phidotSR$, Hubble friction can be neglected  and fragmentation is effective if \Eq{eq:slope bound (H=0)} is satisfied. In this case, the relaxion stops its evolution after a field range $\Delta \phi_{\textrm{frag}}$, given in \Eq{eq:masterfieldexcursion}. This must correspond to a final value of the Higgs VEV equal to $v_\ew$. Therefore, as in the previous section, we can use \Eq{eq:masterfieldexcursion} to relate the scale $f$ to the other parameters $\{\Lambda, g, \Lambda_b, \dot\phi_0 \}$, to obtain Eq.~(\ref{eq:fixingf}).
In the following, we will approximate $\log{( 32 \pi^2 f^4/\dot{\phi}_0^2}) \approx 50$ and drop the logarithmic dependence. The difference compared to Sec.\,\ref{sec:during inflation} is that here  $\dot \phi_0$ is not the slow-roll velocity. 
The parameter space consistent with this scenario for three different velocities is presented in Fig.\,\ref{fig:gpLambdaNOINFLATION}. For each  of these cases we consider one benchmark:
\begin{equation}
\begin{aligned}
\label{eq:benchmarksNoInflation}
\text{Benchmark \textit{\textbf{e}}:} \qquad & \Lambda = 8 \TeV, \, &g'&=2\times10^{-14}, \\
\text{Benchmark \textit{\textbf{f}}:} \qquad & \Lambda = 15 \TeV, \, &g'&= 10^{-12},\\
\text{Benchmark \textit{\textbf{g}}:} \qquad & \Lambda = 60 \TeV, \, &g'&= 3 \times 10^{-15}.
\end{aligned}
\end{equation}
In the appendix \ref{sec:detailedplots} we explore in detail the parameter space for the benchmarks in (\ref{eq:benchmarksNoInflation}). In  Fig.\,\ref{fig:gpLambNOInflationApp} we specify the constraints on the plane $g', \Lambda$ and  in Fig.\,\ref{fig:gpLambContoursApp} we show the contours for the minimal allowed value of the scale $\Lambda_b$ and maximal value of the decay constant $f$.  The maximal     $\Lambda_{b}$ and  minimum $f$  saturate  to the extreme allowed values for all the three different velocities, such that   $\Lambda_{b, \textrm{max}}= \sqrt{4 \pi} \,\vEW$ (\Eq{eq:stability})  and  $f_{\textrm{min}}= \Lambda$ (\Eq{eq:eftvalidity}).  The allowed ranges of $\mphi$ in \Eq{eq:massappr} for the benchmarks above are:
\begin{align} \nonumber
\text{Benchmark \textit{\textbf{e}}:} \qquad & \mphi \in [139, \, 169\GeV] \\ \nonumber
\text{Benchmark  \textit{\textbf{f}}:} \qquad & \mphi \in [14, \, 37 \GeV] \\ \label{eq:mass range}
\text{Benchmark  \textit{\textbf{g}}:} \qquad & \mphi \in [3, \, 9\GeV] 
\end{align}
The contours of $\mphi$ for the allowed range of $f$ and $\Lambda_b$ are shown in Fig.\,\ref{fig:contoursMassApp} in App.\,(\ref{sec:detailedplots}).

The summary is that the relaxion can stop itself without the need for inflation. We find that for natural values of the initial velocity $\dot{\phi}_0\sim {\cal O}(\Lambda^2)$, this can be achieved for a cutoff of $\cal{O}$(10) TeV. The fragmentation stopping mechanism can in principle be effective also for much higher cutoff, $\cal{O}$(100) TeV, however, this would require  tuning the initial velocity. In the corresponding parameter space, the relaxion mass is rather large, from a few GeV up to the EW scale. 
For the largest cutoff, the value of $f$ is $\sim 10-100$ TeV.

%\newpage 
\subsubsection{Cosmological History and relaxion abundance}
\label{subsub:cosmohisto}

In this subsection we discuss the cosmological consequences in the scenarios in which relaxation is realized without inflation, which can happen before or after an inflationary period, see also Appendix \ref{sec:cosmologicalhistory}.
In the case the relaxation dynamics happens before inflation, the imprints from
this period are  diluted away due to the expansion. This implies  that the details about how the equation of state of the universe evolves during relaxation are not important for the dynamics after  inflation.
In the scenario in which the scanning of the Higgs mass parameter  is realized after inflation, we have to investigate the relaxion relic abundance.  The scanning  ends when  most of the relaxion's kinetic energy is transferred to relaxion fluctuations through fragmentation, which then red-shifts as radiation. Assuming that the dominant contribution to the energy density in the fluctuations comes from the mode with $k_{\textrm{cr}}\sim \dot{\phi}_0/f$ (see Sec.~\ref{sec:summaryfragmentation}), we can estimate the  relaxion number density just after particle production as
\begin{equation} \label{eq:number density phi}
n_{\phi} \sim \dot{\phi}_0^2 \, \left(\frac{\dot{\phi}_0}{f}\right)^{-1},
\end{equation}
where we used that the energy density of the relaxion field is $\sim \dot{\phi}_0^2$ and that the typical energy carried by a particle is $\mathcal{O}(k_{\textrm{cr}})$. Therefore, using $m_\phi \approx \mphi = \Lambda_b^2/f$, the relaxion energy  density in the fluctuations is 
\begin{equation} \label{eq:energy density phi}
\rho_{\phi,0}  \sim \frac{g_{s0} T_0^3}{g_{s*} T_*^3} \Lambda_b^2 \dot\phi_0 \,.
\end{equation}
Here $T_0 = 2.7~{\rm K}$ is the current CMB temperature and $g_{s0} = 43/11$ is the effective degrees of freedom which contribute to entropy density.
$T_*$ and $g_{s*}$ are the temperature and the effective degrees of freedom of the thermal plasma after the relaxation.
In deriving Eq.~(\ref{eq:energy density phi}) we used the fact that, today, the relaxion particles are non-relativistic. Indeed, their initial momentum is peaked around $k_\mathrm{cr}\sim \dot\phi/f \sim \Lambda^2/f < \Lambda$. Since relaxation must take place before BBN, this momentum will be redshifted at least by a factor $T_0/T_\mathrm{BBN}\sim 1\,\mathrm{K}/10\MeV\sim 10^{-11}$, making the relaxion non-relativistic in our scenario.
Note that once the energy density has red-shifted enough, relaxion    coherent oscillations  will  contribute to the total energy density as matter, with $\rho_{\phi}^{\textrm{osc}} \sim (\Lambda_b^4/f^2) f^2$.   We will neglect this contribution since $\rho_{\phi}^{\textrm{osc}}$  is very suppressed compared (\ref{eq:energy density phi}).
Therefore, by assuming $\dot\phi_0 \sim \Lambda^2$,  one can estimate the relaxion abundance as 
\begin{equation} \label{eq:relaxion abundance}
\Omega_\phi h^2 \sim 6 \times 10^4 \left(\frac{\Lambda_b}{10 \GeV}\right)^2 \left(\frac{\Lambda}{10 \TeV}\right)^2 \left( \frac{g_{s*}}{100} \right)^{-1} \left(\frac{T_*}{10 \TeV}\right)^{-3}.
\end{equation}
The result in  Eq.\,(\ref{eq:relaxion abundance}) shows that, if the relaxion dominates the energy density of the universe at the time of relaxation and in the absence of a dilution mechanism, the  relaxion is overabundant in the allowed parameter space. Under these assumptions,  in the scenarios where the Higgs mass scanning happens after inflation, the  relaxion has to be unstable to avoid overclosing the universe. The relaxion can decay to the Standard Model  particles via its  mixing with the Higgs. The mixing angle can be computed from Eq.~(\ref{eq:barrierscaling}) as
\begin{equation}\label{eq:mixing angle}
\sin 2\theta_{\phi,h} \approx \frac{4 g \Lambda^3}{\sqrt{(m_h^2-\cos\theta_0 \, \mphi^2)^2v_\ew^2 + 16 g^2 \Lambda^6}} \,,
\end{equation}
where $\theta_0 = \phi_\mathrm{min}/f$ is the position of the minimum at which the $\phi$ field stops, and
\begin{equation}
\sin \theta_0 = -\frac{f}{\Lambda_b^4}(g\Lambda^3 - \frac{1}{2}g' \Lambda v_\ew^2) \,.
\end{equation}
In the case under study, in which the field stops due to fragmentation, we assume $\Lambda_b^4/f\gg g\Lambda^3$, thus we can take $\cos\theta_0\approx 1$ in Eq.~(\ref{eq:mixing angle}). In the `GKR' scenario, instead, $\Lambda_b^4/f\approx g\Lambda^3$, and $\sin\theta_0\approx 1$.%
\footnote{For $\Lambda_b^4/f\approx g\Lambda^3$, and assuming $\mphi\ll m_h$, $g=g'$ and $g \Lambda^3\ll v_\ew^3$, the mixing angle can be approximated as
\[
\sin \theta_{\phi,h} \approx \frac{2 \Lambda_b^4}{f m_h^2 v_\ew} \sin\theta_0 \,,
\]
which matches the expressions given in Refs.~\cite{Choi:2016luu, Flacke:2016szy}. }

We checked that the mixing angle $\theta_{\phi,h}$ is large enough to make the relaxion decay before Big Bang Nucleosynthesis (BBN) in  all the three benchmarks (e), (f), and (g) of Eq.~(\ref{eq:benchmarksNoInflation}). However, this is not the case in a small portion of parameter space with low cutoff $\Lambda$ and small coupling $g'$. In this case, the lifetime can be shortened by adding a coupling to the Standard Model photon.

A coupling of the relaxion to the Standard Model photon may be added in general, irrespectively of the lifetime of the relaxion particle. The Lagrangian is complemented by a term
\begin{equation} \label{eq:phiFFdual}
\mathcal{L} \supset \frac{\phi}{4f_\gamma} F \widetilde{F}\,.
\end{equation}
We now look at the implications of such coupling.
The term in Eq.~(\ref{eq:phiFFdual}) turns out to be the dominant decay channel of $\phi$ into Standard Model particles. This sets  an upper bound on $f_\gamma$, allowing relaxion lifetime to be shorter than  $\tau_\phi < 1 \sec$ for
\begin{equation}\label{eq:bound fgamma}
f_\gamma < 10^{12} \left(\frac{m_\phi}{1 \GeV}\right)^{3/2} \GeV.
\end{equation}
In general, such a term will receive a contribution from the relaxion-Higgs mixing. Such a contribution will be suppressed by both the mixing angle $\theta_{\phi,h}$ and the SM Jarlskog invariant $J\sim10^{-5}$, which appears when the CP-violating $h F \widetilde F$ term is generated at the loop level.

Since the relaxion mass is typically larger than $\mathcal{O}(1) \GeV$ (see \Eq{eq:mass range}), our scenario evades astrophysical and experimental constraints, see e.g. \cite{Choi:2016luu, Bauer:2017ris, Craig:2018kne, Bauer:2018uxu}. The relaxion will also contribute to the electric-dipole moment (EDM) of light SM fermions, through its coupling to photons of Eq.~(\ref{eq:phiFFdual}) and its mixing with the Higgs (see e.g. \cite{Choi:2016luu}). Existing bounds on fermions EDMs do not constrain our model further, due to the smallness of the mixing angle and the large allowed value of $f_\gamma$.

Once the relaxion rolls down its potential, photons can be produced via the coupling in Eq.\,(\ref{eq:phiFFdual}) due to a tachyonic instability in the equation of motion of the photon $A_\mu$ (see e.g. \cite{Anber:2009ua}). These photons produce a thermal bath which modifies the dispersion relation of the $A_\mu$ vector, making  particle production less efficient. Taking this effect into account, the timescale for photon production is estimated to be $\Delta t_\gamma \sim~ T^2 f_\gamma^3/(\dot{\phi}^3)$ \cite{Bellac:2011kqa}. For simplicity,  we want to assure that the production of photons during the relaxation dynamics  does not interfere in our stopping mechanism.  To this end,  the timescale  for photon production should be longer than a Hubble time
\begin{equation} \label{eq:no photons}
\Delta t_{\gamma} > H^{-1},
\end{equation}
which is equivalent to require that particle creation is slow compared to the dilution given by cosmic expansion.
From Eq.\,(\ref{eq:no photons}) we can derive a lower bound on the scale $f_\gamma$ using the fact that relaxion does not slow roll (Eq.\,(\ref{eq:no slow-roll})) and that $\dot{\phi}\sim \Lambda^2$, implying  $f_\gamma \gtrsim \Lambda/g^{1/3}$. In the cases in Fig.\,\ref{fig:afterinflation relaxion dom} in App.\,\ref{sec:cosmologicalhistory}, $\phi$ dominates the energy density such that $H\sim \Lambda^2/(\sqrt{3} \MPl)$.  Therefore, the condition (\ref{eq:no photons}) simply translates to   $f_\gamma \gtrsim (\Lambda^2 \MPl)^{1/3}$.  

One should notice that the coupling in (\ref{eq:phiFFdual}) introduces  another portal between the relaxion sector and the SM besides the Higgs mixing. If the temperature of the thermal plasma, at the end of relaxation, is larger than the confinement scale of the non-abelian gauge group that gives rise to the cosine potential, then the barriers can disappear allowing the relaxion to roll down once again.  This additional $\phi$  displacement  may ruin the mechanism as the field can move away from the correct value of the EW scale. On the other hand, such situation can be avoided if the sector that generates the period potential does not enter in equilibrium with the Standard Model bath. To this end, we can impose that the interaction rate of the strong sector with the photons $\Gamma \sim T^5/(f^2 f^2_\gamma)$ is small compared to the Hubble expansion, resulting in another condition the scale $f_\gamma$ should fulfill, $f_\gamma\gtrsim (\Lambda^3 \MPl)^{1/2}/f$.

We stress again that all the above discussion is relevant if $H=\Lambda^2/\sqrt{3}\MPl$ when relaxation starts. If on the other hand the hidden sector radiation dominates the energy density of the universe during relaxation, the relaxion abundance can be accordingly diluted and subdominant.

%\newpage ~ \newpage

\subsection{Relaxion inflating the Universe}
\label{sec:relaxioninflating}

As a last possibility, we consider the one in which the relaxion dominates the energy density (as most of the cases above) but, instead of \Eq{eq:relaxion velocity}, we assume that the relaxion slow rolls, thus driving a period of inflation
\begin{equation}
\dot\phi_0 = \phidotSR = \frac{g\Lambda^3}{3 H} \,, \qquad \textrm{with} \quad H = \frac{\Lambda^2}{\sqrt{3}\MPl}.
\end{equation}
Note that to be consistent with the effective field  approach, $\phidotSR$ has to satisfy
\begin{equation} \label{eq:phisrconsistent}
\frac{1}{2}\phidotSR^2 < \Lambda^4.
\end{equation}

\begin{figure}
\centering
\includegraphics[width=.45\textwidth]{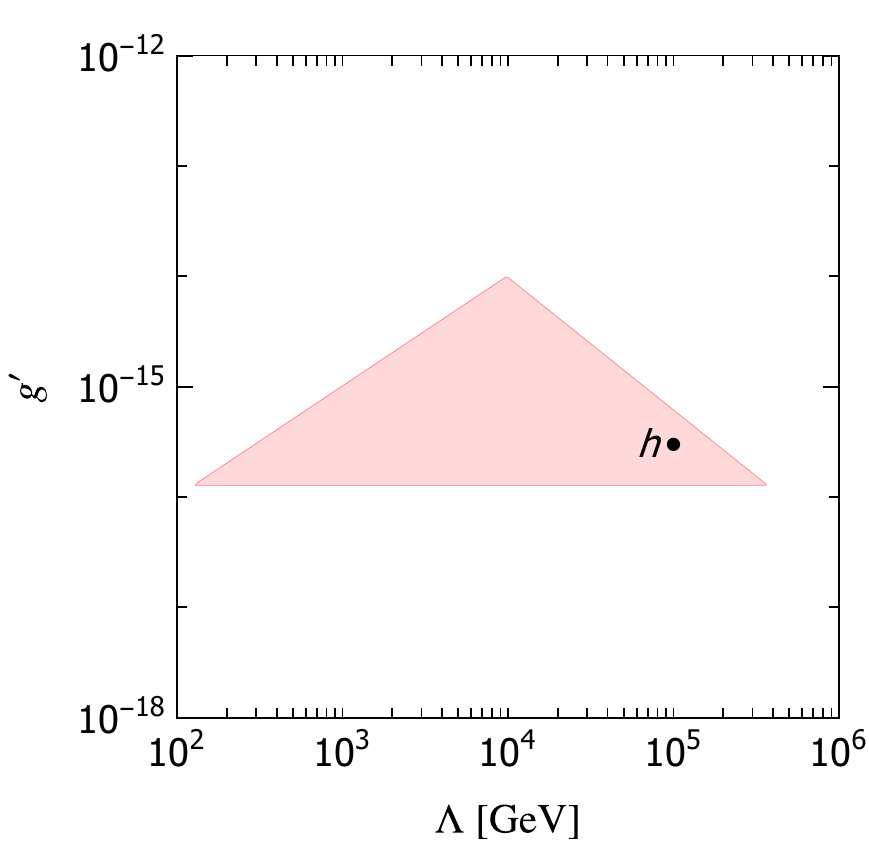}
\caption{\label{fig:gpLamb-relaxion dominates} Allowed parameter space for self-stopping relaxion in the case where the relaxion dominates the energy density of the universe and  slow-rolls during relaxation.}
\end{figure}

In \Fig{fig:duringinflation rel dom}
 in App.\,\ref{sec:cosmologicalhistory} we illustrate an example of such a case and the parameter space consistent with this scenario in the $g',\Lambda$ plane is depicted in \Fig{fig:gpLamb-relaxion dominates}, where we assume $g=g'$. The benchmark point corresponds to
\begin{equation}
\begin{aligned}
\label{eq:benchmarkDrivesInflation}
\text{Benchmark \textit{\textbf{h}}:} \qquad & \Lambda = 100 \TeV, \, &g'&=3\times10^{-16}.
\end{aligned}
\end{equation}
The allowed $\mphi$ range for such benchmark is given by
\begin{align}  \label{eq:mass range dominates}
\text{Benchmark \textit{\textbf{h}}:} \qquad & \mphi \in [0.2, \, 3\GeV].
\end{align}
We refer to Appendix \ref{subsec:relaxioninflaton} for more details about the  parameter space.
 
Comparing the parameter space in \Fig{fig:gpLamb-relaxion dominates} with the case without inflation in \Fig{fig:gpLambdaNOINFLATION}, we see that this case allows for a larger cutoff, while constraining much more the range of the coupling $g'$. 
This scenario has similarities with the third case {\it relaxion fragmentation} discussed in section \ref{sec:during inflation}.

A scenario in which the relaxion is the inflaton was discussed in Ref.\,\cite{Tangarife:2017rgl}. At least two non-trivial additions were required.
First, in~\cite{Tangarife:2017rgl}, a Chern-Simons coupling between the relaxion and a dark photon is included. The dark photons carry very low momentum at the end of inflation such that the particles cannot be thermalized through perturbative scatterings. On the other hand, such low momentum  photons have large occupation number  generating  a strong electromagnetic field, which then  allows for  the vacuum production of electron-positron pairs  via the Schwinger effect. In order to reheat the universe at temperatures above the electron mass using the mechanism in \cite{Tangarife:2017rgl}, it is crucial  to couple the relaxion to a dark photon, which in turn has a  kinetic mixing with the Standard Model photon.
Second, the curvature perturbations in the simplest model are suppressed. Therefore,  the relaxion only addresses the horizon and flatness problems but not the origin of cosmological perturbations and the addition of a curvaton field is necessary. In our case, we simply assume that the relaxion-driven inflation precedes the inflationary period that ends with the Big Bang Nucleosynthesis and originates the cosmic microwave background curvature perturbations. As shown in Fig.~\ref{fig:gpLambINFrelaxiondominates}
the corresponding number of efolds varies from ${\cal O}(1)$ to ${\cal O}(10^5)$ depending on the cutoff scale $\Lambda$ and coupling $g'$. Alternatively, if the number of efolds is smaller than $\mathcal{O}(10)$, the period of relaxion-driven inflation can take place after reheating.

%%%%%%%%%%%%%%%%%%%%%%%%%%%%%%%%%%%%%%%%%%%%%%%%%%%%%%%%%%%%%%%%%%%%%%%%%%%%%%%%%%%
%%%%%%%%%%%%%%%%%%%%%%%%%%%%%%%%%%%%%%%%%%%%%%%%%%%%%%%%%%%%%%%%%%%%%%%%%%%%%%%%%%%

%\clearpage
%\newpage
\section{Consequences II: relaxation with vector boson production}
\label{sec:HMT}

We now examine another class of relaxion models which has triggered interest. The stopping mechanism is qualitatively different as the potential barriers are Higgs-independent and relaxation starts in the EW broken phase.
The idea of using the tachyonic production of Standard Model gauge boson as a source of friction to slow down and stop the relaxion evolution was introduced in~\cite{Hook:2016mqo}, and further studied in various aspects in~\cite{Son:2018avk, Fonseca:2018xzp, Craig:2018kne, Fonseca:2018kqf}. Here we will only summarize the model and its constraints, referring the reader to~\cite{Hook:2016mqo, Fonseca:2018xzp} for further details.
As we will see, relaxion fragmentation poses a serious threat to this scenario: Instead of being active only close to the critical point, thus helping the stopping of the relaxion, the growth of perturbations happens at all times, and can stop the evolution at a position such that the EW scale is closer to the cut-off of the theory. In order to avoid this possibility we will impose that the fragmentation does not take place. This requirement excludes the setup in which the relaxation happens after inflation, which was studied in details in~\cite{Fonseca:2018xzp}. On the contrary, relaxation during inflation is still allowed, and the relaxion dark matter model of~\cite{Fonseca:2018kqf} is not affected.

The model can be characterized by the following Lagrangian:
\begin{equation}
\mathcal{L} = \mathcal{L}_\mathrm{SM} + \frac{1}{2}\partial_\mu \phi \partial^\mu \phi - \frac{\phi}{4\mathcal{F}} \left(g_2^2 W^a_{\mu\nu}\widetilde{W}^{a\,\mu\nu} - g_1^2 B_{\mu\nu}\widetilde{B}^{\mu\nu} \right) - V(\phi, h)
\label{eq:lagrangian vectors}
\end{equation}
with
\begin{equation}
V(\phi, h) = \Lambda^4 - g \Lambda^3 \phi +\frac{1}{2} \left(-\Lambda^2 + g'\Lambda \phi\right)h^2+ \frac{\lambda}{4} h^4 + \Lambda_b^4\cos\left(\frac{\phi}{f}\right)\,,
\label{eq:relaxion_potential}
\end{equation}
where $g_1$ and $g_2$ are the $\textrm{U}(1)$ and $\textrm{SU}(2)$ coupling constants, respectively. The coupling to gauge bosons above forbids the coupling of the relaxion to the photon's $F_{\mu\nu} \widetilde F^{\mu\nu}$ term at tree level.
The initial conditions are such that, initially, $\left(-\Lambda^2 + g'\Lambda \phi\right)<0$, thus the Higgs has a large VEV of order $\Lambda$. The barriers are now independent on the Higgs VEV.

In the EW broken phase, the relevant part of \Eq{eq:lagrangian vectors} can be rewritten in terms of the mass eigenstates $A_\mu$, $Z_\mu, W^\pm_\mu$
\begin{align}\label{eq:lagbroken}
\mathcal{L} \supset\, & m_W^2(h)W^-_\mu W^{+\,\mu} + \frac{1}{2}m_Z^2(h)Z_\mu Z^\mu \nonumber \\
& - \frac{\phi}{\mathcal{F}} \epsilon^{\mu\nu\rho\sigma} \left(
2 g_2^2 \partial_\mu W^-_\nu \partial_\rho W^+_\sigma +
(g_2^2-g_1^2) \partial_\mu Z_\nu \partial_\rho Z_\sigma
+ 2g_1 g_2 \partial_\mu Z_\nu \partial_\rho A_\sigma
\right)\,,
\end{align}
where the masses of the gauge bosons are $m_W(h) = g_2 h/2$ and $m_Z(h) = \sqrt{g_2^2+g_1^2}h/2$.
We will neglect the contribution of the $WW$ and of the $ZA$ terms, and concentrate on the $Z$ dependent one.
The coupling to photons $\phi F_{\mu\nu} \widetilde F^{\mu\nu}$ is generated through the small mixing with the Higgs, controlled by the small parameter $g'$, and also from a $W$ loop through the interaction in \Eq{eq:lagbroken}~\cite{Bauer:2017ris, Craig:2018kne}. We will discuss below how this coupling is harmful for the model, and what are the conditions to make it ineffective.

Absorbing the gauge couplings in the definition of the scale $\fZZ$,
\begin{equation} \label{eq:f convention}
\frac{1}{\fZZ} = \frac{(g_2^2 -g_1^2)}{\mathcal{F}},
\end{equation}
the equations of motion for the relaxion zero mode and the transverse modes of the $Z$ field are:
\begin{align}
\ddot\phi - g\Lambda^3 + g'\Lambda h^2 + \frac{\Lambda_b^4}{\fcos}\sin\frac{\phi}{\fcos} + \frac{1}{4\fZZ}\langle Z\widetilde{Z} \rangle &= 0 \,,\label{eq:phieomvector} \\
\ddot{Z}_\pm +(k^2 + m_Z^2 \mp k\frac{\dot\phi}{\fZZ})Z_\pm &= 0 \,.\label{eq:Aeomvector}
\end{align}
When the Higgs VEV (and consequently the mass of the gauge bosons) decreases, \Eq{eq:Aeomvector} exhibits a tachyonic instability for the $Z_+$ polarization (assuming $\dot\phi>0$), which starts as soon as $m_Z < \dot\phi/(2 \fZZ)$. After this point (which we call $t_c$), the kinetic energy of the relaxion's zero mode is converted into the helical $Z$ field in a timescale $\sim m_Z^{-1}$.
The parameters of the model must be chosen in such a way that, at $t_c$, the relaxion field is as close as possible to the critical value $\Lambda/g'$, thus generating the hierarchy between the cutoff $\Lambda$ and the EW scale.%
\footnote{An important caveat to this discussion would come from the inclusion of SM fermions. In the background of strong hypercharge fields, SM fermions are copiously produced, and backreact on the gauge fields themself~\cite{Domcke:2018eki}. As a result, the amplification of the $Z$ and of the photon fields can be suppressed, depending on the coupling of the axion field to gauge bosons, the velocity of the field and the Hubble rate. This effect would possibly change the parameter space discussed here and in Refs.~\cite{Hook:2016mqo, Fonseca:2018xzp}. A detailed analysis, that takes into account the evolution of the gauge and fermion fields, together with the varying velocity of the $\phi$ field, is beyond the scope of this work.}

After particle production starts, the Higgs field potential is altered in the presence of the thermal bath of vector bosons, and EW symmetry is temporarily restored, making the tachyonic growth even faster.
On the other hand, the presence of the thermal bath modifies the dispersion relation of the $Z$ boson, making the process less efficient. Once this effect is included, the timescale for particle production can be estimated as 
\begin{equation}
\Deltatpp \sim \frac{9\pi g_\ew^2}{16} \frac{T^2 \fZZ^3}{\dot \phi^3} \,,
\end{equation}
where $g_\ew^2 \approx 0.2$ and $T$ is the temperature of the plasma.
The temperature of the plasma can be estimated by using energy conservation. Assuming all the kinetic energy of the relaxion is transferred to the thermal plasma, we have
\begin{equation}
\frac{\pi^2}{30} g_* T^4 = \frac{\dot\phi_0^2}{2} \,.
\end{equation}

A very important caveat to the above discussion comes from the coupling to photons $ (\phi/\fFF)  F_{\mu\nu} \widetilde F^{\mu\nu}$.
This coupling must be suppressed for this mechanism to work, for at least three reasons. First, if photon production is efficient, the corresponding  friction term is always active and could slow down the relaxion evolution irrespectively of the value of the Higgs VEV. Second, if these photons thermalize they could deconfine the strong sector which generates the potential barriers. Finally, thermal corrections to the Higgs potential will make the relaxion scanning the Higgs thermal  mass instead  of the vacuum mass parameter $\mu_h^2$.
Neglecting the contribution suppressed by the mixing with the Higgs, such a coupling arises at one and two loops from the Lagrangian in \Eq{eq:lagbroken}, accompanied by a coupling to the Standard Model fermions~\cite{Bauer:2017ris, Craig:2018kne}:
\begin{equation}\label{eq:coupling fermions photons}
\frac{\partial_\mu \phi}{F_\psi} (\bar{\psi} \gamma^\mu \gamma_5 \psi)
\quad\text{and}\quad
\frac{\phi}{4\fFF} F_{\mu\nu} \widetilde F^{\mu\nu}
\end{equation}
where
\begin{equation}\label{eq:fermions}
\frac{1}{F_\psi} = \frac{3 \alpha_{\textrm{em}}^2}{4 \mathcal{F}}\left[  \frac{Y_{\psi_L}^2 +Y_{\psi_R}^2}{\cos^4\theta_W} -\frac{3}{4 \sin^4\theta_W}
\right]\log\frac{\Lambda^2}{m_W^2},
\end{equation}
and
\begin{equation} \label{eq:fgamma}
\frac{1}{\fFF} =  \frac{8 \alpha_{\rm em}^2}{  \sin^2\theta_W \mathcal{F}} B_2\left(x_W\right)+\sum_\psi \frac{2 \alpha_{\rm em} N_c^\psi Q_\psi^2}{\pi F_\psi}   B_1\left( x_\psi\right),
\end{equation}
where $N_c^\psi$ and $Q_\psi$  are respectively  the color multiplicity  and the electric charge of the fermion $\psi$ with mass $m_\psi$, and the $x_i$ is defined as $x_i \equiv 4 m_i^2/m_\phi$. The functions $B_{1,2}$ are written as follows:
\begin{equation}
\begin{aligned}
    B_1(x) &= 1 - x [b(x)]^2\\
    B_2(x) &= 1 - (x - 1) [b(x)]^2
  \end{aligned}
 \qquad  \qquad b(x)=\begin{cases}
    \arcsin \frac{1}{\sqrt{x}} & x \geq 1\\
    \frac{\pi}{2} + \frac{i}{2} \log \frac{1 + \sqrt{1-x}}{1 - \sqrt{1-x}} & x < 1.
  \end{cases}
\end{equation}
These functions tend to  $B_1(x_\psi)\rightarrow -m_\phi^2/(12 m_\psi^2)$ and $B_2(x_W)\rightarrow m_\phi^2/(6 m_W^2)$ for $m^2_\phi\rightarrow 0$. These  contributions to the effective coupling  to photons are  suppressed  by the spurion $m_\phi^2$ and  are absent  in the massless limit.

\medskip

Following the same logic as in Sec.\,\ref{sec:relaxion growing barriers}, we list here the conditions that the model should satisfy to guarantee a successful relaxation of the EW scale. Additional conditions will be introduced below, when discussing the realization of this model during or after inflation.
\begin{itemize}

\item \textbf{Prediction for the EW scale:} As we discussed above, dissipation starts when
\begin{equation} \label{eq:phistop}
m_Z \approx \frac{\dot\phi_0}{2 \fZZ} \,.
\end{equation}
Imposing that $m_Z$ matches the measured value of $91\GeV$, we will use \Eq{eq:phistop} to relate the scale $\fZZ$ to the other parameters of the model.

\item \textbf{Size of the barriers:}
During its rolling phase, the relaxion must be able to jump over the barriers, thus $\dot\phi^2/2 > \Lambda_b^4$. On the other hand, for the validity of the EFT we will always assume $\dot\phi^2/2 < \Lambda^4$, and therefore
\begin{equation}
\Lambda_b < \Lambda \, .
\end{equation}

\item \textbf{Overcome the wiggles:}
The field must be able to overcome many wiggles before stopping, thus
\begin{align}
\frac{1}{2}\dot\phi_0^2 & > \Lambda_b^4 \label{eq:HMT large velocity}\\
\Delta t_1 & = \frac{2\pi \fcos}{\dot\phi_0} < \frac{1}{H} \,.
\end{align}
 
\item \textbf{Efficient energy dissipation:} The energy lost by the relaxion due to particle production must be larger than the one gained by rolling down the potential slope,
\begin{equation}
\Delta K_\textrm{rolling}\, <\, \Delta K_\textrm{pp}.
\end{equation}
Using $\Delta K_\textrm{pp}\sim \dot\phi^2 /2$ and $\Delta K_\textrm{rolling} \sim \frac{dK}{dt}  \Deltatpp$, with $dK/dt = -dV/dt \sim g \Lambda^3 \dot\phi$, we get
\begin{equation}\label{eq:condition energy dissipation}
g \Lambda^3 \dot\phi_{\rm{stop}} \Deltatpp < \frac{1}{2}\dot\phi_{\rm{stop}}^2 \,.
\end{equation}
To obtain the most stringent bound, we  evaluate this condition for $\dot\phi^2/2 = \dot\phi_{\rm{stop}}^2/2 \sim\Lambda_b^4$, the maximum velocity the relaxion can have after it has been trapped.

\item \textbf{Small variation of the Higgs mass:} The particle production phase must be fast enough, so that the variation of the Higgs mass during this time is less than a fraction of the EW scale:
\begin{equation}\label{eq:f42}
\Delta m_h \sim \frac{\Delta m_h^2}{m_h} \sim  \frac{1}{m_h} g' \Lambda \, \dot\phi \, \Deltatpp < m_h\,,
\end{equation} 
which, again, we evaluate at $\dot\phi = \dot{\phi}_{\rm{stop}}\sim\Lambda_b^2$ to derive the most stringent bound.

\item \textbf{Shift symmetry not restored:}
After the relaxion has been trapped, the temperature may be larger than the condensation scale of the cosine potential. To avoid this scenario, we impose that
\begin{equation}
T < \Lambda_b \,.
\label{eq:TsmallerLambdab}
\end{equation}
This condition applies when the sector generating the barriers is in equilibrium with the Standard Model. Assuming that the former interacts with the relaxion through a term $\phi G'\widetilde G' /f$, we naively  estimate the rate for $gg\leftrightarrow ZZ$ interactions as $\Gamma \sim T^5/(\fZZ^2f^2)$, which must be larger than the Hubble rate $H$. Thus, we impose
\begin{equation} \label{eq:H12}
T < \max \left\{ \Lambda_b, (H F^2 f^2)^{1/5}\right\} \, .
\end{equation}
As already noted above, this constraint can be avoided in a scenario in which, after reheating, the barriers disappear and the relaxion rolls for a short amount of time, not overshooting the value of the Higgs VEV (see Refs.~\cite{Choi:2016kke, Banerjee:2018xmn, Abel:2018fqg}.)

\item \textbf{Particle production faster than Hubble expansion:}
Our analysis was conducted in Minkowski space. A large Hubble rate would suppress the production of gauge boson. Thus we impose that
\begin{equation}
\label{eq:HMT-no-inflation-depend-on-H-1}
\Deltatpp < H^{-1}\,.
\end{equation}

\item \textbf{Suppressed coupling to photons:}
Contrarily to the production of massive bosons, we assume that the photons are efficiently diluted away, \ie
\begin{equation} \label{eq:photon dilution}
\Delta t_{\gamma} > H^{-1},
\end{equation}
where  $\Delta t_{\gamma} \sim  T^2 \fFF^3/\dot{\phi}^3$.

\item \textbf{Hierarchy of the effective scales:} For the validity of the EFT we assume that the scale $F$, which controls the coupling of the relaxion to the SM fields, is larger than the cut-off scale:
\begin{equation}\label{eq:HMT F>Lambda}
F>\Lambda,
\end{equation}
where $F$ is given in \Eq{eq:f convention}.
\item Finally, we list here some conditions that are identical to those imposed in Sec.~\ref{sec:relaxion growing barriers}, referring the reader to it for their discussion:%
\footnote{In order to avoid fine-tuning of the initial 
conditions, the relaxion field excursion has to be
larger than $\Lambda/g'$. Similarly to Sec.\,\ref{sec:relaxion growing barriers}, here we assume for definiteness  that initially $\phi=0$, then the total field range is given as in \Eq{eq:HMT field range}.
Moreover, differently to Eq.~(\ref{eq:Higgs tracking}), and following Ref.~\cite{Fonseca:2018xzp} we impose in Eq.~(\ref{eq:Higgs tracking 2018}) a condition on the mass term $\mu_h$ instead of the vev $v$. The difference is a factor of $\lambda$ at the denominator.}
\begin{align}
\left| \frac{\dot\mu_h}{\mu_h^2} \right|_{v=v_\ew}
< &~ 1 \Longleftrightarrow \frac{g' \Lambda \dot\phi}{2\lambda^{3/2}v_\ew^3} <~ 1 & \text{Higgs tracking the minimum} \label{eq:Higgs tracking 2018}\\
g\Lambda^3 < &~ \frac{\Lambda_b^4}{f} & \text{Large barriers} \label{eq:HMT large barriers} \\
g' \Lambda (2\pi f) < & ~\frac{m_h^2}{2} & \text{Precision of the mass scanning} \label{eq:HMT scanning precision}\\
f > & ~\Lambda & \text{Consistency of the EFT} \label{eq:HMT eftvalidity} \\
f < & ~\MPl & \text{Sub-Planckian decay constant} \\
\Delta\phi ~= &~ \frac{\Lambda}{g'} & \text{Field range} \label{eq:HMT field range}
\end{align}

\end{itemize}

To constrain the parameter space, we will apply the same reasoning as in Sec.~\ref{sec:relaxion growing barriers}. The free parameters are $\{\Lambda, g, g', \Lambda_b, f, \fZZ, \dot\phi, H\}$. Unless otherwise specified, we assume $g=g'$, and we use \Eq{eq:phistop} to fix the scale $f$ in terms of the other parameters of the model.

\subsection{Relaxation after inflation}
\label{subsectionHMTafter}

Let us first consider the possibility of relaxation after inflation with the tachyonic production of Standard Model gauge bosons, which was discussed in~\cite{Hook:2016mqo} and, in greater details, in~\cite{Fonseca:2018xzp}. In addition to Eqs.~(\ref{eq:phistop})-(\ref{eq:HMT field range}), we assume that the relaxion dominates the energy density
\begin{equation}\label{eq:HMT H relaxion}
H = \frac{\Lambda^2}{\sqrt 3 \MPl} \,.
\end{equation}
Moreover, we assume that the relaxion does not drive a secondary period of inflation, in which the curvature perturbations generated during inflation would be erased. Thus we impose~\cite{Fonseca:2018xzp}
\begin{equation}
g' > 0.18\frac{\Lambda}{\MPl}\,,
\end{equation}
where the numerical factor comes from requiring that, if a short period of relaxion-driven inflation takes place, this does not exceed $20$ efolds. A similar bound can be obtained by imposing that the velocity $\Lambda^2$ is smaller than the slow-roll velocity $g\Lambda^3/(3H)$, with $H$ as in \Eq{eq:HMT H relaxion}. Under this condition, it is safe to neglect Hubble in the equation of motion for the relaxion field and its fluctuations.

Due to the constant barriers $\Lambda_b^4\cos\phi/f$, relaxion fragmentation is always active in this construction, and it can slow down the field evolution at a position which is not related to the Higgs VEV. To avoid this scenario, we assume that either the effect of fragmentation is subdominant compared to the acceleration due to the large slope, or that, if present, the fragmentation time-scale is longer than the time needed to complete a full field excursion $\Lambda/g'$:
\begin{equation}\label{eq:HMT slope fragmentation}
g \Lambda^3 > (g\Lambda^3)_\mathrm{max} \equiv \frac{\pi \Lambda_b^8}{2 f \dot\phi_0^2} \left(W_0\left(\frac{32\pi^2f^4}{e\,\dot\phi_0^2}\right)\right)^{-1}
\end{equation}
or
\begin{equation}\label{eq:HMT frag not efficient}
g \Lambda^3 < (g\Lambda^3)_\mathrm{max}
\qquad\mathrm{and}\qquad
\Delta\phi < \Delta\phi_\mathrm{frag} = \frac{f \dot\phi_0^4}{2\pi\Lambda_b^8} \log\left(\frac{32\pi^2f^4}{\dot\phi_0^2}\right),
\end{equation}
where the upper bound for the slope  $g\Lambda^3$ is given in \Eq{eq:slope bound (H=0)} and the excursion $\Delta\phi_\mathrm{frac}$ is written in \Eq{eq:masterfieldexcursion}. 
Once Hubble friction is neglected, the velocity of the relaxion after it travels a field range starting from rest is $\Delta\phi=\Lambda/g'$ is $\dot\phi = \sqrt{2g'/g}\Lambda^2$. Assuming $g=g'$, the first and second inequalities in~\ref{eq:HMT frag not efficient} become incompatible, \ie if fragmentation takes place then it always happens within a short time. Thus we apply \Eq{eq:HMT slope fragmentation}, and combine it with Eqs.~(\ref{eq:phistop})-(\ref{eq:HMT field range}) to obtain bounds in the plane $\Lambda, g'$.

\begin{figure}
\centering
\includegraphics[width=.55\textwidth]{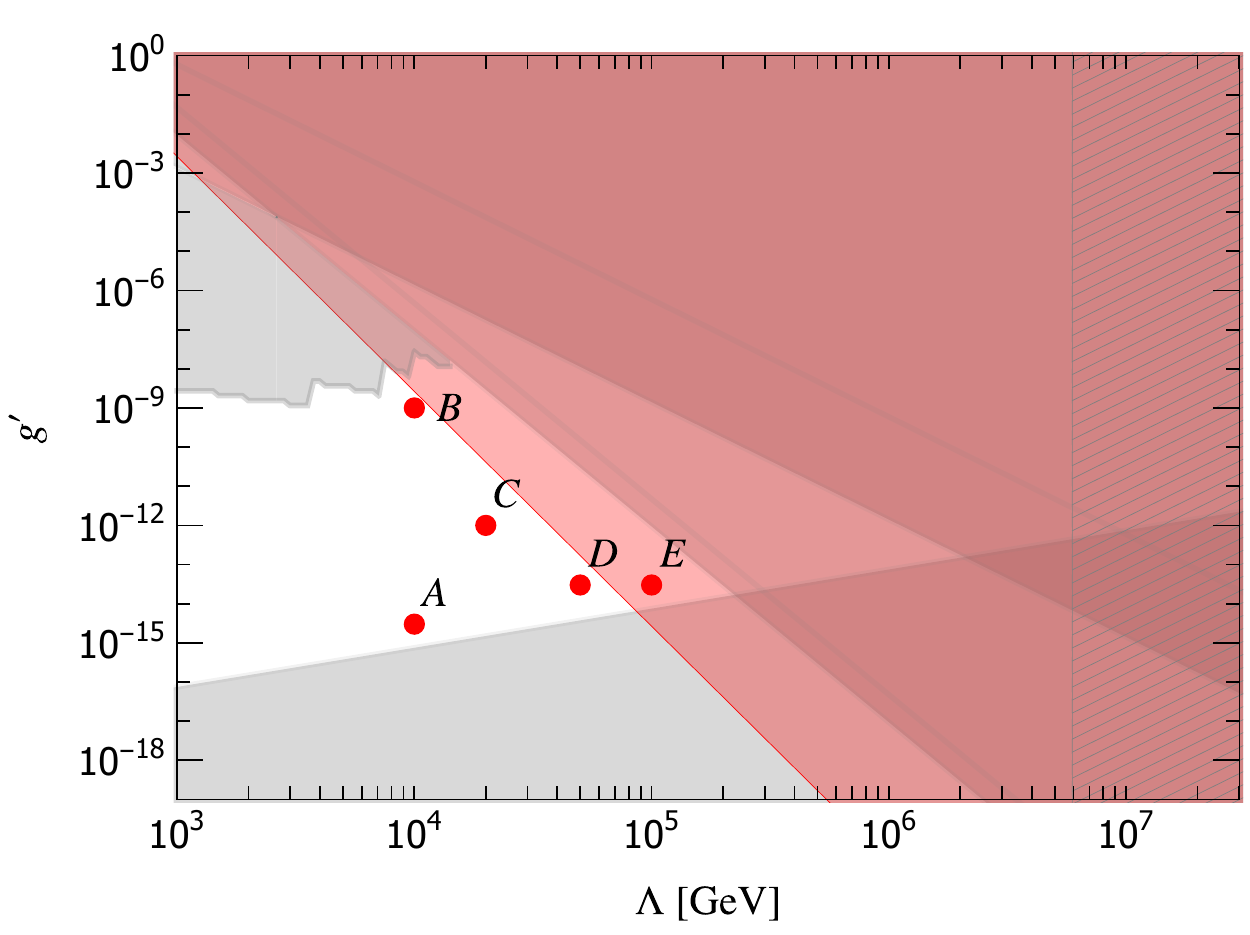}
\caption{\label{fig:gpLambda-HMT-after} Case of relaxation AFTER inflation with EW gauge boson production  as described in Section \ref{subsectionHMTafter}.  White region is the allowed region that was derived in Ref.~\cite{Fonseca:2018kqf} without including the effect of relaxion fragmentation. The red region is the new exclusion due to relaxion fragmentation. }
\end{figure}

\begin{figure}
\centering
\includegraphics[width=.4\textwidth]{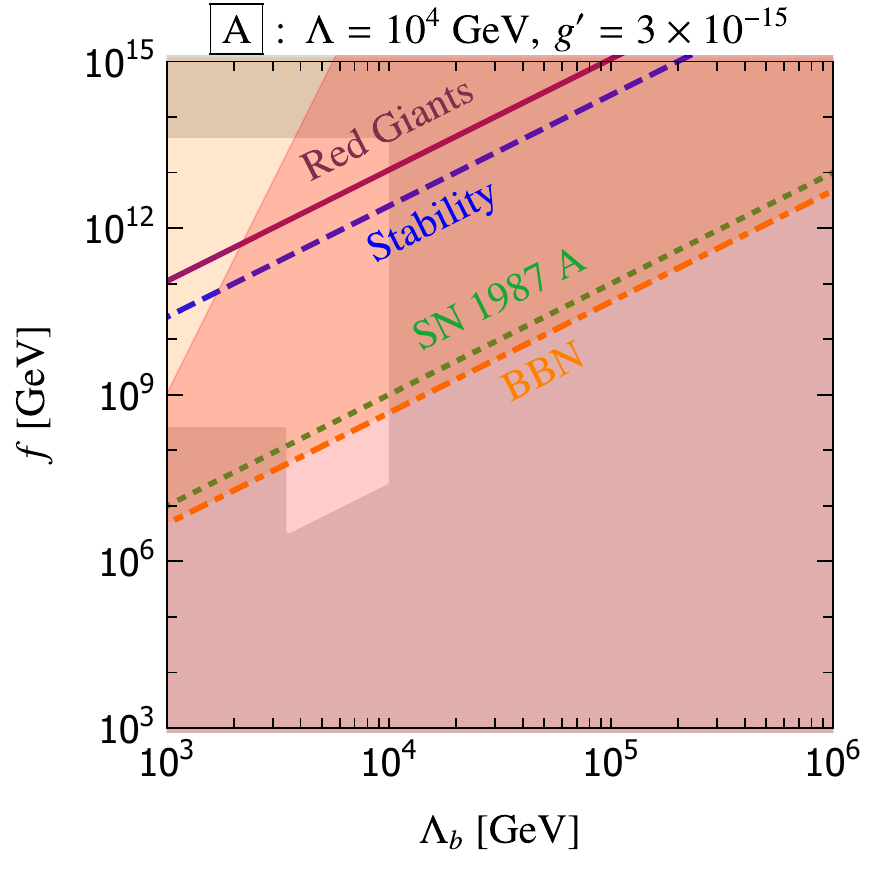}
\includegraphics[width=.4\textwidth]{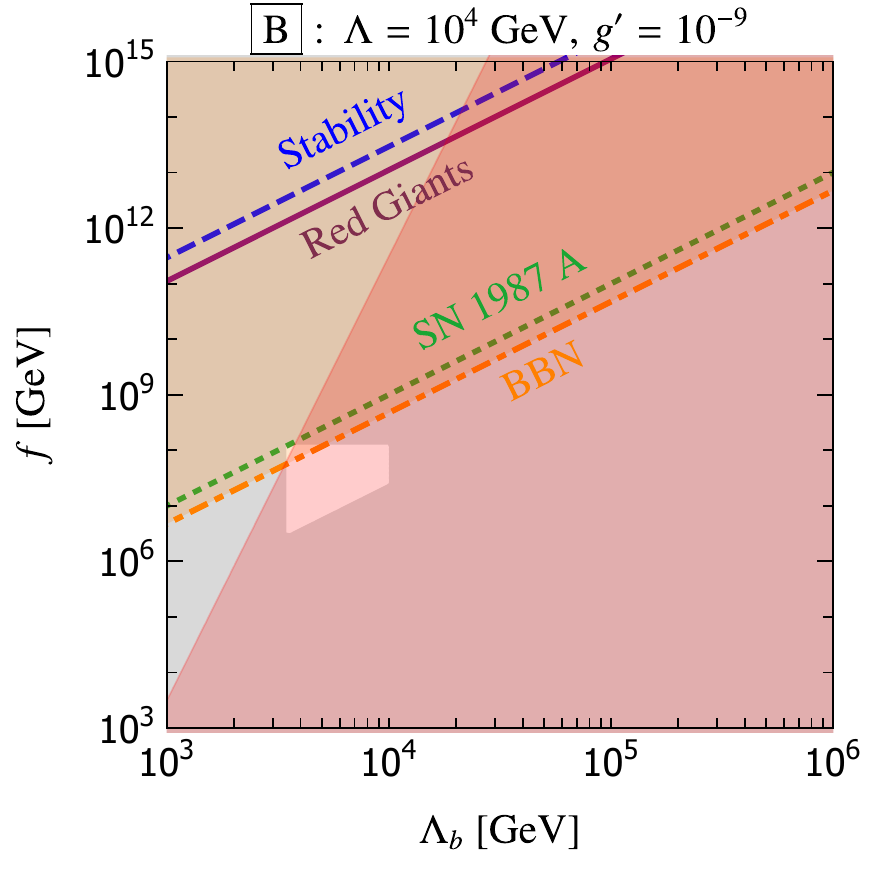}
\includegraphics[width=.4\textwidth]{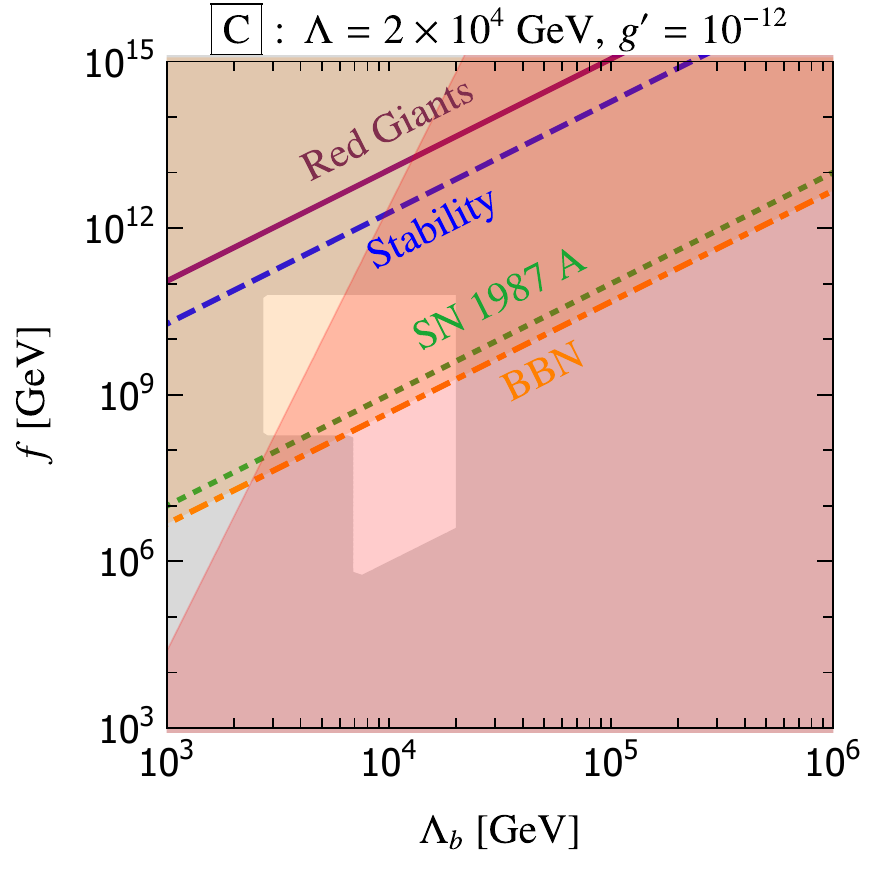}
\includegraphics[width=.4\textwidth]{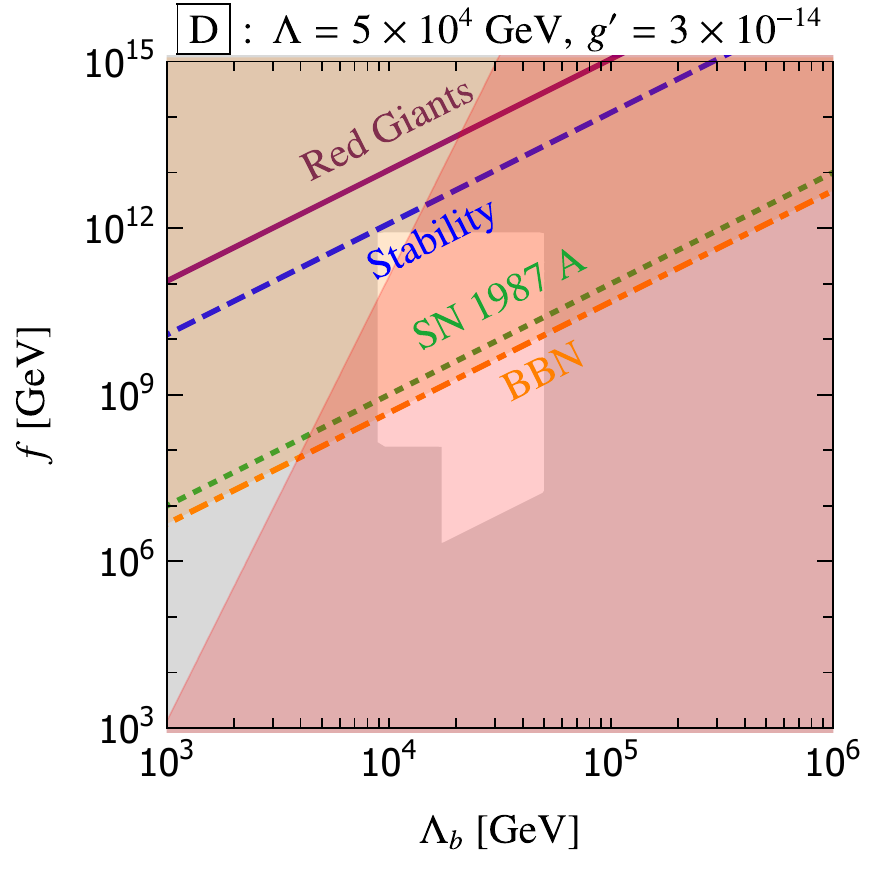}
\includegraphics[width=0.4\textwidth]{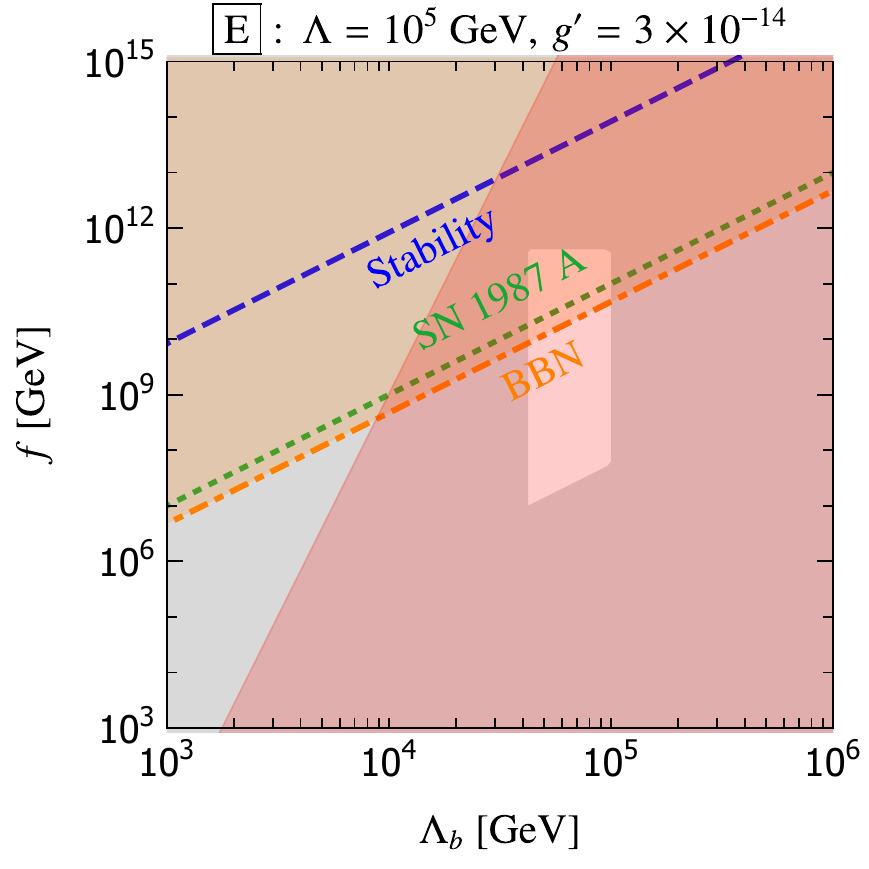}
\caption{\label{fig:fpLambdac-HMT-after} Case of relaxation AFTER inflation with EW gauge boson production as described in Section \ref{subsectionHMTafter}.  Pale regions were the allowed region derived in Ref.~\cite{Fonseca:2018xzp} without including the effect of relaxion fragmentation. The red region is the new exclusion due to relaxion fragmentation, which rules out everything under the assumption that the relaxion dominates the energy density of the universe during relaxation (Eq.~\ref{eq:HMT H relaxion}).}
\end{figure}

 Fig.~\ref{fig:gpLambda-HMT-after} and~\ref{fig:fpLambdac-HMT-after}  show the relaxion
parameter space including the new exclusion   due to relaxion fragmentation in red. 
 The choice of benchmarks for Fig.~\ref{fig:fpLambdac-HMT-after} is the same as in Ref.~\cite{Fonseca:2018xzp}. In Fig.~\ref{fig:fpLambdac-HMT-after} we also show the upper bound on $f$ derived in~\cite{Fonseca:2018xzp}. The purple and green lines are derived from Red Giants and SN 1987 A; the blue line excludes the region in which the relaxion (which is always overabundant in this model) is cosmologically stable and overcloses the universe, and the orange line excludes the case in which relaxion decays during or after Big-Bang nucleosynthesis, distorting the abundance of light elements.
The above constraints are derived using the decay widths listed in~\cite{Craig:2018kne, Fonseca:2018xzp} and the mixing angle of~\cite{Fonseca:2018xzp}, which differs in this case from the one of Eq.~(\ref{eq:mixing angle}) because here the barriers are independent of the Higgs:
\begin{equation}
\left.\theta_{\phi,h}\right|_{\text{const. barr.}} = \frac{2 g' v_\ew \Lambda}{\sqrt{(m_h^2-m_\phi^2)^2+4g'^2 v_\ew^2 \Lambda^2}}
\end{equation}
An additional constrain comes from the electric dipole moment of light SM fermions, which is generated through the $\phi Z \widetilde \gamma$ coupling and the $\phi-h$ mixing. We checked that the relaxion contribution to the EDMs is always smaller than the experimental bound, hence there is no relevant constraint on the relaxion parameter space.
As it can be seen in Fig.~\ref{fig:fpLambdac-HMT-after},  astrophysical and cosmological constraints are very severe and thus exclude all these benchmarks. 
Despite the impression that there is some space open in the plane $(\Lambda,g')$ in Fig.~\ref{fig:gpLambda-HMT-after}, it seems this scenario is essentially ruled out due to the other constraints in the plane $(\Lambda_b,f)$ in \Fig{fig:fpLambdac-HMT-after}.
We therefore conclude that, if no new element is added to modify the late-time evolution of the relaxion population, this realization of the relaxion mechanism is excluded.

Note that this conclusion remains unchanged if we relax equation (\ref{eq:HMT H relaxion})  and instead impose $H> \Lambda^2/\sqrt{3}\MPl$, assuming that the relaxion energy density is subdominant compared to hidden sector radiation (which eventually decays into the SM before BBN). The only conditions which depend on the precise value of the Hubble rate are (\ref{eq:H12}), 
(\ref{eq:HMT-no-inflation-depend-on-H-1}) and (\ref{eq:photon dilution}). The conditions (\ref{eq:HMT-no-inflation-depend-on-H-1}) and (\ref{eq:photon dilution})  affect the constraints in the plane $(\Lambda, g')$. The condition (\ref{eq:photon dilution}) depends on the relaxion mass indirectly, so also affects the $ (\Lambda_b,f)$.
By increasing $H$, \Eq{eq:photon dilution} becomes easier to satisfy, then  we could reopen a region in the $(\Lambda, g')$ plane which could be probed by SHiP/CHARM. However,  the whole parameter space is still excluded in the $(\Lambda_b,f)$ plane. While the BBN constraints can be evaded since the relaxion abundance would now be diluted, the astrophysical bounds are still excluding the full region. The only way to save the `relaxion after inflation' scenario in models with Higgs-independent barriers would require to question the application of supernovae bounds to axions, 
following for instance \cite{Bar:2019ifz}. Alternatively, one can try to combine  this construction with the double scanner mechanism presented in \Ref{Espinosa:2015eda}, where a second scalar field scans the barriers's amplitude. In order to suppress the fragmentation effect, this extra scanner field would have to cancel the amplitude of the cosine potential during the scanning of the Higgs mass parameter. Here we do not pursue this possibility further and leave these studies for future work.

%\clearpage

\subsection{Relaxation during inflation}
\label{subsectionHMTduring}

We now consider the case in which the relaxion slowly rolls down its potential during inflation, with a velocity $\dot\phi=\phidotSR = g\Lambda^3/(3H)$. As before, we assume that the following conditions hold:
\begin{align}
H > & \,\frac{\Lambda^2}{\sqrt{3}\MPl} & \text{Inflaton dominates the energy density} \label{eq:HMT relaxion subdominant}\\
H < & \,f & \text{Relaxion present during inflation} \\
H < &\, \Lambda & \text{Shift-symmetry broken during inflation} \\
\frac{\phidotSR}{H} > & \,\frac{H}{2\pi} & \text{Classical evolution}
\end{align}
We assume that either fragmentation is suppressed by Hubble friction, or that, if present, its time-scale is longer than the time needed to complete a full field excursion $\Lambda/g'$:
\begin{equation}\label{eq:HMT infl H fragmentation}
\phidotSR^3 H > \frac{\pi\Lambda_b^8}{2 f} \left(W_0\left(\frac{32\pi^2f^4}{e\, \phidotSR^2}\right)\right)^{-1}
\end{equation}
or, if Eq.~(\ref{eq:HMT infl H fragmentation}) is violated,
\begin{equation}\label{eq:HMT infl fragmentation not efficient}
\Delta\phi < \Delta\phi_\mathrm{frag} = \frac{f \phidotSR^4}{2\pi\Lambda_b^8} \log\left(\frac{32\pi^2f^4}{\phidotSR^2}\right),
\end{equation}
where  Eq.~(\ref{eq:HMT infl H fragmentation}) comes from violating \Eq{eq:condition 3}, $\Delta\phi = \Lambda/g'$ as usual, and we neglect the logarithmic dependence taking $W_0(\dots)\approx\log(\dots)\approx50$. The above conditions can be rewritten in a compact form as
\begin{equation}\label{eq: HMT infl no fragmentation}
H < \max \left\{ \frac{1}{3} \left(\frac{2 f g^3 \Lambda^9}{3\pi \Lambda_b^8}W_0(\dots) \right)^{1/2}, \frac{1}{3} \left(\frac{f g^4 g' \Lambda^{11}}{2\pi \Lambda_b^8}\log(\dots)\right)^{1/4} \right\} \,.
\end{equation}

\begin{figure}
\centering
\includegraphics[width=.45\textwidth]{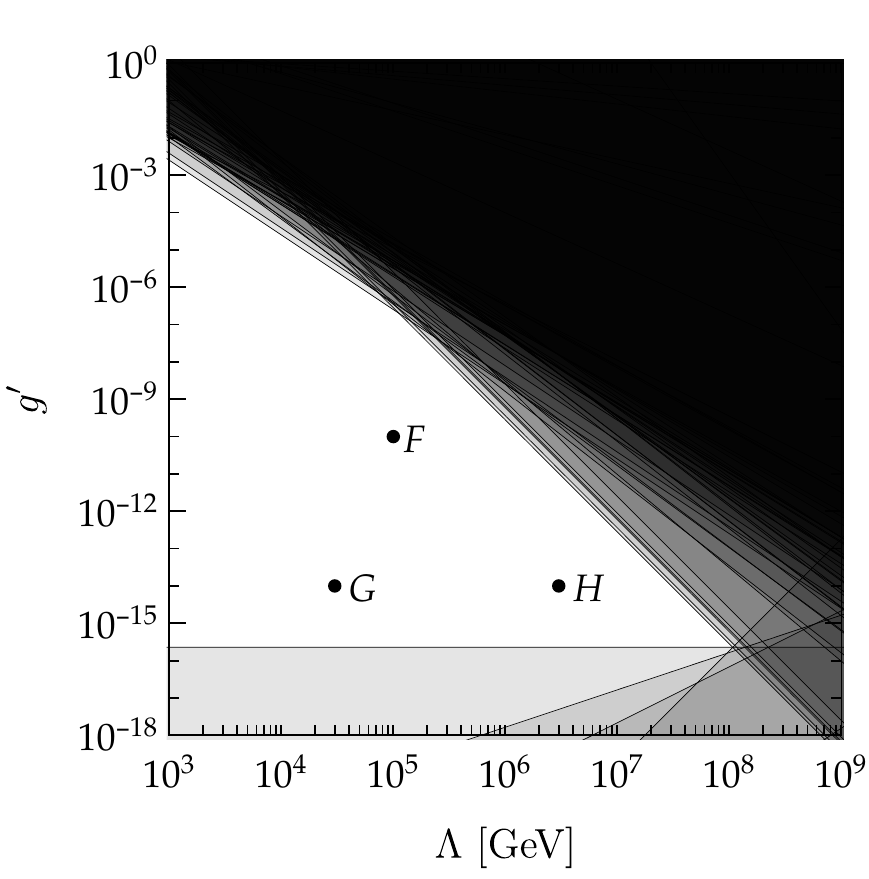}
\includegraphics[width=.45\textwidth]{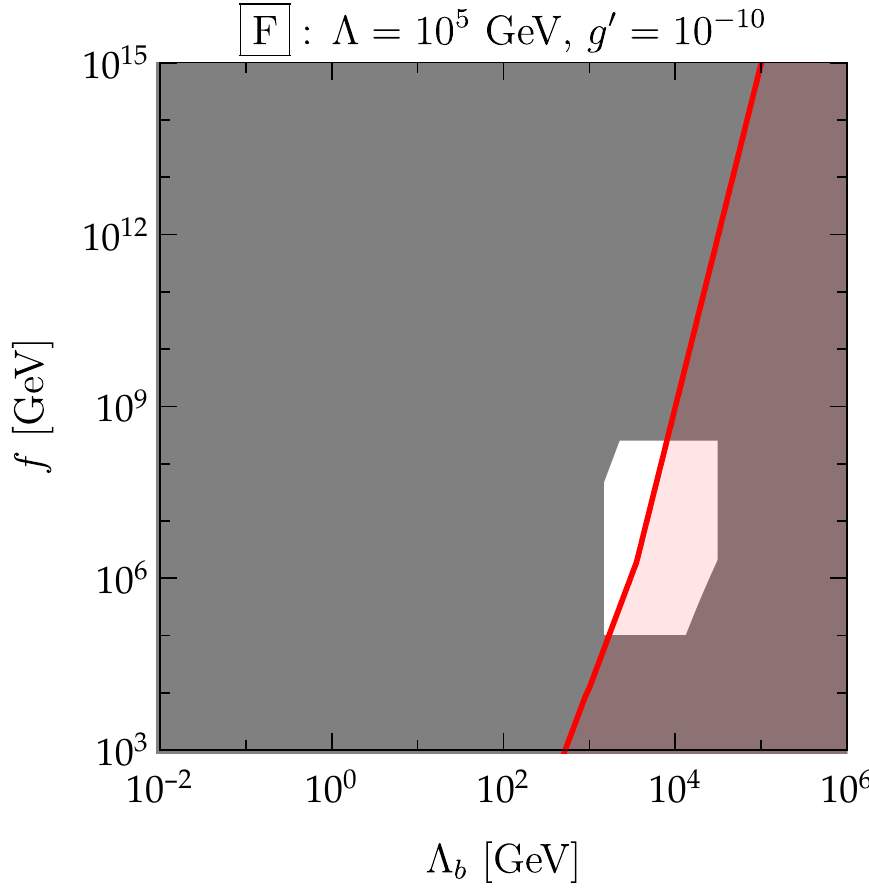}
\includegraphics[width=.45\textwidth]{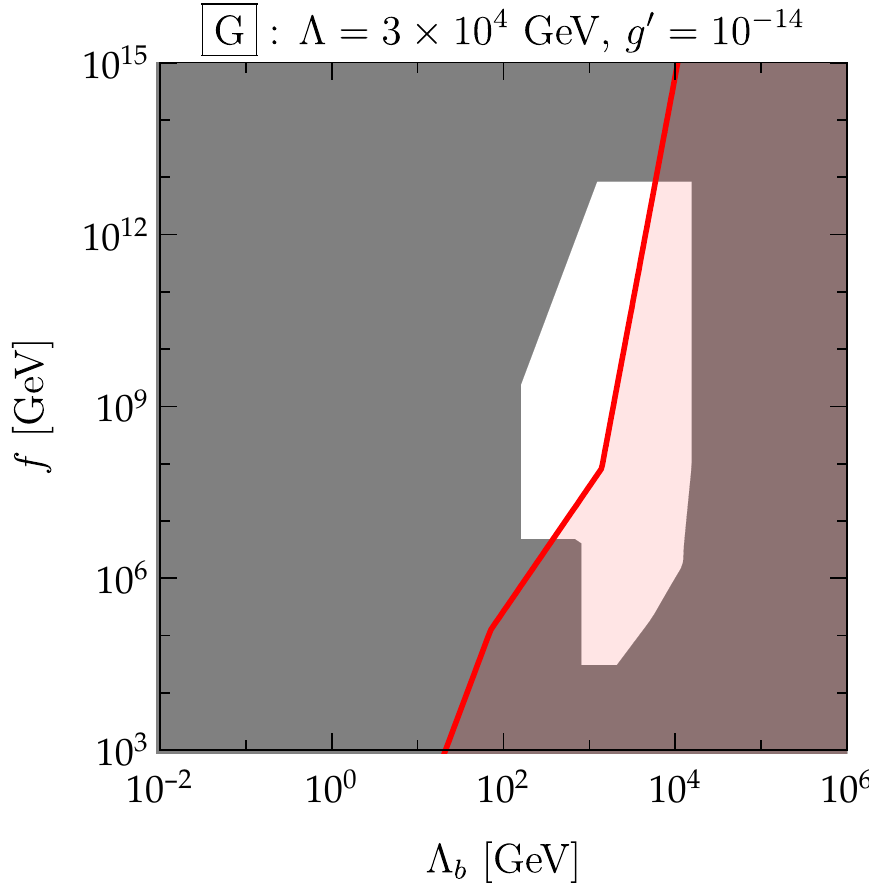}
\includegraphics[width=.45\textwidth]{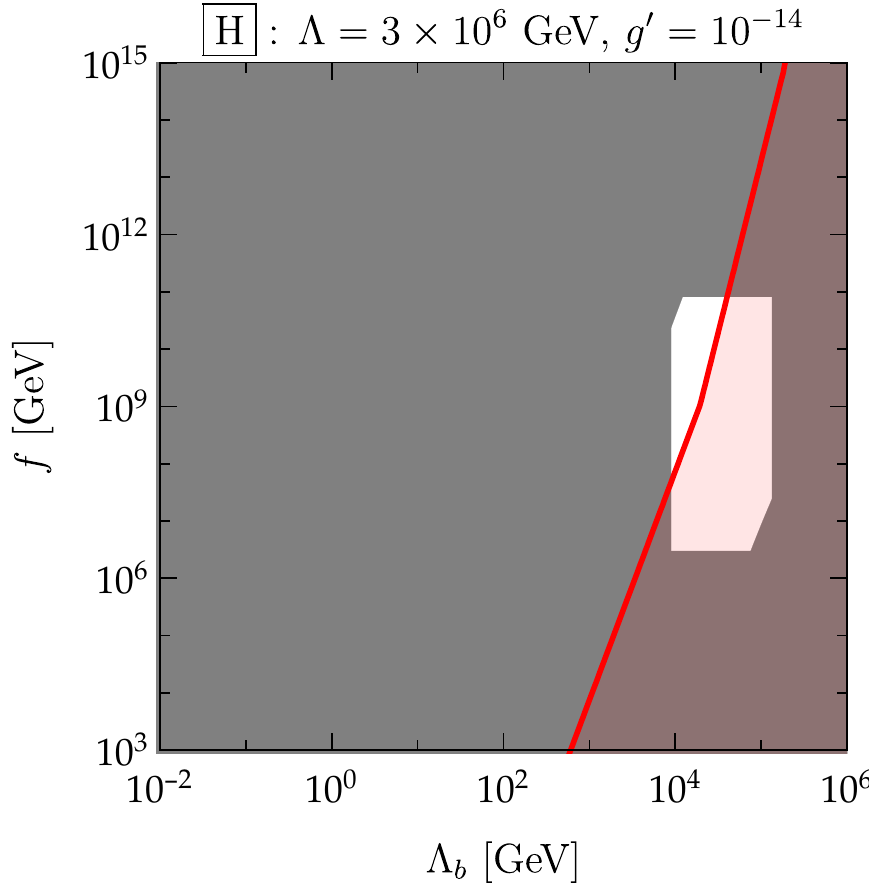}
\caption{\label{fig:HMT inflation}Relaxation with EW gauge boson production during inflation.  White region is the would-be allowed region  that was derived in Ref.~\cite{Fonseca:2018kqf}, not including the effect of relaxion fragmentation. The red region is the new exclusion due to relaxion fragmentation. For more details on the delimitating lines, see Figs.~\ref{fig:HMT inflation contours gp Lambda}, \ref{fig:HMT inflation contours f Lambdab}, \ref{fig:HMT inflation contours f Lambdab frag} in Appendix  \ref{sec:AppendixHMTduring}. Figure~\ref{fig:HMT inflation contours HI Ne f Lambdab} shows the contours of $H_I$ and $N_e$ in the same plane.}
\end{figure}

The allowed parameter space for this scenario is shown in Fig.~\ref{fig:HMT inflation}. The contours in the plane ($\Lambda_b, f$) do not depend on any choice of $H$, and they are obtained by using the conditions of the form $H<\dots$ and those $H>\dots$ to get inequalities independent of $H$.
Details on the origin of the constraints delimitating the allowed regions are provided in Figs.~\ref{fig:HMT inflation contours gp Lambda}--\ref{fig:HMT inflation contours f Lambdab frag} of Appendix \ref{sec:AppendixHMTduring}.
Fig.~\ref{fig:HMT inflation} shows the relaxion parameter space including the new exclusion due to relaxion fragmentation in red. 

\subsubsection{Relaxion Dark Matter}

As it was discussed in Ref.~\cite{Fonseca:2018kqf}, after inflation has ended the relaxion is  produced from scattering with the thermal Standard Model plasma, and its relic abundance can match the observed DM one (see~\cite{Banerjee:2018xmn} for a different construction in which the relaxion constitutes the DM component of the universe in the GKR model). The main production channel is Compton scattering off electrons $\gamma + e \leftrightarrow \phi + e$, which is active until electrons become non-relativistic. Imposing that the relaxion is stable on cosmological timescales, enough long-lived to avoid indirect detection constraints on decaying DM, and heavy enough to avoid the Ly-$\alpha$ constraints on hot DM, the relaxion is a viable DM candidate for $m_\phi\approx 2-17\keV$.
We repeated the analysis of~\cite{Fonseca:2018kqf}, imposing the additional constraint of \Eq{eq: HMT infl no fragmentation}, and we found that the conditions on the relaxion's lifetime are always stronger than \Eq{eq: HMT infl no fragmentation}. In other words, fragmentation is never efficient in the parameter space of relaxion DM, and the results of~\cite{Fonseca:2018kqf} are unchanged.

Fragmentation opens up another interesting possibility for relaxion DM. If relaxation takes place after inflation, and the relaxion is only a subdominant component of the energy density at that time, the relaxion particles produced by fragmentation can in principle have the correct relic abundance to represent a good warm DM candidate, similarly to Ref.~\cite{Berges:2019dgr}. We postpone the study of this effect and of the details of the relaxion cool down to a future work.

Irrespectively of fragmentation, the computation of the relic abundance performed in~\cite{Fonseca:2018kqf} could be used to look for regions of parameter space in which the relaxion is overabundant. Such regions, if they exist, would then be excluded, unless the cosmological history is modified including a phase of dilution, or additional decay channels are added to make the relaxion unstable. This would affect the parameter space displayed in Fig.~\ref{fig:HMT inflation}. We do not pursue this aspect any further here.

%\clearpage

\section{Summary and outlook}
\label{secsummary}

The production of relaxion particles during the evolution of the homogeneous  relaxion field while rolling down its potential had so far been ignored in the relaxion literature. 
Technical details about this effect, dubbed {\it axion fragmentation}, are presented in a companion paper ~\cite{Fonseca:2019ypl}.
Here we showed that axion fragmentation can act as an efficient source of  friction and eventually stop the relaxion field.
This opens parameter space for the relaxion mechanism, especially in the original implementation of the relaxion mechanism of Ref.~\cite{Graham:2015cka} (referred as {\it GKR}). This can also severely reduce the parameter space in the second class of models  \cite{Hook:2016mqo,Fonseca:2018xzp} where the potential barriers are Higgs-independent and relaxation starts in the EW broken phase.
The parameter space comprises the cutoff scale $\Lambda$, the relaxion coupling $g'$, the size of the periodic potential barrier $\Lambda_b$, and, in the case where we invoke inflation, the value of the Hubble scale during inflation $H_I$. These parameters can also be traded for 
$\Lambda$,  $g'$, $m_{\phi}$ and $M_I$, where $m_{\phi}$ is the relaxion mass and $M_I$ is the scale of inflation.

We have worked out in detail the precise regions of parameter space when the relaxion mechanism is successful. In particular, an important question is whether cosmological relaxation of the EW scale can occur without inflation, as this clearly modifies the perspectives and constraints for model building. We have shown that this is possible in the case of Higgs-dependent barriers.

 For  given values of the cutoff scale $\Lambda$ and the relaxion coupling $g'$, $\Lambda_b$, there are three ways by which a relaxion with Higgs-dependent barriers can be stopped during an inflation era: From Hubble friction (as in \cite{Graham:2015cka}), from large barriers and low Hubble friction, from relaxion particle production and low Hubble friction. These last two cases were not considered in  Ref.~\cite{Graham:2015cka}. They correspond to distinct values of $H_I$ and $\Lambda_b$ (equivalently of  $M_I$ and $m_{\phi}$). This is summarised in Fig.~\ref{fig:summaryINFL} and \ref{fig:comparisonINFL}.
 If instead the relaxion has Higgs-independent barriers and an additional coupling to EW gauge bosons 
 $\phi W \tilde{W}$ and $\phi B \tilde{B}$,  it can still be stopped during inflation, as shown in Fig.~\ref{fig:HMT inflation}.  Our results are summarised in the points below.

\begin{itemize}

\item  Relaxation via Higgs-dependent barriers \cite{Graham:2015cka}:

\begin{itemize}

\item During inflationary stage not driven by the relaxion (Section \ref {sec:during inflation}, benchmark points $a,b,c,d$). Cutoff scale can be as high as $10^9$ GeV as in the original proposal.
Interestingly, relaxion fragmentation opens the parameter space towards smaller inflationary scale ${\cal O}(100)$ TeV and heavier relaxion ${\cal O}(1)$ GeV. The inflationary stage can be much shorter, ${\cal O}(10)$ e-folds, and therefore does not have to be tied to the inflationary stage responsible for cosmological perturbations. Besides, a larger range of barrier sizes $\Lambda_b$ are now allowed, up to a TeV. In this new parameter space, the cutoff scale can be as high as several hundreds of TeV and the field excursion is mostly subplanckian. 
The relaxation can also be stopped simply because of larger barriers. In this case, the relaxion mass is also larger than in the original proposal.

\item Without inflation (either before or after), the relaxion may dominate or not the energy density of the universe (Section \ref{sec:withoutinflation}, benchmark points $e,f,g$). 
Interestingly, the relaxion  can stop without the need for inflation, only from the fragmentation effect!
Our analysis shows that the cut-off scale $\Lambda$ cannot be pushed very high. 
The main obstacle comes from the initial velocity of the relaxion $\dot{\phi}_0$. If it is slightly lower than the cutoff scale squared, the fragmentation effect can very efficiently stop the relaxion and relaxation without inflation can work even for large cutoff values $\Lambda$. However, the natural value expected for the relaxion initial velocity is of order $\Lambda^2$ and in this case,  the cutoff scale can be pushed to $\sim 20$ TeV.

\item During an inflationary stage driven by the relaxion (Section \ref{sec:relaxioninflating}, benchmark point $h$). Another economical relaxion model does not require any other scalar field driving inflation. The relaxion itself could  trigger an inflationary stage and then self-stops because of fragmentation. Interestingly, the cutoff scale can be pushed to $\sim 300$ TeV in this case. The associated number of efolds varies between ${\cal O}(10)$  to ${\cal O}(10^5)$ depending on the cutoff scale.

\end{itemize}

\item Relaxation via Higgs-independent barriers and EW gauge boson production \cite{Hook:2016mqo,Fonseca:2018xzp}:

\begin{itemize}

\item After inflation (Section \ref{subsectionHMTafter}, benchmark points $A,B,C,D,E$). While 
Ref.\cite{Fonseca:2018xzp} had 
stressed the possibility to achieve relaxation without inflation, the present work shows that the effect of fragmentation is actually so efficient in this scenario that it completely kills it. In fact, the relaxion always stops too early before it reaches the correct EW scale. 
\item During inflation (Section \ref{subsectionHMTduring}, benchmark points $F,G,H$). We showed that the only way to save the mechanism of \cite{Hook:2016mqo} is to have it happen during inflation. 
\end{itemize}

\end{itemize}

In all cases where relaxion fragmentation is responsible for stopping the relaxion, we predict a relaxion mass heavier than in GKR, $m_{\phi}\gtrsim {\cal O}(1)$ GeV.
The various scenarios we have looked at are compiled in Tables \ref{tab:comparison} and \ref{tab:bench}.
They would deserve some further attention. In each case, there are specific cosmological histories (summarized in Appendix \ref{sec:cosmologicalhistory}) and it remains to look in more details at the model building and phenomenological aspects of each of them. In particular, it will be interesting to revisit the scenario of relaxion dark matter.

In this paper we have discussed two main classes of relaxion models.
One may wonder whether relaxion fragmentation has any implication for the QCD axion model discussed first in  \cite{Graham:2015cka}. In this case, the Higgs-dependent barrier is much smaller, $\Lambda_b^4\sim\Lambda_{QCD}^3 m_u$. Then, from \Eq{eq:fixingf} where $f$ scales as $\Lambda^8_b$, 
it is clear that for fragmentation to effectively  stop the relaxion without inflation, we would need an unacceptably small value for $f$ unless the initial relaxion velocity is tuned to be somewhat smaller than $\Lambda^2$. 
So there are no implications for the QCD relaxion model. The non-QCD model with  Higgs-dependent barriers requires the introduction of new weak scale fermions charged under $SU(2)$, see equation (\ref{eq:boundonmL}) in Appendix \ref{sec:explicit model}.
This is good from the point of view of testability \cite{Beauchesne:2017ukw}, as future high energy colliders can in principle test this scenario. However, a compelling possibility would be that the relaxion mechanism does not require the introduction of the weak scale by hand. In fact, such `coincidence' problem was solved in \cite{Espinosa:2015eda}, which showed that the relaxion mechanism does not require any new weak scale physics and thus does not predict any signals at collider experiments. The new physics  responsible for solving the hierarchy problem may instead only feature very weakly coupled and light axion-like particles. In \cite{Espinosa:2015eda}, the weak scale is generated dynamically though a double-scanner-field mechanism. It would be interesting to investigate the implications of relaxion fragmentation on the parameter space of this scenario.
Further phenomenological aspects of relaxion particle production such as dark matter and gravitational waves remain to be investigated. Besides, another  potential signature in our framework could come from domain walls. We leave  these topics for future work.

To conclude, the  fragmentation of the relaxion  is a generic effect present in all relaxion constructions, even beyond the scenarios discussed above, and should be taken into account in all future relaxion implementations as this can alter predictions substantially.

\vspace{.5cm}

\begin{table}[h]
\centering
\begin{tabular}{|l|l|l|c|c|c|c| c| c| c| c| c| c| c| c|}
\hline
&&&\\
Inflationary  {\it a la} GKR& $\dot{\phi}_0=\dot{\phi}_{SR}$& $\frac{1}{2}\dot{\phi}_0^2 \lessgtr \Lambda_b^4  $ & $\Delta t_1 > H^{-1}$\\
(Sec.~\ref{sec:during inflation}.1)&&&\\
\hline
&&&\\
Inflationary with large barriers& $\dot{\phi}_0=\dot{\phi}_{SR}$& $\frac{1}{2}\dot{\phi}_0^2 = \Lambda_b^4 $ &$\Delta t_1 < H^{-1}$\\
(Sec.~\ref{sec:during inflation}.2)
&&&\\
\hline
&&&\\
Inflationary with particle production& $\dot{\phi}_0=\dot{\phi}_{SR}$& $\frac{1}{2}\dot{\phi}_0^2 > \Lambda_b^4 $ &$\Delta t_1 < H^{-1}$\\
(Secs.~\ref{sec:during inflation}.3, \ref{sec:relaxioninflating}, \ref{subsectionHMTduring})&&&\\
\hline
&&&\\
Non-inflationary &$\dot{\phi}_0 \sim {\cal O}(\Lambda^2)<\dot{\phi}_{SR}$& $\frac{1}{2}\dot{\phi}_0^2 > \Lambda_b^4  $ &$\Delta t_1 < H^{-1}$ \\
(Secs.~\ref{sec:withoutinflation}, \ref{subsectionHMTafter})&&&\\
&&&\\
\hline
\end{tabular}
\caption{Comparison of relaxion models examined in this paper, in terms of the initial relaxion velocity compared to the slow-roll velocity, the initial kinetic energy compared to the final potential barriers and the 1-period rolling time compared to the Hubble time. We stress that the choice $\dot{\phi}_0^2/2 = \Lambda_b^4$ in the `large barrier' case is not a tuning in the initial conditions, but rather a consequence of our \textit{a posteriori} definition of $\Lambda_b$ as the amplitude of the barriers when the relaxion stops and the Higgs vev reaches its measured value (see the comments around Eq.~(\ref{eq:barrierscaling}).}
\label{tab:comparison}
\end{table}

%\clearpage

\begin{table}[t]
\centering
\begin{tabular}{|l|l|l|c|c|c|c| c| c| c| c| c| c| c|}
\hline
&&\\
&Stopping mechanism&Relevant sections\\
Relaxion model &   and corresponding & and figures \\
& benchmark points & \\
&&\\
\hline
\hline
\hline
%&General overview:& Sec.\ref{sec:during inflation} and Fig.\ref{fig:sketchinflation}--\ref{fig:comparisonINFL} \\
&& \\
During inflation with&General overview:&  Sec.\,\ref{sec:during inflation} and Figs.\,\ref{fig:sketch}--\ref{fig:comparisonINFL}\\
Higgs-dependent barriers   & $\boxed{a}$ : Fragmentation: &    Figs.\,\ref{fig:gpLambdafraginfl}--\ref{fig:gpLambdamphifraginfl}    \\
and subdominant $\rho_{\phi}$ & $\boxed{a},\boxed{b}$ : Large barriers: & Figs.\,\ref{fig:gpLambdaINFLBARR},\ref{fig:fLambdabINFLBARR}  \\
(non-QCD model of \cite{Graham:2015cka}) & $\boxed{a},\boxed{b},\boxed{c},\boxed{d}$ : Hubble friction:& Fig.\,\ref{fig:HILambdabINFLGKR} \\
&&\\
\hline
&& \\
After or before inflation, & &     \\
Higgs-dependent barriers & $\boxed{e},\boxed{f},\boxed{g}$: Fragmentation: & Sec.\,\ref{sec:withoutinflation}  \\
$\rho_{\phi}$ can dominate&&Figs.\,\ref{fig:gpLambdaNOINFLATION} and \ref{fig:gpLambNOInflationApp}--\ref{fig:contoursMassApp}, \ref{fig:afterinflation relaxion dom}--\ref{fig:afterinflation relaxion NOT dom} \\
(non-QCD model of \cite{Graham:2015cka})  &&\\
&&\\
\hline
&& \\
During inflation with & &     \\
Higgs-dependent  barriers & $\boxed{h}$: Fragmentation: & Sec.\,\ref{sec:relaxioninflating}\\
$\rho_{\phi}$ dominating && Figs.\,\ref{fig:gpLamb-relaxion dominates} and \ref{fig:gpLambINFrelaxiondominates}--\ref{fig:gpLambINFrelaxiondominatesCONTOURSmphi}, \ref{fig:duringinflation rel dom}\\
(non-QCD model of \cite{Graham:2015cka})  &&\\
&& \\
\hline
\hline
\hline
&& \\
After inflation with & &     \\
Higgs-INdependent  & $\boxed{A},\boxed{B},\boxed{C},\boxed{D},\boxed{E}$: & Sec.\,\ref{subsectionHMTafter}\\
 barriers, $\rho_{\phi}$ dominating &Fragmentation excludes &Figs.\,\ref{fig:gpLambda-HMT-after},\ref{fig:fpLambdac-HMT-after}\\
(model of  \cite{Hook:2016mqo,Fonseca:2018xzp})&the model &\\
&& \\
\hline
&& \\
During inflation, & &     \\
Higgs-INdependent  & $\boxed{F},\boxed{G},\boxed{H}$: & Sec.\,\ref{subsectionHMTduring} \\
 barriers, $\rho_{\phi}$ sub-dominant &EW gauge boson production&Figs.\,\ref{fig:HMT inflation} and \ref{fig:HMT inflation contours gp Lambda}--\ref{fig:HMT inflation contours f Lambdab frag} \\
(model of  \cite{Hook:2016mqo,Fonseca:2018xzp})&&\\
&& \\
\hline
\end{tabular}
\caption{Summary of relaxion models examined in this paper with their  benchmark points analysed in the respective figures.  \label{tab:bench}}
\end{table}

\clearpage

\section*{Acknowledgements}
GS thanks the organisers of the String Pheno'18 conference for providing a stimulating environment relevant for this work. EM and GS are grateful to Sven Krippendorf for important  discussions in the early stages of this project. We thank Valerie Domcke, Yohei Ema, Hyungjin Kim, Kyohei Mukaida, Gilad Perez, and Adam Scherlis for useful discussions.
This work is supported by the Deutsche Forschungsgemeinschaft under Germany’s Excellence Strategy - EXC 2121 ``Quantum Universe'' - 390833306, and by the Cluster of Excellence ``Precision Physics, Fundamental Interactions, and Structure of Matter'' (PRISMA+ EXC 2118/1) funded by  the  German  Research  Foundation (DFG)  within  the German  Excellence  Strategy  (Project  ID  39083149).

%\clearpage

\appendix

\section{Origin of the backreaction term}
\label{sec:explicit model}

Here we discuss the simple UV completion which leads to Higgs-dependent barriers for the relaxion potential used in Section \ref{sec:relaxion growing barriers}.
Let us assume that the relaxion couples to the field strengths $G \widetilde G$ of a new strongly interacting gauge group, and that new fermions $L, L^c, N, N^c$ are charged under this group. Under the Standard Model gauge group, the fermions $L,L^c$ have the same quantum numbers as left- and right-handed leptons respectively, while $N,N^c$ are singlet. The Lagrangian of this model is:
\begin{equation}
\mathcal L = - m_N N N^c - m_L L L^c + y H L N^c + \tilde y H^\dagger L^c N + \frac{\phi}{f}G \widetilde{G} + \mathrm{h.c.}
\end{equation}
With a chiral rotation of the new fermion phases, the last term can be cancelled and the field $\phi$ appears as a phase in the mass terms.
Let us assume that $m_L\gg 4\pi f_\pi \gg m_N$, where $f_\pi$ is the confinement scale.
Integrating out the $L$ fermions one gets
\begin{equation}
\mathcal L = - \left(m_N + y\tilde y\frac{|H|^2}{m_L}\right)N N^c \cos\frac{\phi}{f} \,.
\end{equation}
Below the confinement scale, one can replace $NN^c$ with $\langle NN^c\rangle = 4\pi f_\pi^3$. After EW symmetry breaking, the Higgs can be expanded as $H = \langle h\rangle + h$, where we denote by $\langle h\rangle$ the Higgs VEV. Hence
\begin{equation}
\label{eq:contributionstobarrier}
\mathcal{L} = - \left(m_N + y\tilde y\frac{\langle h\rangle^2}{m_L} + y\tilde y\frac{h^2}{m_L} + y\tilde y\frac{2 \langle h\rangle h}{m_L} \right) (4\pi f_\pi^3) \cos\frac{\phi}{f}
\end{equation}
The mass $m_N$ contains a tree level term and a loop correction,
\begin{equation} \label{eq:mN}
m_N = m_N^0 + \frac{y\tilde y}{16\pi^2}m_L \log\frac{\Lambda}{m_L} \,.
\end{equation}
The key point is  that the third term in (\ref{eq:contributionstobarrier}) generates, when closing the Higgs loop, a contribution to the relaxion potential. This loop has a natural cut-off at $4\pi f_\pi$. The potential is then
\begin{equation} \label{eq:rel pot}
V = \left( m_N^0 + \frac{y\tilde y}{16\pi^2}m_L \log\frac{\Lambda}{m_L} + y \tilde y\frac{\langle h\rangle^2}{m_L} + \frac{y \tilde y}{16\pi^2}\frac{(4\pi f_\pi)^2}{m_L} \right) (4\pi f_\pi^3) \cos\frac{\phi}{f}
\end{equation}
Finally, we impose that the wiggles are dominated by the term proportional to the Higgs VEV $\langle h\rangle^2$. The tree level mass $m_N^0$ can be set to $0$, while comparison with the other terms give
\begin{align}
f_\pi & < \langle h\rangle \\
m_L & < \frac{4\pi \, \langle h\rangle}{\sqrt{\log (\Lambda/m_L)}}
\label{eq:boundonmL}
\end{align}
The scale $f_\pi$ must be below the EW scale, while $m_L$ can go up to the TeV. This strongly constrains the model, because the $N,L$ fermions (or at least one of them) are charged under the Standard Model, and cannot be too light. On the other hand, this feature makes the model testable. Experimental bounds on this model have been discussed in \cite{Beauchesne:2017ukw}.
The backreaction term thus reads $V_\mathrm{br} = \Lambda_\mathrm{br}^2 \langle h\rangle^2 \cos\phi/f$ with
\begin{equation}
\Lambda_\mathrm{br}^2 \approx  \frac{y \tilde y}{m_L} (4\pi f_\pi^3)  < y \tilde y \, f_\pi^2 < y \tilde y \, v_\mathrm{EW}^2 < 16\pi^2 v_\mathrm{EW}^2
\end{equation}
or, using our notation $\Lambda_b^4 \langle h\rangle^2/v_\mathrm{EW}^2 \cos\phi/f$,
\begin{equation}
\Lambda_b^2 < 4\pi v_\mathrm{EW}^2.
\end{equation}

%%%%%%%%%%%%%%%%%%%%%%%%%%%%%%%%%%%%
%%%%%%%%%%%%%%%%%%%%%%%%%%%%%%%%%%%%
%%%%%%%%%%%%%%%%%%%%%%%%%%%%%%%%%%%%

\section{Stopping condition for Higgs-dependent wiggles}
\label{sec:GKR rolling}

\begin{figure}
\includegraphics[width=0.49\hsize]{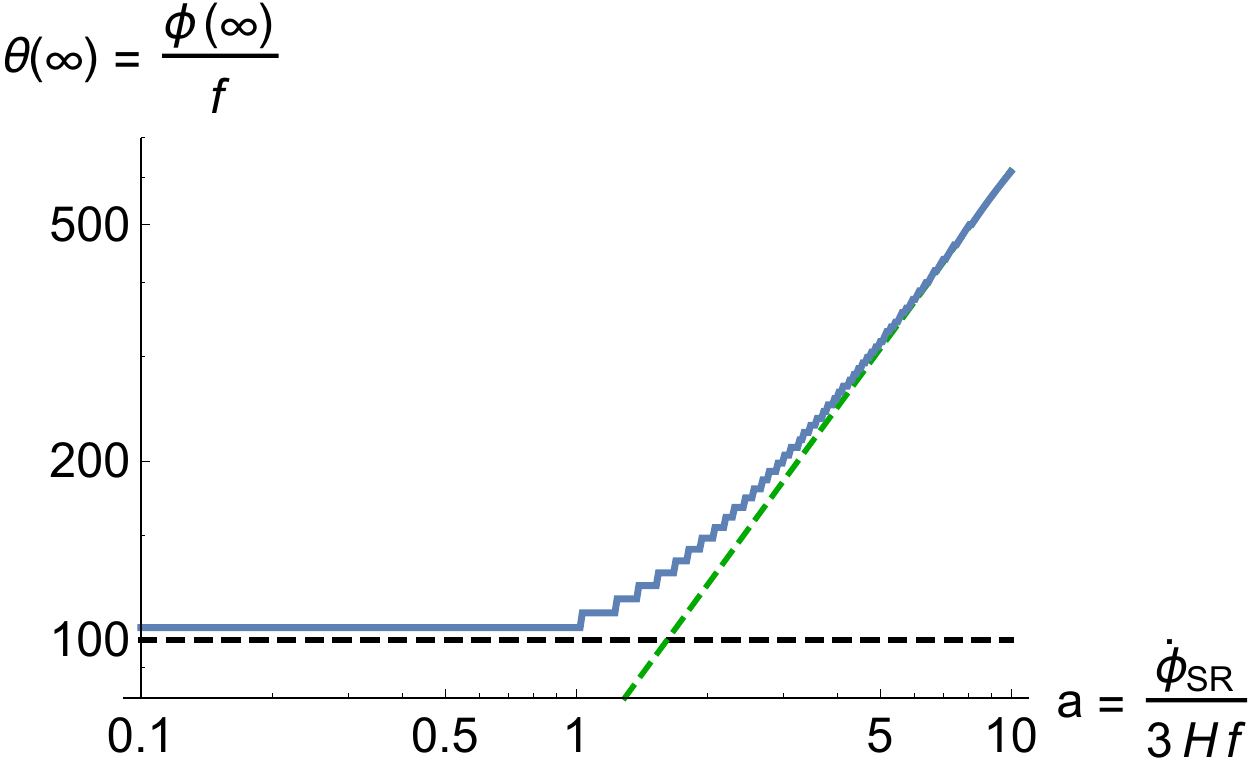}
\includegraphics[width=0.49\hsize]{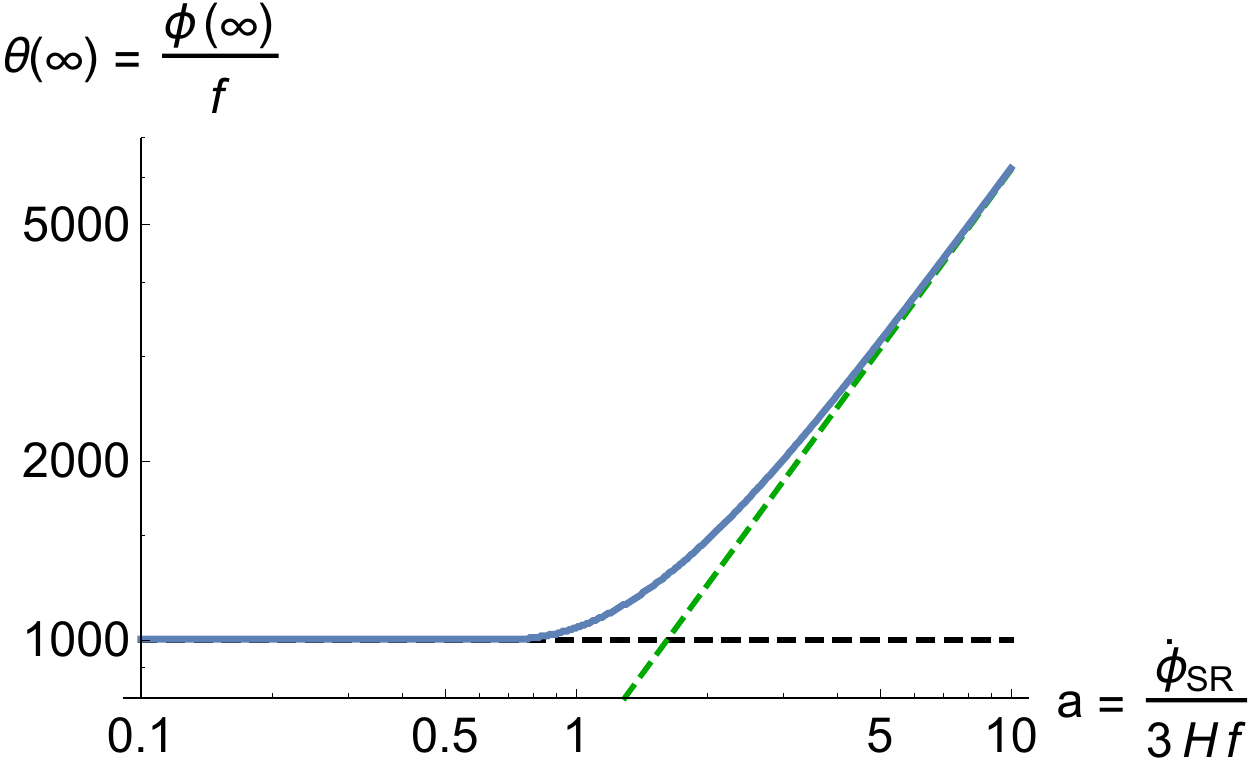}
\caption{\label{fig:GKR rolling}
$\lim_{\tau\to\infty} \theta(\tau)$ of the solution of Eq.~(\ref{eq:GKR EOM 2}).
We take $b = 100$ in the left panel, and 1000 in the right panel.
The black dashed line shows $\theta(\tau) = b$, and the green dashed line shows $\theta(\tau) = (\pi^2/16) ab$.
}
\end{figure}

In this Appendix, we discuss the stopping condition of the relaxion in the case of Higgs-dependent barriers and negligible particle production.
For this, let us solve the following equation of motion:
\begin{align}
\ddot\phi + 3H \dot\phi - \frac{1}{f} \Lambda_b^4(\phi) \sin\frac{\phi}{f} - g\Lambda^3 = 0. \label{eq:GKR EOM}
\end{align}
We consider the evolution from the time when EW symmetry gets broken  and we assume that the Higgs field always tracks its VEV at the minimum of its potential such that $\Lambda_b^4(\phi) = {\tilde{\Lambda}}^3 (\phi - \Lambda/g')$.
The initial condition is $\phi(0)=\Lambda/g'$, $\dot\phi(0)=g\Lambda^3/3H \equiv \dot\phi_{\rm SR}$.
Let us define the dimensionless variables and parameters:
\begin{align}
\theta \equiv \frac{1}{f}\left( \phi - \frac{\Lambda}{g'} \right), \qquad
\tau \equiv \frac{\dot\phi_{\rm SR}}{f} t, \qquad
a \equiv \frac{\dot\phi_{\rm SR}^2}{g\Lambda^3 f} = \frac{\dot\phi_{\rm SR}}{3Hf}, \qquad
b \equiv \frac{g\Lambda^3}{\tilde\Lambda^3}.
\end{align}
Here $a$ controls the size of the Hubble friction.  Small (large) $a$ means strong (weak) Hubble friction.
$b$ controls the shape of the potential. The first local minimum appears around $\phi - \Lambda/g'\simeq bf$.
%In the non-QCD model in Ref.~\cite{Graham:2015cka}, $b$ is calculated as
%\begin{align}
%b = \frac{g}{g'} \frac{m_L \Lambda^2}{y \tilde y \Lambda_s^3}.
%\end{align}
In the UV completion described in Appendix \ref{sec:explicit model}, $b$ is calculated as
\begin{align}
b = \frac{g}{g'} \frac{\lambda}{y\tilde y} \frac{\Lambda^2 m_L}{4\pi f_\pi^3}.
\end{align}
$f_\pi$ should be below the EW scale while $m_L$ should be above ${\cal O}(100)$ GeV because of collider constraints.
Hence, $b \gtrsim \Lambda^2 / v_{\rm EW}^2 \gg 1$ if $g\sim g'$.
Eq.~(\ref{eq:GKR EOM}) can be rewritten as
\begin{align}
\frac{d^2\theta}{d\tau^2} 
+ \frac{1}{a} \left( \frac{d\theta}{d\tau} - 1\right)
- \frac{1}{ab} \theta \sin\theta = 0. \label{eq:GKR EOM 2}
\end{align}
which we solve numerically. The initial condition  is $\theta(0)=0$, $\theta'(0)=1$.
The numerical value of $\lim_{\tau\to\infty} \theta(\tau)$ is shown in Fig.~\ref{fig:GKR rolling}.
We can see $\theta(\infty) = (1/f)(\phi(\infty)-\Lambda/g')$ becomes constant if $a = \dot\phi_{\rm SR} / 3Hf \ll 1$, and $\theta(\infty)$ becomes proportional to $a$ if $a\gg 1$.

In the case $a\ll 1$ ($\dot\phi_{\rm SR} \ll 3Hf$), $\dot\phi$ always tracks $V'/3H$ because the friction term $3H\dot\phi$ is much larger than $\ddot\phi \sim \dot\phi/\Delta t_1$,
where $\Delta t_1  \sim f/\dot\phi_{\rm SR}$ is the time scale to traverse on one wiggle and $H \Delta t_1 \gg 1$.
As a result, the relaxion stops immediately when $V' = 0$ is satisfied.
The terminal value of $\theta$ is $\lim_{\tau\to\infty} \theta(\tau) \simeq a$.
In the language of $\phi$,
\begin{align}
\lim_{t\to\infty} \phi(t) \simeq f \times \frac{g\Lambda^3}{\tilde\Lambda^3}.
\end{align}
This means the rolling of $\phi$ stops immediately when $\Lambda_b^4(\phi) = g\Lambda^3 f$ is satisfied,
\textit{i.e.}, the relaxion stops immediately when the first local minimum appears.
This is the stopping condition which is discussed in the original paper \cite{Graham:2015cka} and labelled as 'GKR' in the main text.

On the other hand, in the case of $a\gg 1$ ($\dot\phi_{\rm SR} \gg 3Hf$),
$\dot\phi$ does not track $V'/3H$ because $H\Delta t_1 \ll 1$.
Hubble friction is inefficient during one wiggle and the energy is almost conserved.
However, the Hubble friction reduces the energy of the relaxion after traversing a large number of wiggles.
Thus, we can define $\phi$-dependent ``energy'' $\varepsilon(\phi)$ and $\dot\phi$ can be written as
\begin{align}
\dot\phi = \sqrt{ 2\varepsilon(\phi) - 2\Lambda_b^4(\phi) \cos\frac{\phi}{f}},
\end{align}
After traversing on a large number of wiggles, 
$\varepsilon(\phi)$ is set to cancel to the Hubble friction and the acceleration by the slope during the time to traverse on one wiggle:
\begin{align}
%\frac{1}{2\pi f} \int_0^{2\pi f} d\delta\phi \sqrt{ 2\varepsilon(\phi) - 2\Lambda_b^4(\phi) \cos\frac{\delta\phi}{f}}  = 3H\dot\phi_{\rm SR}.
3H \int_0^{2\pi f} d\delta\phi \sqrt{ 2\varepsilon(\phi) - 2\Lambda_b^4(\phi) \cos\frac{\delta\phi}{f}}  = g\Lambda^3 \times 2\pi f.
\end{align}
There is no solution for $\varepsilon$ with $\phi > (\pi^2/16)ab f$ in the above equation.
Thus, the relaxion is stopped at
\begin{align}
\lim_{t\to\infty} \phi(t) = \frac{\pi^2}{16} ab f = \frac{\pi^2}{16} \frac{\dot\phi_{\rm SR}^2}{\tilde\Lambda^3}.
\end{align}
At this point, the height of the barrier is roughly equal to the kinetic energy with the slow roll velocity $\dot\phi_{\rm SR}^2/2$.
This stopping condition has not been discussed in Ref.~\cite{Graham:2015cka}.

%%%%%%%%%%%%%%%%%%%%%%%%%%%%%%%%%%%%%%%%%%
%%%%%%%%%%%%%%%%%%%%%%%%%%%%%%%%%%%%
%%%%%%%%%%%%%%%%%%%%%%%%%%%%%%%%%%%%
%%%%%%%%%%%%%%%%%%%%%%%%%%%%%%%%%%%%
%%%%%%%%%%%%%%%%%%%%%%%%%%%%%%%%%%%%
%%%%%%%%%%%%%%%%%%%%%%%%%%%%%%%%%%%%

\section{Effect of quantum fluctuations}\label{sec:quantum fluctuations}

\begin{figure}[h]
%\label{fig:regionB}
\centering
\includegraphics[width=5cm]{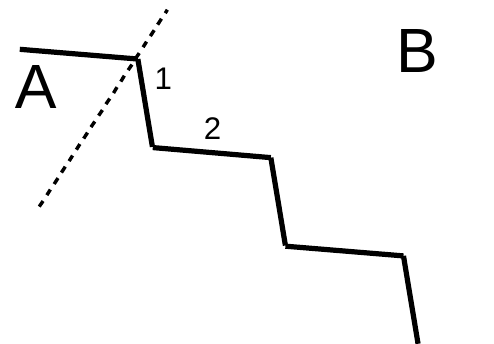}
\caption{
\label{fig:AppendixQuantumFlucSimplifiedPotential}
Simplified potential}
\end{figure}

Reference~\cite{Graham:2015cka} (see Fig.~(2) and the text) points out that, before actually stopping, there is a short phase in which the relaxion dynamics is dominated by quantum jumps of order $H/(2\pi)$. Referring to the notation in that figure, we distinguish 4 periods:
\begin{itemize}
\item[A:] Classical rolling dominates
\item[B:] Classically unstable, large quantum fluctuations
\item[C:] Classically stable, large quantum fluctuations
\item[D:] Classically stable and suppressed quantum fluctuations (lifetime larger than the age of the Universe)
\end{itemize}
Appendix.~\ref{sec:bounce} shows that period C is extremely short. Here we estimate the duration of period B.
To simplify the discussion, we approximate the potential with discrete steps, as in Fig.~\ref{fig:AppendixQuantumFlucSimplifiedPotential}.
In region 1, the velocity of the field is $\dot\phi_1 = (g\Lambda^3+\Lambda_b^4/f)/(3H)$, while in region 2, it is
\begin{equation}
\dot\phi_2 = \frac{g\Lambda^3+\Lambda_b^4/f}{3H} \,.
\end{equation}
We impose $\dot\phi_2 = H/(2\pi)$, and using $\Lambda_b^4 = M^2 h^2$ we obtain the Higgs vev at the end of period B:
\begin{equation}
h_B^2 = \frac{g\Lambda^3 f - \frac{3}{2\pi}H^3 f}{M^2} = v_\ew^2 \left(1-\frac{\frac{3}{2\pi}H^3}{g\Lambda^3} \right) \,,
\end{equation}
where the second term in the parenthesis is $\ll1$ because of the condition of classical rolling.

%
%%%%%%%%%%%%%%%%%%%%%%%%%%%%%%%%%%%%
%%%%%%%%%%%%%%%%%%%%%%%%%%%%%%%%%%%%
%%%%%%%%%%%%%%%%%%%%%%%%%%%%%%%%%%%%

%%%%%%%%%%%%%%%%%%%%%%%%%%%%%%%%%%%%
%%%%%%%%%%%%%%%%%%%%%%%%%%%%%%%%%%%%
%%%%%%%%%%%%%%%%%%%%%%%%%%%%%%%%%%%%
%%%%%%%%%%%%%%%%%%%%%%%%%%%%%%%%%%%%
%%%%%%%%%%%%%%%%%%%%%%%%%%%%%%%%%%%%
%%%%%%%%%%%%%%%%%%%%%%%%%%%%%%%%%%%%

\section{Bounce action for the relaxion}\label{sec:bounce}

Because of the slope term, the vacuum is metastable in the relaxion potential.
The lifetime of the vacuum should be sufficiently longer than the age of the universe.
In this appendix, we discuss the bounce action for the quantum tunneling to lower vacuum.
For simplicity, we assume the height of the barrier is constant and take the following potential:
\begin{align}
V(\phi) = \Lambda_b^4 \cos \frac{\phi}{f} - g\Lambda^3 \phi.
\end{align}
The local minima of this potential are at
\begin{align}
\phi = (2n-1)\pi f - f \arcsin \kappa, \qquad
\end{align}
Here $\kappa$ is defined as
\begin{align}
\kappa \equiv \frac{g\Lambda^3}{\Lambda_b^4/f}.
\end{align}
$\kappa$ should satisfy $|\kappa|<1$ to have local minima. In this note, we assume $\kappa$ is positive.
The bounce equation for the vacuum decay is
\begin{align}
\frac{d^2 \phi}{dr^2} + \frac{3}{r} \frac{d\phi}{dr} + \frac{\Lambda_b^4}{f} \sin \frac{\phi}{f} + g\Lambda^3 = 0. \label{eq:bounce eq}
\end{align}
The boundary conditions for $\phi$ are
\begin{align}
\lim_{r\to\infty} \phi(z) = f(-\pi + \arcsin\kappa), \qquad
\frac{d\phi}{dr}\biggr|_{r=0} = 0. \label{eq:bounce bc}
\end{align}
The bounce action is given as
\begin{align}
S_E = 2\pi^2 \int_0^\infty dr~r^3 \left[ \frac{1}{2}\left( \frac{d\phi}{dr} \right)^2
+ \Lambda_b^4 \cos \frac{\phi}{f} - g\Lambda^3 \phi 
- \left[ \Lambda_b^4 \cos \frac{\phi(\infty)}{f} - g\Lambda^3 \phi(\infty) \right] \right]. \label{eq:bounce SE}
\end{align}

For later convenience, let us define dimensionless parameters $\theta$ and $z$:
\begin{align}
\theta \equiv \frac{\phi}{f}, \qquad
z \equiv \frac{\Lambda_b^2 r}{f}.
\end{align}
By using Eqs.~(\ref{eq:bounce eq}, \ref{eq:bounce bc}), 
we obtain the EOM and the boundary condition for $\theta$:
\begin{align}
\frac{d^2 \theta}{dz^2} + \frac{3}{z} \frac{d\theta}{dz} + \sin \theta + \kappa = 0, \\
\lim_{z\to\infty} \theta(z) = -\pi + \arcsin\kappa, \qquad
\frac{d\theta}{dz}\biggr|_{z=0} = 0.
\end{align}
The bounce action Eq.~(\ref{eq:bounce SE}) can be rewritten by using $\theta(z)$:
\begin{align}
S_E = \frac{2\pi^2 f^4}{\Lambda_b^4} \int_0^\infty dz~z^3 \left[ \frac{1}{2}\left( \frac{d\theta}{dz} \right)^2 + \cos\theta - \kappa \theta - 
\left[ \cos\theta(\infty) - \kappa\theta(\infty) \right]  \right],
\end{align}

\subsection{Large barrier limit ($\kappa \ll 1$)}
Let us discuss the case with $\kappa \ll 1$, \textit{i.e.}, the case with large barrier.
In this case, we can use thin wall approximation for the bounce calculation.

Let us assume there is a wall at $z=z_0$. Then, the bounce action is
\begin{align}
S_E \times \frac{\Lambda_b^4}{2\pi^2 f^4}
&\simeq
-\frac{1}{4}z_0^4 \times 2\pi \kappa
+ z_0^3 \int_{-\pi}^{\pi} d\theta \frac{d\theta}{dz}. \nonumber\\
&=
-\frac{1}{4}z_0^4 \times 2\pi \kappa
+ 8z_0^3. \label{eq:bounce SE thin wall}
\end{align}
Here we used ``energy conservation law'' around the wall:
\begin{align}
\frac{1}{2} \left(\frac{d\theta}{dz}\right)^2 = 1 + \cos\theta
\end{align}
The solution is
\begin{align}
\frac{d\theta}{dz} = 2\cos\frac{\theta}{2}.
\end{align}
The bounce action $S_E$ given in Eq.~(\ref{eq:bounce SE thin wall}) is extremized at
\begin{align}
z_0 = \frac{12}{\kappa \pi}.
\end{align}
Thus, the Euclidean action is given as
\begin{align}
S_E = \frac{6912}{\kappa^3 \pi} \frac{f^4}{\Lambda_b^4}.
%,
%\qquad
%S_E \times \frac{\Lambda_b^4}{2\pi^2 f^4}
%= \frac{3456}{\kappa^3 \pi^3}.
\label{eq:SE formula thin wall}
\end{align}

\subsection{Small barrier limit ($\kappa \simeq 1)$}
Let us discuss the case with $\kappa \simeq 1$, \textit{i.e.}, the case with small barrier.
This situation is realized in the original GKR scenario \cite{Graham:2015cka} which is discussed in Section \ref{sec:during inflation}.
In this case, we can expand the potential around the local minimum.
\begin{align}
\cos\theta - \kappa\theta \simeq  \cos\theta_0 - \kappa\theta_0 + \frac{1}{2} \sqrt{1-\kappa^2} (\theta-\theta_0 )^2 - \frac{1}{6}(\theta-\theta_0)^3,
\end{align}
where $\theta_0 = -\pi + \arcsin \kappa$. Let us define $\delta\theta = \theta-\theta_0$. Then,
the bounce action $S_E$ given in Eq.~(\ref{eq:bounce SE thin wall}) is
\begin{align}
S_E
&\approx \frac{2\pi^2 f^4}{\Lambda_b^4} \int_0^\infty dz~z^3 \left[ \frac{1}{2}\left( \frac{d\delta\theta}{dz} \right)^2 + \frac{1}{2}\sqrt{1-\kappa^2} \delta\theta^2 - \frac{1}{6}\delta\theta^3  \right] \nonumber\\
&= \frac{2\pi^2 f^4}{\Lambda_b^4} \sqrt{1-\kappa^2} \times
 \int_0^\infty dw~w^3 \left[ \frac{1}{2}\left( \frac{d\xi}{dw} \right)^2 + \frac{1}{2} \xi^2 - \frac{1}{6}\xi^3  \right]. \label{eq:bounce SE small barrier}
\end{align}
Here $\xi \equiv (1-\kappa^2 )^{-1/2} \delta\theta$ and $w \equiv (1-\kappa^2)^{1/4} z$.
Then, by using Eqs.~(\ref{eq:bounce eq}, \ref{eq:bounce bc}), 
we obtain the EOM and the boundary condition for $\xi$:
\begin{align}
\frac{d^2\xi}{dw^2} + \frac{3}{w} \frac{d\xi}{dw} - \xi + \frac{1}{2}\xi^2 = 0, \\
\lim_{w\to\infty } \xi = 0, \qquad
\frac{d\xi}{dw} \biggr|_{w=0}= 0.
\end{align}
By using the above equations, the integral in Eq.~(\ref{eq:bounce SE small barrier}) can be evaluated numerically:
\begin{align}
 \int_0^\infty dw~w^3 \left[ \frac{1}{2}\left( \frac{d\xi}{dw} \right)^2 + \frac{1}{2} \xi^2 - \frac{1}{6}\xi^3  \right]
=
41.4.
\label{eq:SE formula small barrier}
\end{align}
Therefore, the bounce action with the small barrier is given as
\begin{align}
S_E
&\simeq \frac{2\pi^2 f^4}{\Lambda_b^4} \sqrt{1-\kappa^2} \times 41.4 .
\end{align}

\subsection{Numerical estimation of a lowerbound on $f/\Lambda_b$}
\begin{figure}[h]
\centering
\includegraphics[width=0.52\hsize]{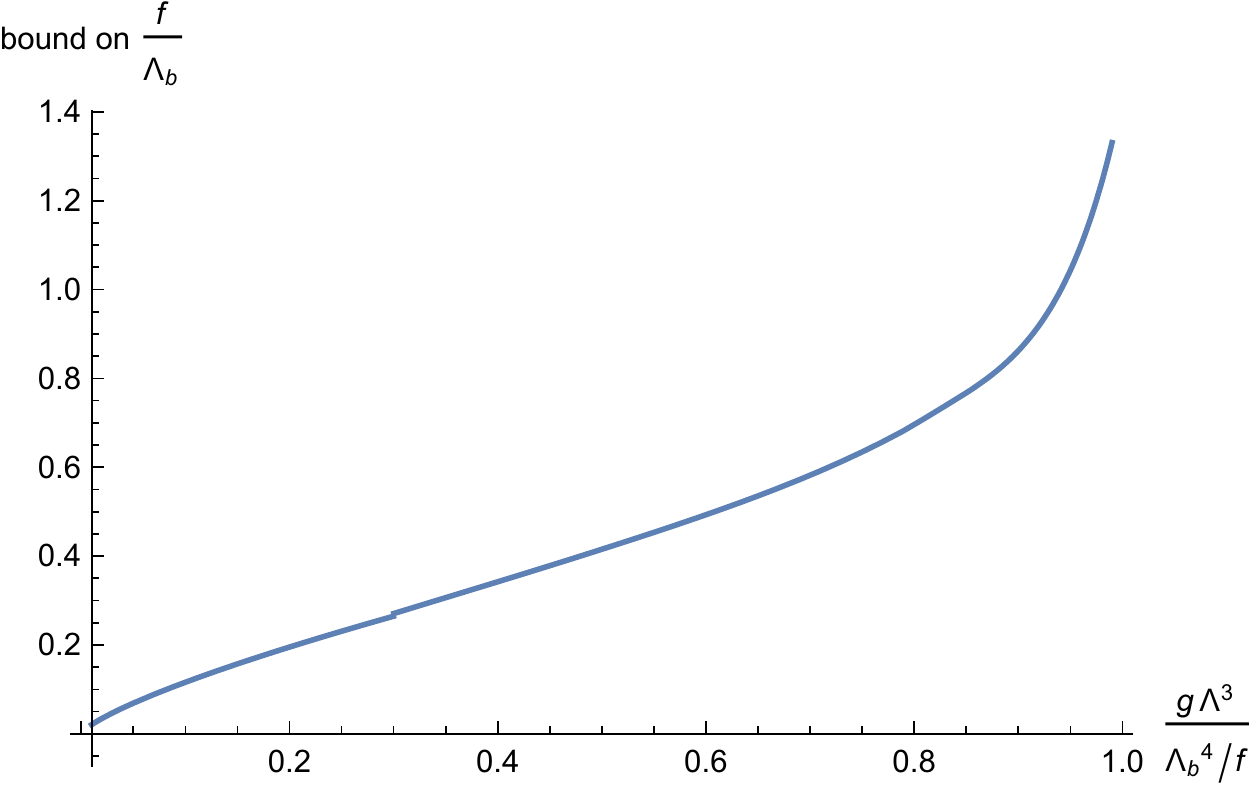}\\[5mm]
\includegraphics[width=0.52\hsize]{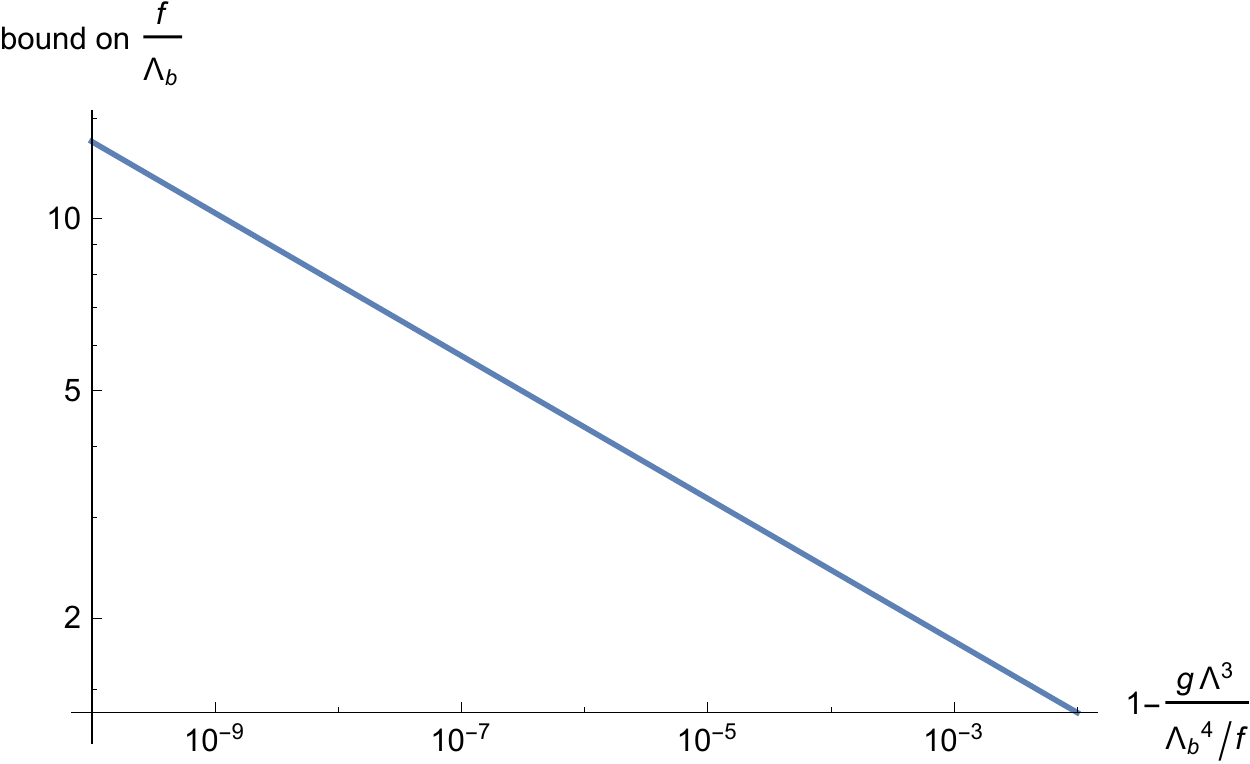}
\caption{The lowerbound on $f/\Lambda_b$ as a function of $\kappa \equiv g\Lambda^3/(\Lambda_b^4/f)$.
The lower panel shows the lowerbound on $f/\Lambda_b$ for $\kappa \simeq 1$.
}
\label{fig:AppendixQuantumFlucBound}
\end{figure}

For given $\kappa$, the bounce action $S_E$ is determined by the value of $f/\Lambda_b$.
The lifetime of our vacuum should be longer than the age of the universe,
and this can be realized if $S_E \gtrsim 400$.
This condition gives a lowerbound on $f/\Lambda_b$ as a function of $\kappa$.
Fig.~\ref{fig:AppendixQuantumFlucBound} show the numerical results on the lowerbound on $f/\Lambda_b$.
For $f/\Lambda_b = 10$, the lifetime of the false vacuum is longer than the age of the universe if $1-\kappa > 1.2 \times 10^{-9}$.
For larger $f/\Lambda_b$, this constraint on $\kappa$ becomes weaker.
Thus, we can see that the vacuum is stable enough if there is a mild hierarchy between $f$ and $\Lambda_b$
unless $\kappa$ is extremely close to $1$.

%%%%%%%%%%%%%%%%%%%%%%%%%%%%%%%%%%%%%%%
%%%%%%%%%%%%%%%%%%%%%%%%%%%%%%%%%%%%%%%%%%%%%%%%%%%%%%%%%%%%%%%%%%%%%%%%%%%%%%%%%%%%%%%%%%%%%%%%%%%%%%%%%%%%%%%%%%%%%

\clearpage

\section{Plots of the parameter space}
%\section{Plots for the relaxion with Higgs-dependent barriers}
\label{sec:detailedplots}

\subsection{Relaxation with Higgs-dependent barriers during Inflation}
%\subsection{Relaxation during Inflation}
\label{sec:GKRcondition}

\begin{figure}[h!]
\centering
\includegraphics[width=.32\textwidth]{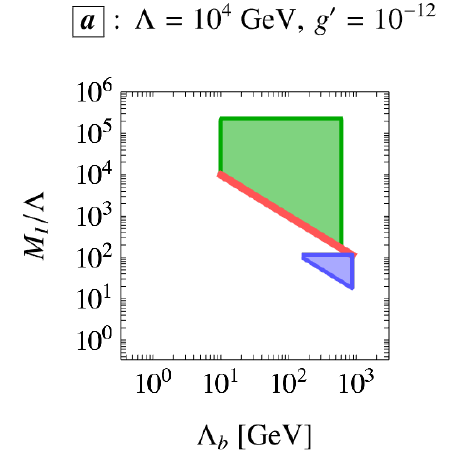}
\includegraphics[width=.32\textwidth]{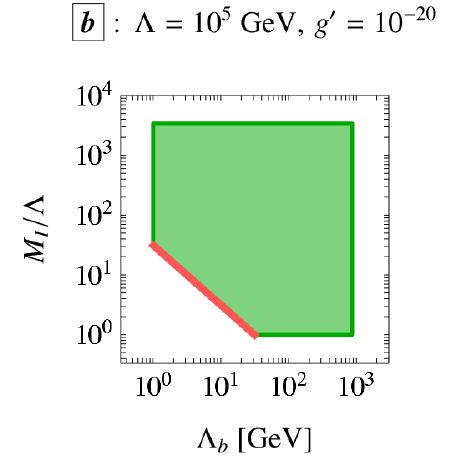}\\
\includegraphics[width=.32\textwidth]{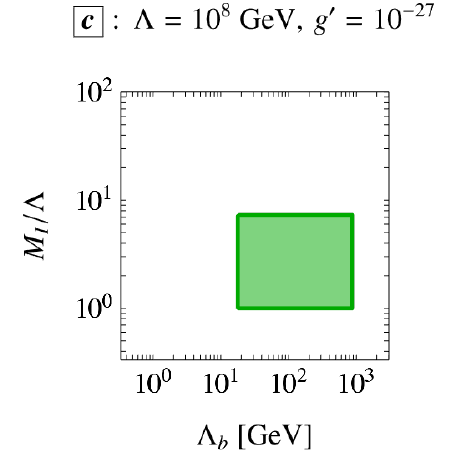}
\includegraphics[width=.32\textwidth]{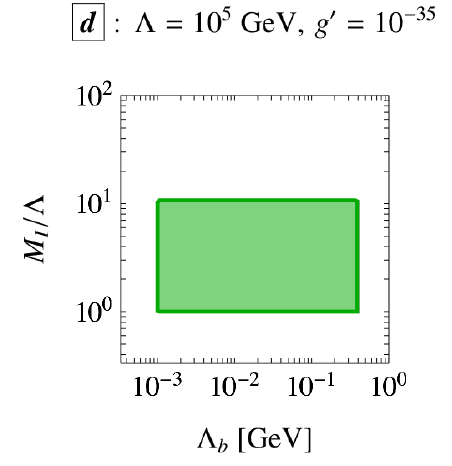}
\caption{\label{fig:comparisonINFL Lb Mi/L}Same as in Fig.~\ref{fig:comparisonINFL}, in the plane $\Lambda_b$, $M_I/\Lambda$. In the Fragmentation case, the scale of inflation $M_I$ is closer to the cut-off $\Lambda$ than in the other scenarios.}
\end{figure}

\begin{figure}[h!]
\centering
\begin{minipage}{.49\textwidth}
\includegraphics[width=\textwidth]{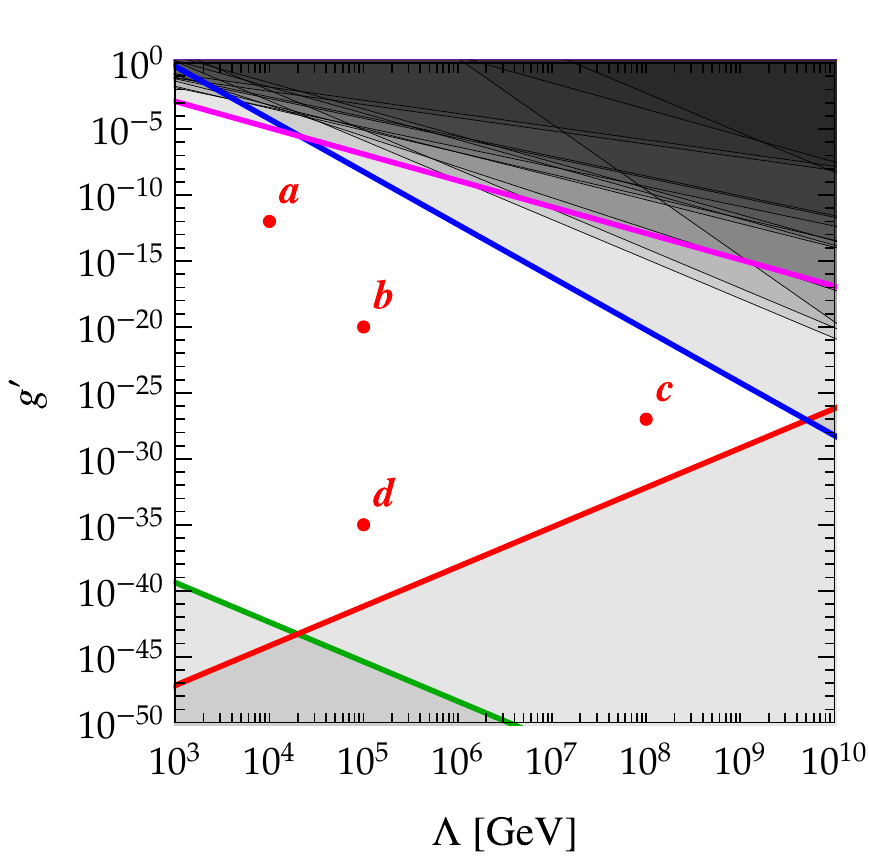}
\end{minipage}
\begin{minipage}{.49\textwidth}
\begin{itemize}
\item[{\tikz\fill[rounded corners = .5mm, blue](0,0)rectangle(0.3,0.2);}] Symmetry breaking pattern \Eq{eq:eftvalidity} and microscopic origin of the barriers \Eq{eq:stability}
\item[{\tikz\fill[rounded corners = .5mm, red](0,0)rectangle(0.3,0.2);}] Classical rolling \Eq{eq:i1} and relaxion subdominant with respect to the inflaton \Eq{eq:i4}
\item[{\tikz\fill[rounded corners = .5mm, Green](0,0)rectangle(0.3,0.2);}] Reheating \Eq{eq:BBN} and sub-Planckian decay constant \Eq{eq:fsubpl}
\item[{\tikz\fill[rounded corners = .5mm, magenta](0,0)rectangle(0.3,0.2);}] \Eq{eq:eftvalidity} and precision of the Higgs mass scanning \Eq{eq:enoughprecision}
\end{itemize}
\end{minipage}
\caption{\label{fig:gpLambdaINFLGKR}Details of the origin of the constraints delimitating the green region presented in Section \ref{sec:during inflation} and Fig.~\ref{fig:summaryINFL} corresponding to the case where relaxation happens through Higgs-dependent barriers, during a stage of inflation, where the relaxion is a subdominant component of the energy density of the universe and where the relaxion stops because of {\it Hubble friction  (so-called  GKR case)}.}
\end{figure}

\begin{figure}
\centering
\includegraphics[width=.4\textwidth]{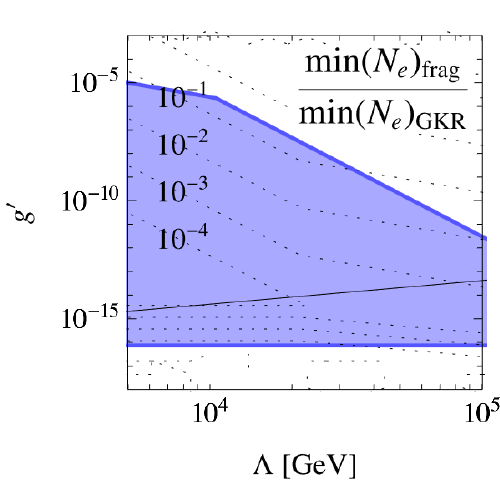}
\caption{\label{fig:efoldsduringinflationGKRfrag}%
In the scenario where relaxation takes place during an inflation era, ratio of the minimal number of efolds in the ``GKR'' and ``fragmentation'' scenarios discussed in Sec.~\ref{sec:during inflation}. The coloured region corresponds to the region where fragmentation can stop the relaxion as  an alternative mechanism to Hubble friction like in the GKR scenario.  This plot shows that a smaller number if efolds is required to stop with fragmentation. Below the solid line the field excursion $\Lambda/g'$ is super-Planckian.}
\end{figure}

\begin{figure}
\centering
\includegraphics[width=.49\textwidth]{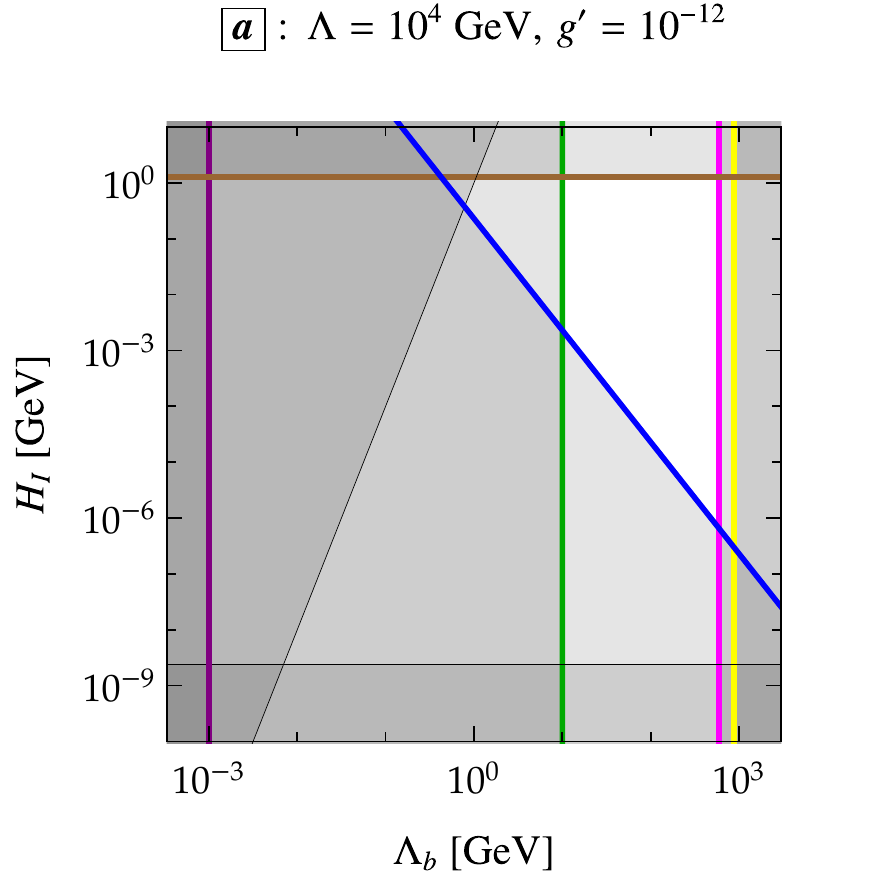}
\includegraphics[width=.49\textwidth]{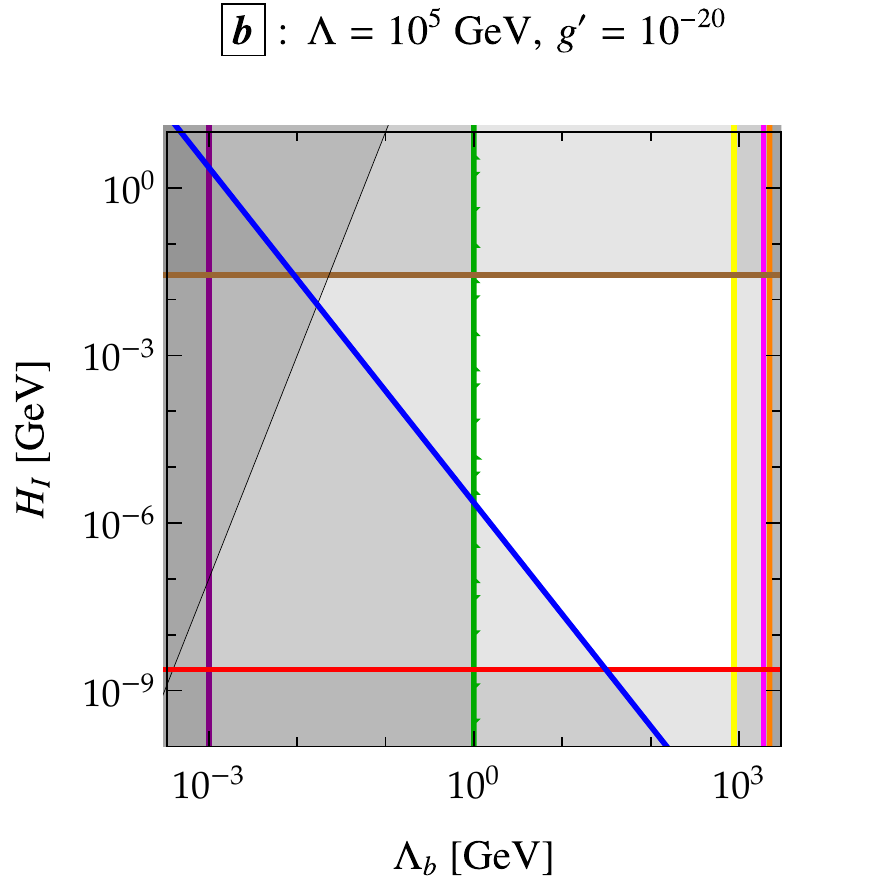}
\includegraphics[width=.49\textwidth]{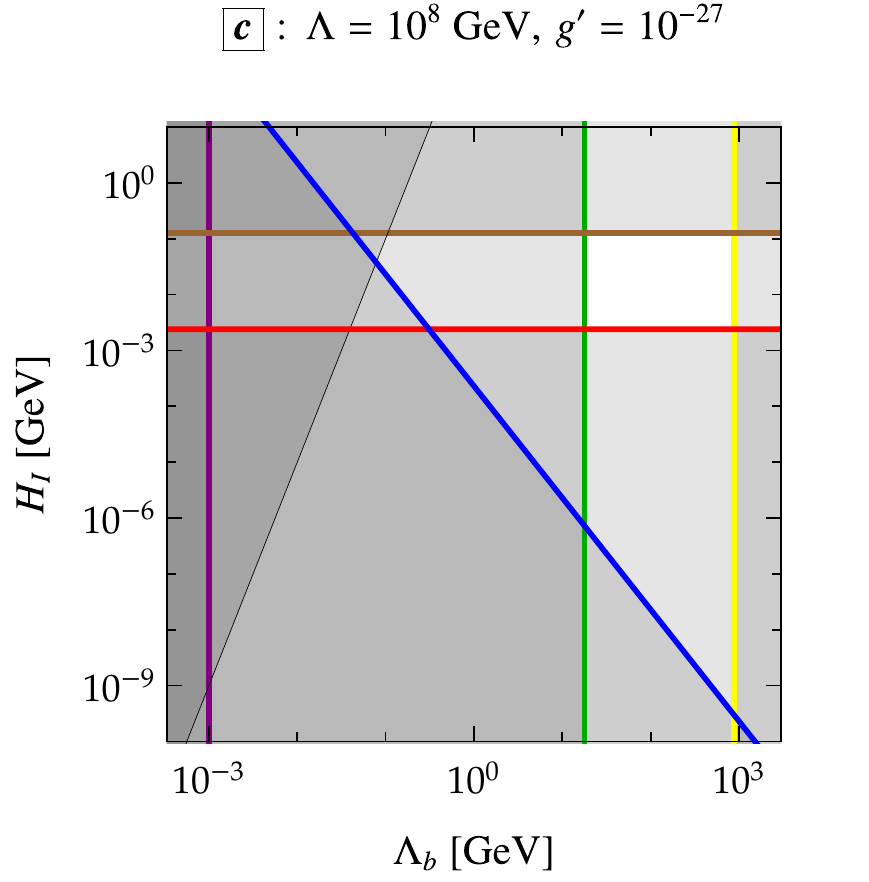}
\includegraphics[width=.49\textwidth]{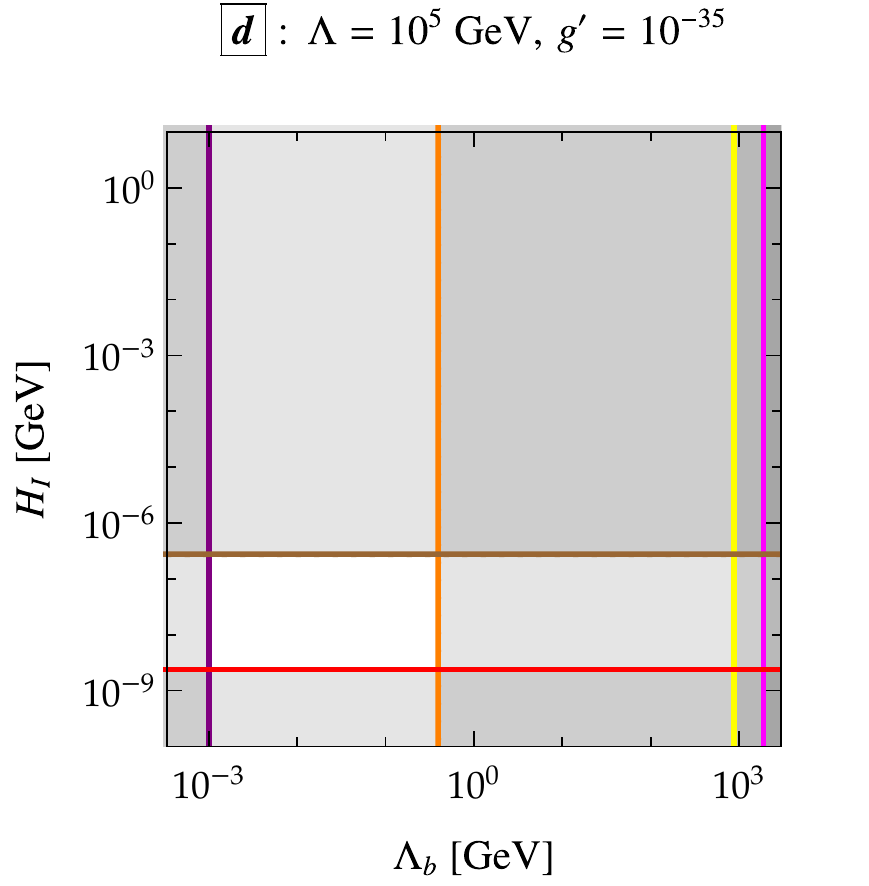}
\begin{minipage}{1.0\textwidth}
\begin{multicols}{1}
\begin{itemize}
\item[{\tikz\fill[rounded corners = .5mm, blue](0,0)rectangle(0.3,0.2);}] Going over 1 wiggle in more than 1 Hubble time Eq.~(\ref{eq:1wiggleslow})
\item[{\tikz\fill[rounded corners = .5mm, Green](0,0)rectangle(0.3,0.2);}] Symmetry breaking pattern Eq.~(\ref{eq:eftvalidity})
\item[{\tikz\fill[rounded corners = .5mm, Brown](0,0)rectangle(0.3,0.2);}] Classical rolling Eq.~(\ref{eq:i1})
\item[{\tikz\fill[rounded corners = .5mm, Red](0,0)rectangle(0.3,0.2);}] Relaxion subdominant with respect to the inflaton \Eq{eq:i4}
\item[{\tikz\fill[rounded corners = .5mm, Yellow](0,0)rectangle(0.3,0.2);}] Microscopic origin of the barriers \Eq{eq:stability}
\item[{\tikz\fill[rounded corners = .5mm, magenta](0,0)rectangle(0.3,0.2);}] Precision of the Higgs mass scanning Eq.~(\ref{eq:enoughprecision})
\item[{\tikz\fill[rounded corners = .5mm, Orange](0,0)rectangle(0.3,0.2);}] Sub-Planckian decay constant \Eq{eq:fsubpl}
\item[{\tikz\fill[rounded corners = .5mm, Purple](0,0)rectangle(0.3,0.2);}] Reheating \Eq{eq:BBN}

\end{itemize}
\end{multicols}
\end{minipage}
\caption{\label{fig:HILambdabINFLGKR} Details of the origin of the constraints delimitating the green regions presented in Section \ref{sec:during inflation} and Fig.~\ref{fig:comparisonINFL}, for each benchmark point $a$, $b$, $c$, $d$, corresponding to the case where relaxation happens through Higgs-dependent barriers, during a stage of inflation, where the relaxion is a subdominant component of the energy density of the universe and where the relaxion stops because of {\it Hubble friction  (so-called  GKR case)}.}
\end{figure}

%%%%%%%%%%%%%%%%%%%%%%%%%%%%%%%%%%%%%%%%%%%%%%%%%%%%%%%%%%%%%%%%%%%%%%%%%%%%%%%%%%%
%%%%%%%%%%%%%%%%%%%%%%%%%%%%%%%%%%%%%%%%%%%%%%%%%%%%%%%%%%%%%%%%%%%%%%%%%%%%%%%%%%%

\begin{figure}
\centering
\begin{minipage}{.49\textwidth}
\includegraphics[width=\textwidth]{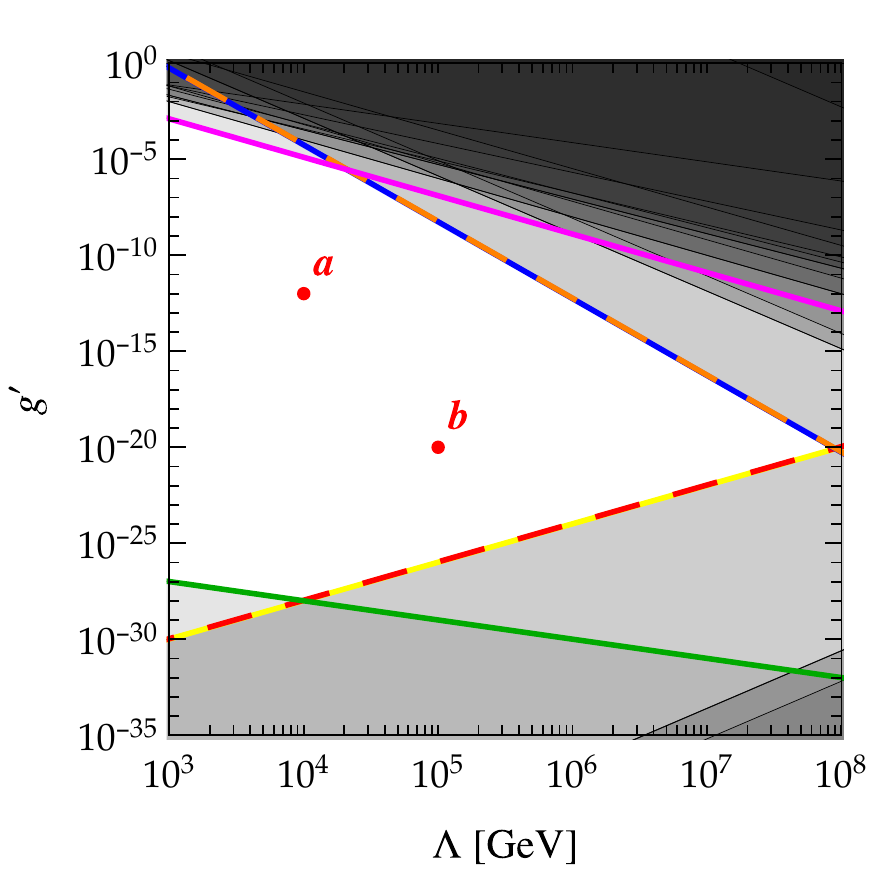}
\end{minipage}
\begin{minipage}{.49\textwidth}
\begin{itemize}
\item[{\tikz\fill[rounded corners = .5mm, blue](0,0)rectangle(0.3,0.2);}] Symmetry breaking pattern \Eq{eq:eftvalidity}, going over 1 wiggle in less than 1 Hubble time \Eq{eq:1wigglefast} and microscopic origin of the barriers \Eq{eq:stability}
\item[{\tikz\fill[rounded corners = .5mm, orange](0,0)rectangle(0.3,0.2);}] \Eq{eq:eftvalidity}, large barriers \Eq{eq:visiblewiggles} and \Eq{eq:stability}
\item[{\tikz\fill[rounded corners = .5mm, red](0,0)rectangle(0.3,0.2);}] \Eq{eq:eftvalidity}, \Eq{eq:1wigglefast} and relaxion subdominant with respect to inflaton \Eq{eq:i4}
\item[{\tikz\fill[rounded corners = .5mm, yellow](0,0)rectangle(0.3,0.2);}] \Eq{eq:eftvalidity}, large barriers \Eq{eq:visiblewiggles} and \Eq{eq:i4}
\end{itemize}
\end{minipage}
\begin{itemize}
\item[{\tikz\fill[rounded corners = .5mm, magenta](0,0)rectangle(0.3,0.2);}] \Eq{eq:eftvalidity} and precision of the Higgs mass scanning \Eq{eq:enoughprecision}
\item[{\tikz\fill[rounded corners = .5mm, Green](0,0)rectangle(0.3,0.2);}] Reheating \Eq{eq:BBN} and \Eq{eq:i4}
\end{itemize}
\caption{\label{fig:gpLambdaINFLBARR} Details of the origin of the constraints delimitating the red region presented in Section \ref{sec:during inflation} and Fig.~\ref{fig:summaryINFL} corresponding to the case where relaxation happens through Higgs-dependent barriers, during a stage of inflation, where the relaxion is a subdominant component of the energy density of the universe and where the relaxion stops because of {\it large barriers}.}
\end{figure}

\begin{figure}
\centering
\includegraphics[width=.49\textwidth]{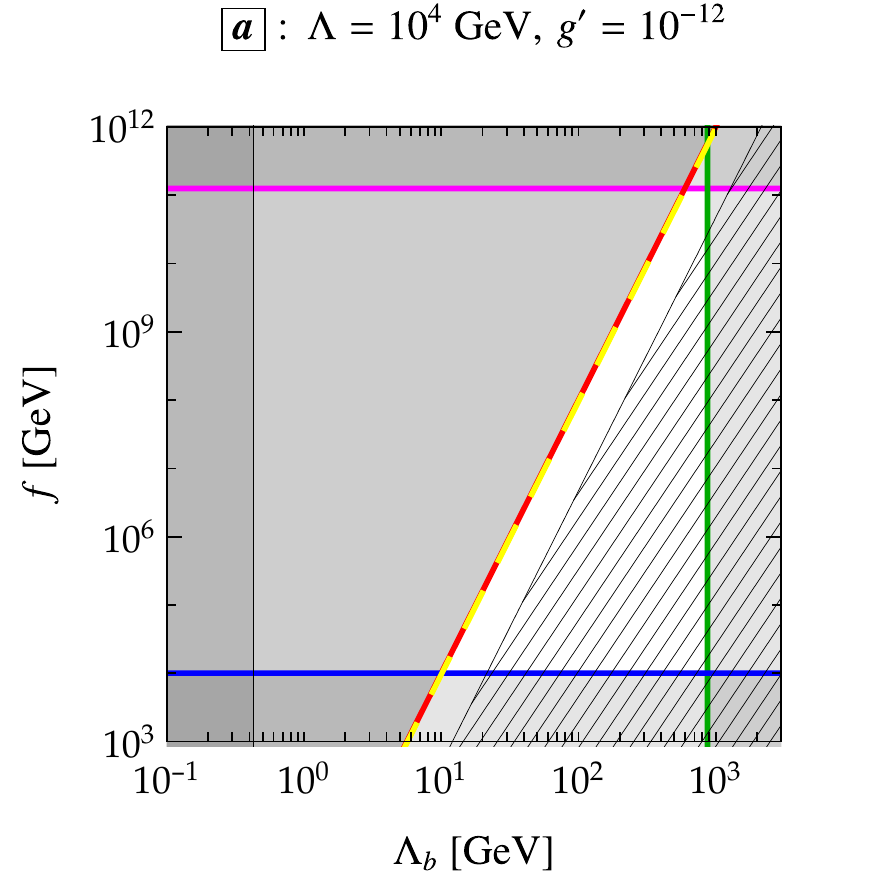}
\includegraphics[width=.49\textwidth]{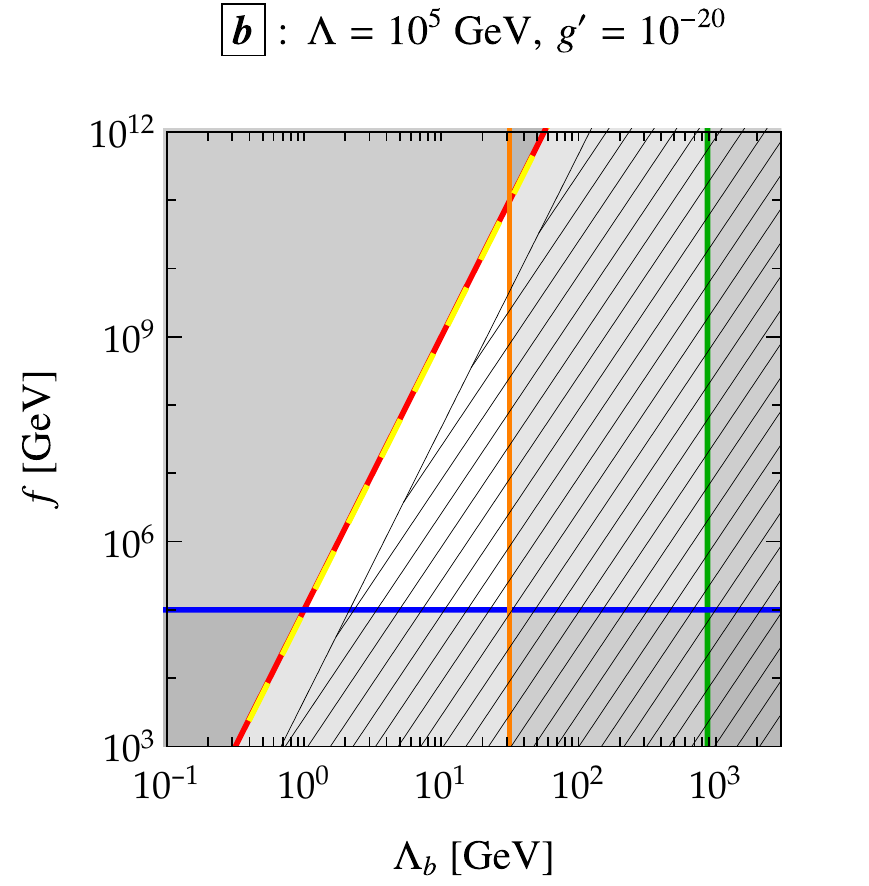}
\begin{minipage}{1.0\textwidth}
\begin{multicols}{1}
\begin{itemize}
\item[{\tikz\fill[rounded corners = .5mm, Blue](0,0)rectangle(0.3,0.2);}] Symmetry breaking Eq.~(\ref{eq:eftvalidity})
\item[{\tikz\fill[rounded corners = .5mm, Green](0,0)rectangle(0.3,0.2);}] Microscopic origin of the barriers \Eq{eq:stability}
\item[{\tikz\fill[rounded corners = .5mm, Red](0,0)rectangle(0.3,0.2);}] Large barriers \Eq{eq:visiblewiggles}
\item[{\tikz\fill[rounded corners = .5mm, Yellow](0,0)rectangle(0.3,0.2);}] Going over 1 wiggle in more than 1 Hubble time Eq.~(\ref{eq:1wiggleslow})
\item[{\tikz\fill[rounded corners = .5mm, magenta](0,0)rectangle(0.3,0.2);}] Precision of the Higgs mass scanning Eq.~(\ref{eq:enoughprecision})
\item[{\tikz\fill[rounded corners = .5mm, Orange](0,0)rectangle(0.3,0.2);}] Relaxion subdominant with respect to inflaton \Eq{eq:i4}
\end{itemize}
\end{multicols}
\end{minipage}
\caption{\label{fig:fLambdabINFLBARR} Details of the origin of the constraints delimitating the red regions presented in Section \ref{sec:during inflation} and Fig.~\ref{fig:comparisonINFL}, for benchmark points $a$, and $b$, corresponding to the case where relaxation happens through Higgs-dependent barriers, during a stage of inflation, where the relaxion is a subdominant component of the energy density of the universe and where the relaxion stops because of {\it large barriers}.}
\end{figure}

%%%%%%%%%%%%%%%%%%%%%%%%%%%%%%%%%%%%%%%%%%%%%%%%%%%%%%%%%%%%%%%%%%%%%%%%%%%%%%%%%%%
%%%%%%%%%%%%%%%%%%%%%%%%%%%%%%%%%%%%%%%%%%%%%%%%%%%%%%%%%%%%%%%%%%%%%%%%%%%%%%%%%%%

\begin{figure}
\centering
\begin{minipage}{.49\textwidth}
\includegraphics[width=\textwidth]{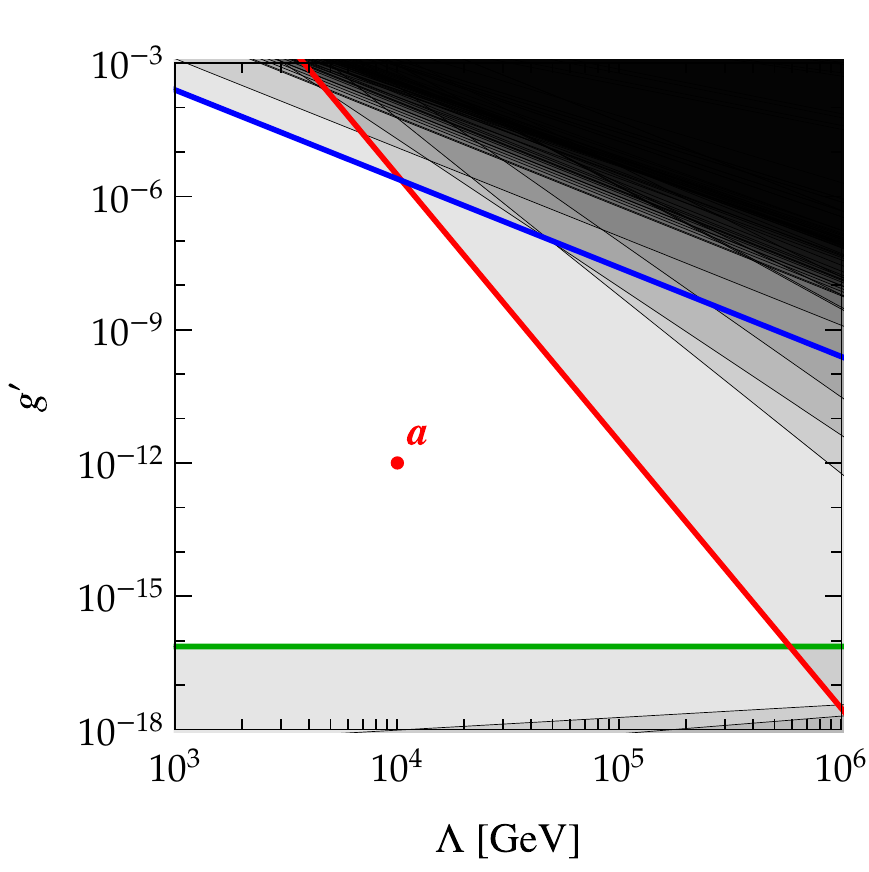}
\end{minipage}
\begin{minipage}{.49\textwidth}
\begin{itemize}
\item[{\tikz\fill[rounded corners = .5mm, blue](0,0)rectangle(0.3,0.2);}] Symmetry breaking pattern \Eq{eq:eftvalidity} and large velocity \Eq{eq:large velocity}
\item[{\tikz\fill[rounded corners = .5mm, red](0,0)rectangle(0.3,0.2);}] \Eq{eq:eftvalidity}, efficient fragmentation \Eq{eq:condition 3} and microscopic origin of the barriers \Eq{eq:stability}
\item[{\tikz\fill[rounded corners = .5mm, Green](0,0)rectangle(0.3,0.2);}] Relaxion subdominant with respect to inflaton \Eq{eq:i4} and \Eq{eq:condition 3}
\end{itemize}
\end{minipage}
\caption{\label{fig:gpLambdafraginfl}Details of the origin of the constraints delimitating the blue region presented in Section \ref{sec:during inflation} and Fig.~\ref{fig:summaryINFL} corresponding to the case where relaxation happens through Higgs-dependent barriers, during a stage of inflation, where the relaxion is a subdominant component of the energy density of the universe and where the relaxion stops because of {\it fragmentation}.}
\end{figure}

\begin{figure}
\centering
\begin{minipage}{.49\textwidth}
\includegraphics[width=\textwidth]{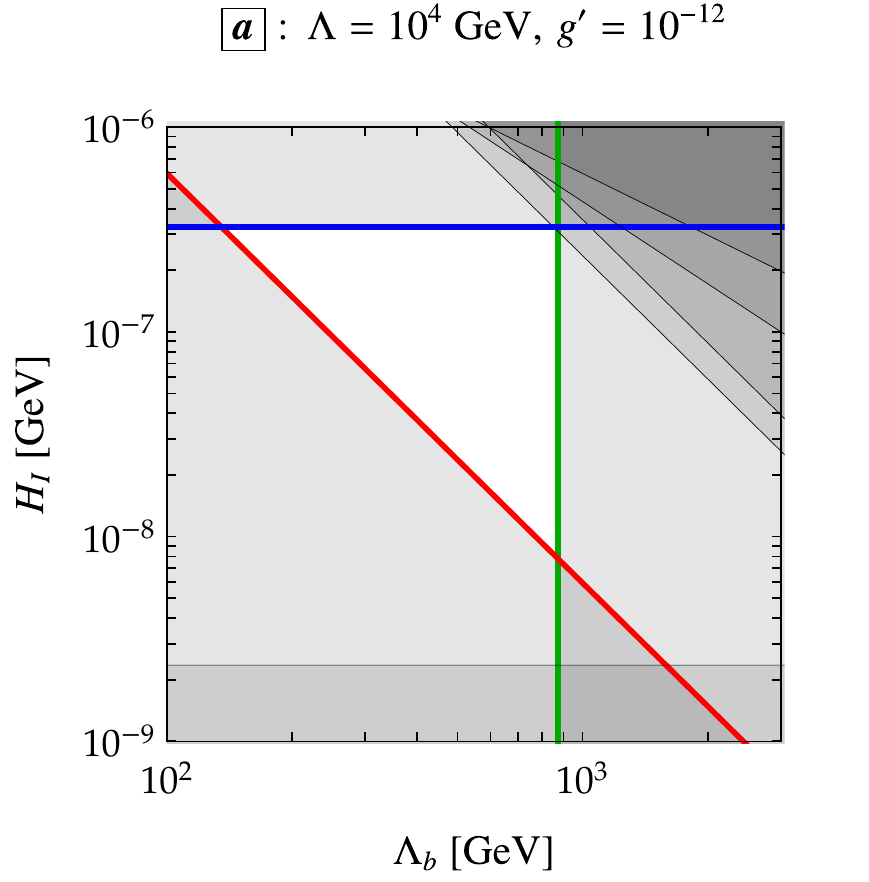}
\end{minipage}
\begin{minipage}{.49\textwidth}
\begin{itemize}
\item[{\tikz\fill[rounded corners = .5mm, blue](0,0)rectangle(0.3,0.2);}] Efficient fragmentation \Eq{eq:condition 3}
\item[{\tikz\fill[rounded corners = .5mm, red](0,0)rectangle(0.3,0.2);}] Symmetry breaking \Eq{eq:eftvalidity}
\item[{\tikz\fill[rounded corners = .5mm, Green](0,0)rectangle(0.3,0.2);}] Microscopic origin of the barriers \Eq{eq:stability}
\end{itemize}
\end{minipage}
\caption{\label{fig:HILambdabfraginfl}Details of the origin of the constraints delimitating the blue regions presented in Section \ref{sec:during inflation} and Fig.~\ref{fig:comparisonINFL}, for benchmark points $a$, corresponding to the case where relaxation happens through Higgs-dependent barriers, during a stage of inflation, where the relaxion is a subdominant component of the energy density of the universe and where the relaxion stops because of {\it fragmentation}.}
\end{figure}

\begin{figure}
\centering
\begin{minipage}{.49\textwidth}
\includegraphics[width=\textwidth]{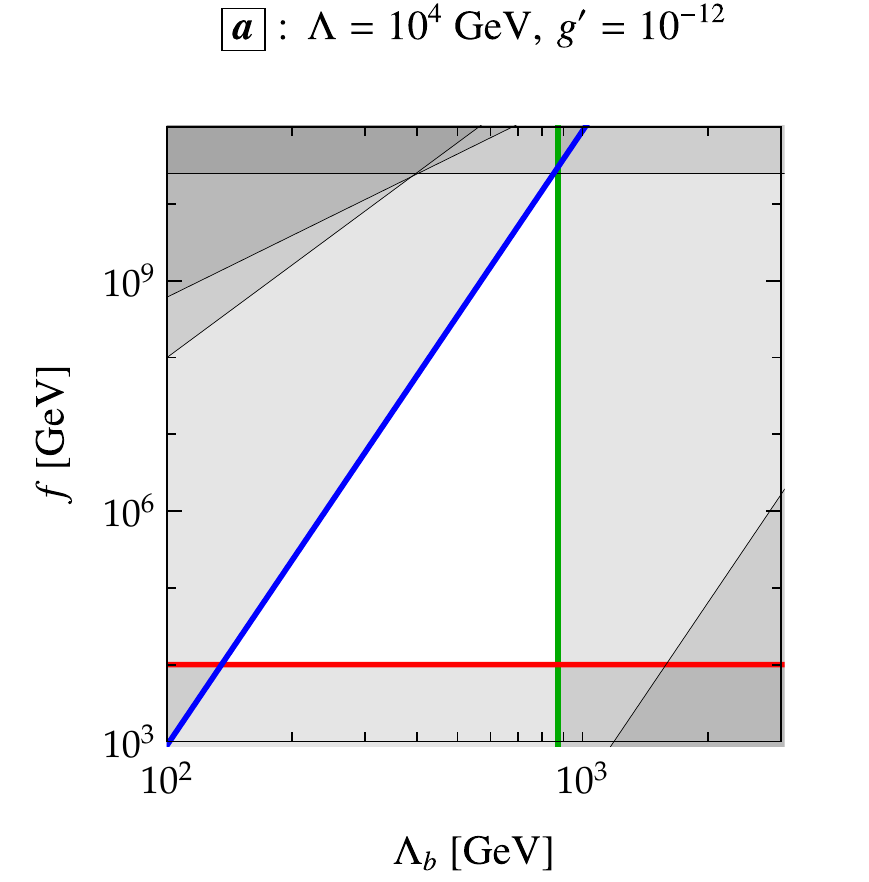}
\end{minipage}
\begin{minipage}{.49\textwidth}
\begin{itemize}
\item[{\tikz\fill[rounded corners = .5mm, blue](0,0)rectangle(0.3,0.2);}] Efficient fragmentation \Eq{eq:condition 3}
\item[{\tikz\fill[rounded corners = .5mm, red](0,0)rectangle(0.3,0.2);}] Symmetry breaking \Eq{eq:eftvalidity}
\item[{\tikz\fill[rounded corners = .5mm, Green](0,0)rectangle(0.3,0.2);}] Microscopic origin of the barriers \Eq{eq:stability}
\end{itemize}
\end{minipage}
\caption{\label{fig:fLambdabfraginfl}Same as Fig.~\ref{fig:HILambdabfraginfl} (fragmentation during inflation) but in the ($\Lambda_b$, $f$) plane.}
\end{figure}

\begin{figure}
\centering
\includegraphics[width=.49\textwidth]{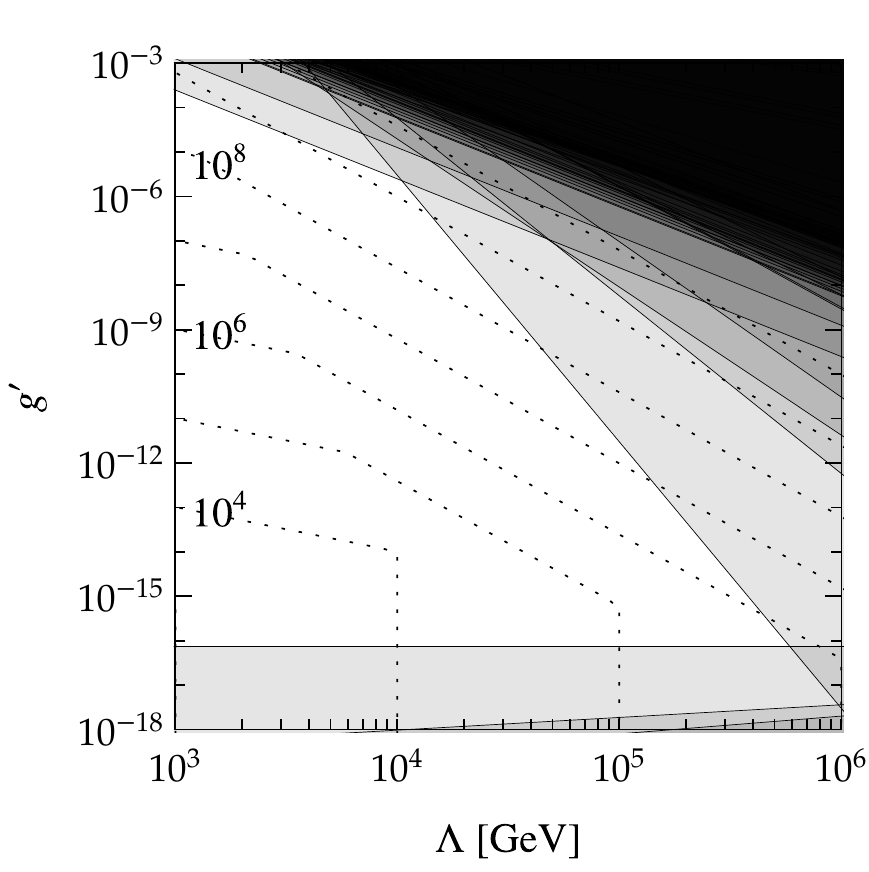}
\includegraphics[width=.49\textwidth]{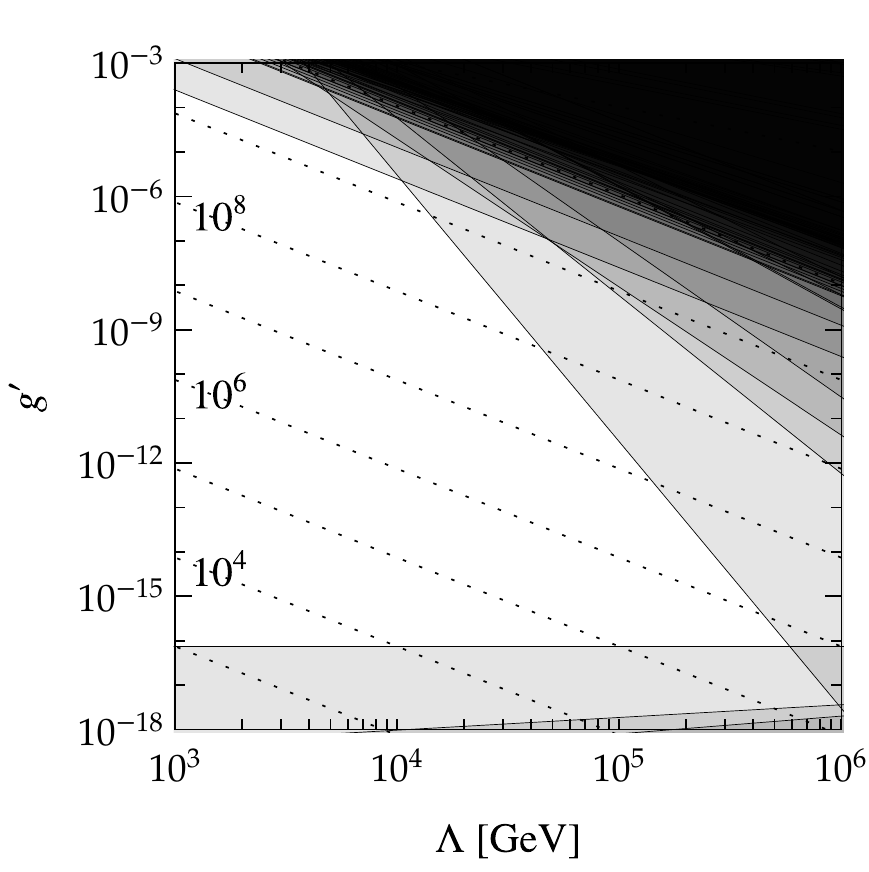}
\caption{\label{fig:gpLambdaMIfraginfl}Same region as in Fig.~\ref{fig:gpLambdafraginfl} (fragmentation during inflation) where we show the contours of the minimum (left) and maximum (right) values  of the inflationary scale $M_I$ in GeV.}
\end{figure}
\begin{figure}
\centering
\includegraphics[width=.49\textwidth]{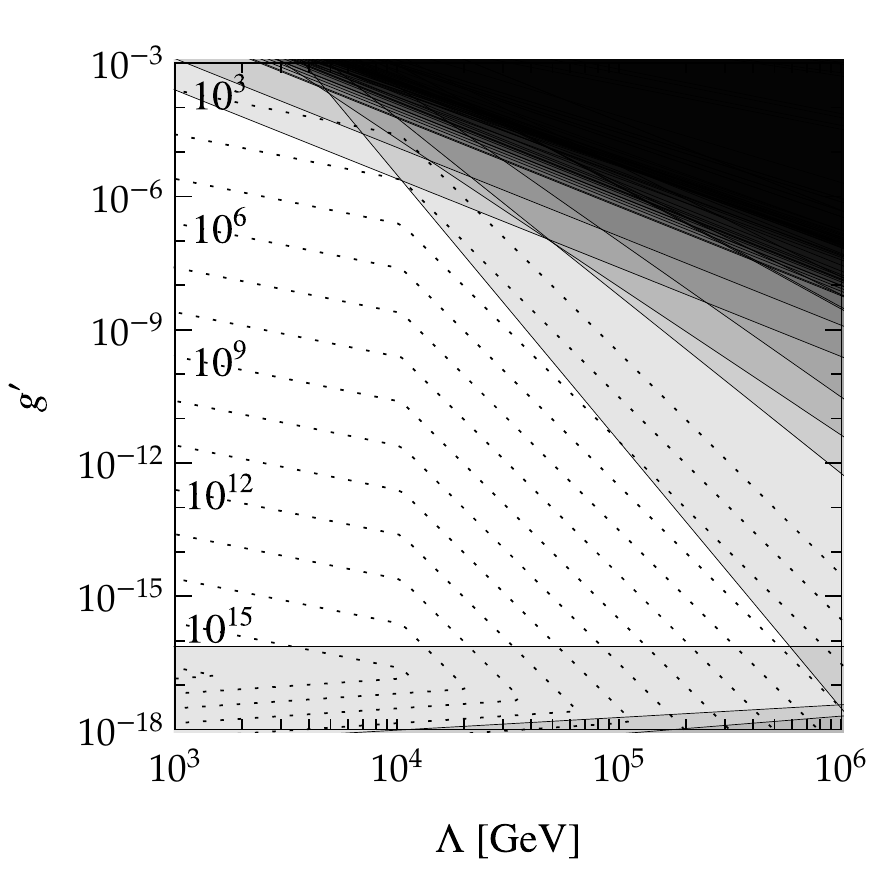}
\caption{\label{fig:gpLambdaffraginfl}Same region as in Fig.~\ref{fig:gpLambdafraginfl} (fragmentation during inflation) where we show the contours of the maximum value  of the scale $f$ in GeV. The minimal value of $f$ is, in the region of interest, always equal to the cut-off scale $\Lambda$, as required by the condition in \Eq{eq:eftvalidity} on the consistency of the symmetry breaking pattern.}
\end{figure}
\begin{figure}
\centering
\includegraphics[width=.49\textwidth]{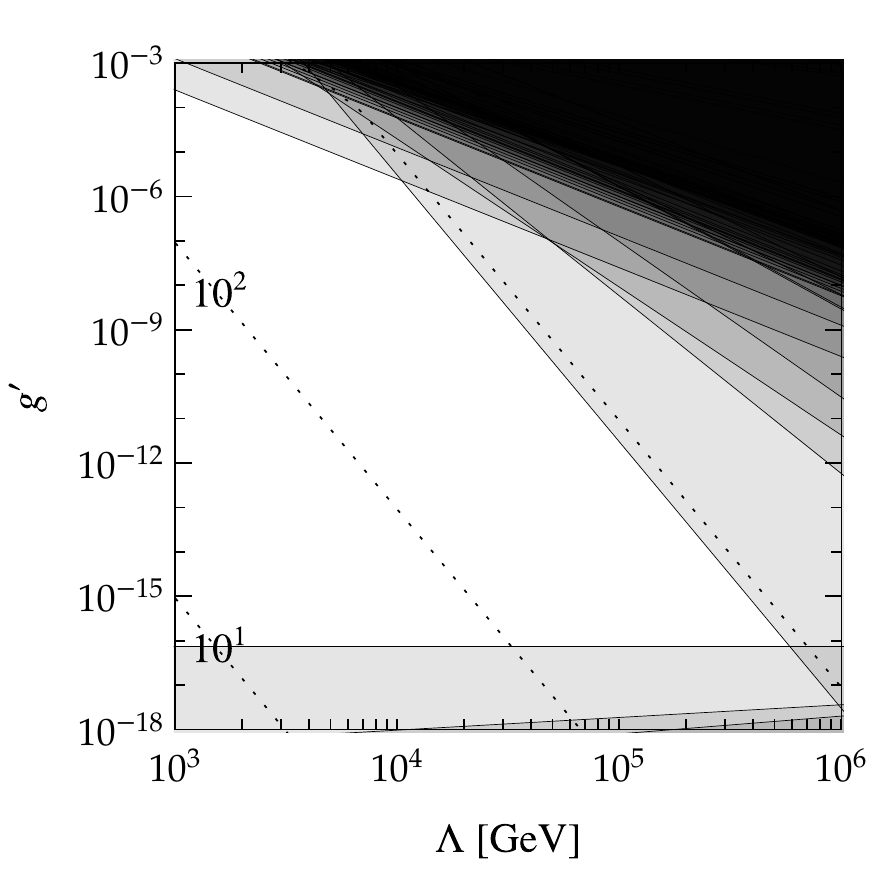}
\caption{\label{fig:gpLambdaLbfraginfl}Same region as in Fig.~\ref{fig:gpLambdafraginfl} (fragmentation during inflation) where we show the contours of the minimum of the scale $\Lambda_b$ in GeV.  The maximal value of $\Lambda_b$ is, in the region of interest, always equal to $\sqrt{4\pi}v_\ew$, as required by \Eq{eq:stability}.}
\end{figure}
\begin{figure}
\centering
\includegraphics[width=.49\textwidth]{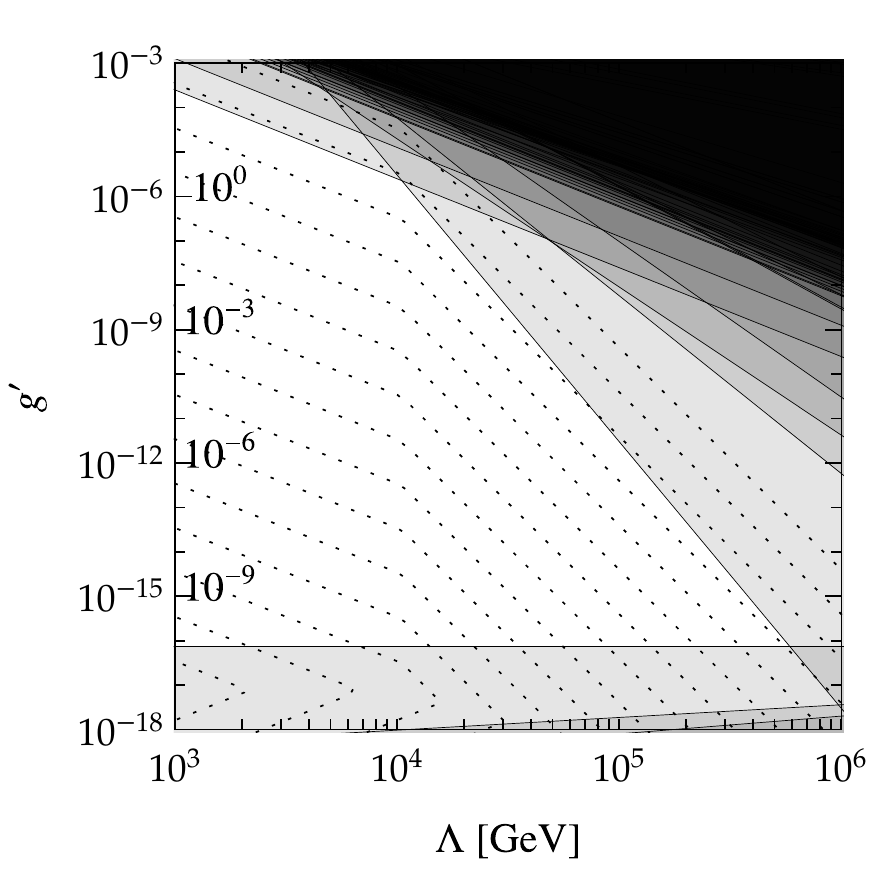}
\includegraphics[width=.49\textwidth]{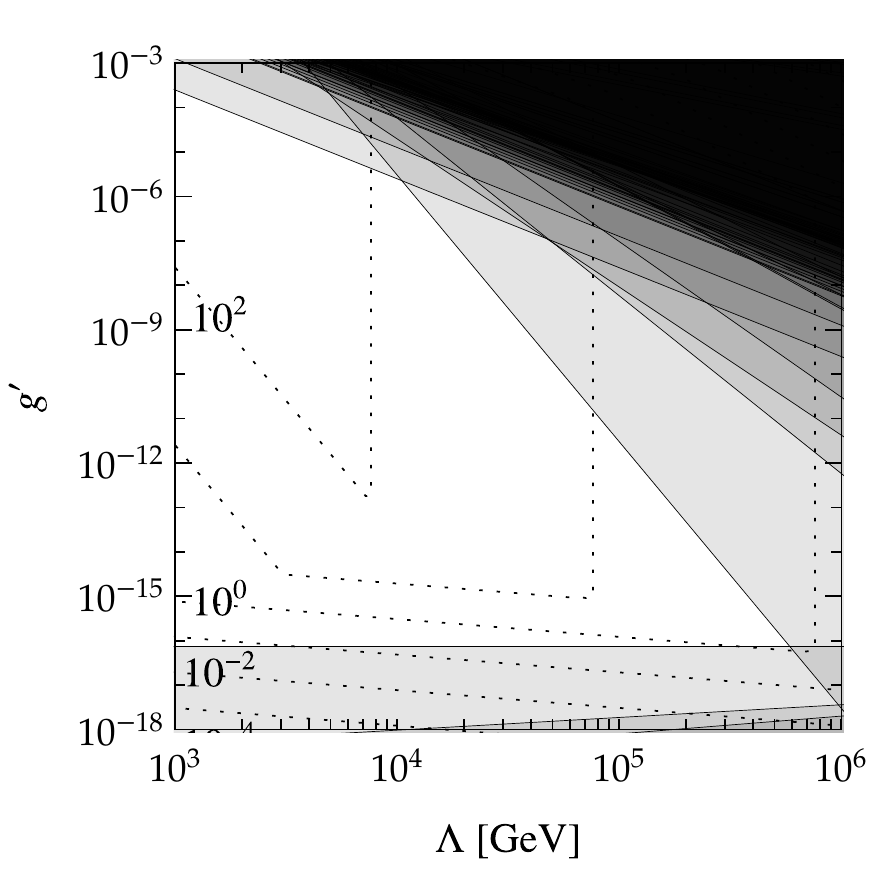}
\caption{\label{fig:gpLambdamphifraginfl} Same region as in Fig.~\ref{fig:gpLambdafraginfl} (fragmentation during inflation) where we show the contours of the minimum (left) and maximum (right) values  of the relaxion mass $\widetilde m_\phi$ in GeV.}
\end{figure}
\begin{figure}
\centering
\includegraphics[width=.49\textwidth]{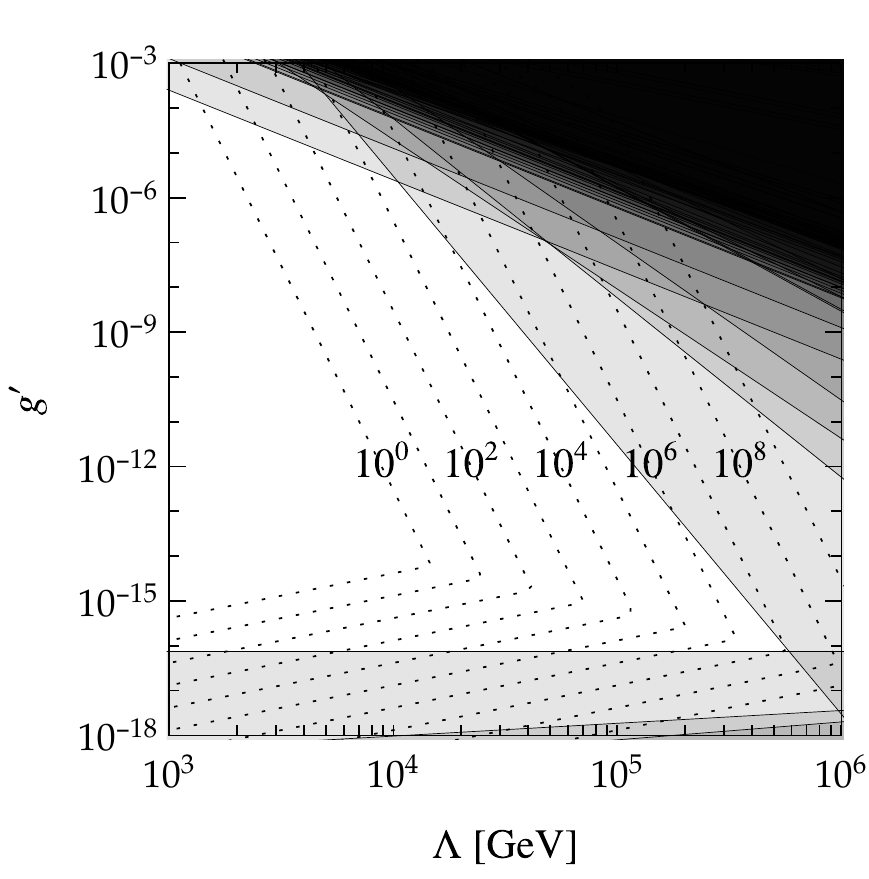}
\includegraphics[width=.49\textwidth]{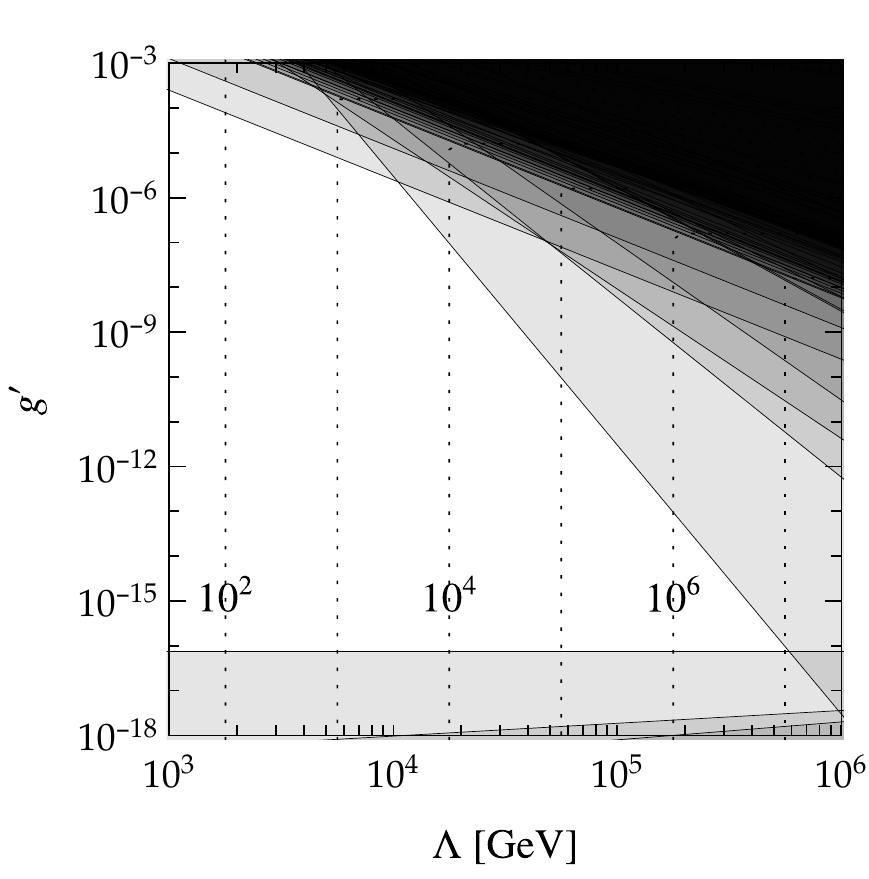}
\caption{\label{fig:gpLambdaNefoldsfraginfl} Same region as in Fig.~\ref{fig:gpLambdafraginfl} (fragmentation during inflation) where we show the contours of the minimum (left) and maximum (right) values  of the number of e-folds $N_e$ required for a successful relaxation. }
\end{figure}

\clearpage

\subsection{Self-stopping relaxion without inflation}
\label{AppendixPlotsSelfStopping}

\begin{figure}[h!]
\centering
\includegraphics[width=.45\textwidth]{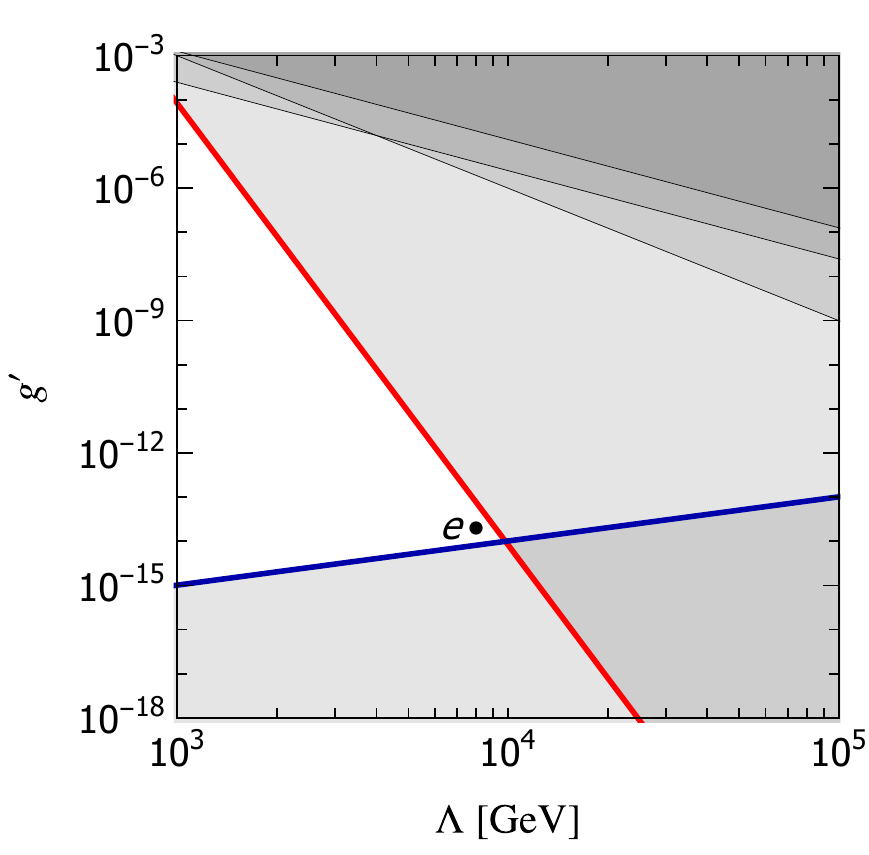}
\includegraphics[width=.45\textwidth]{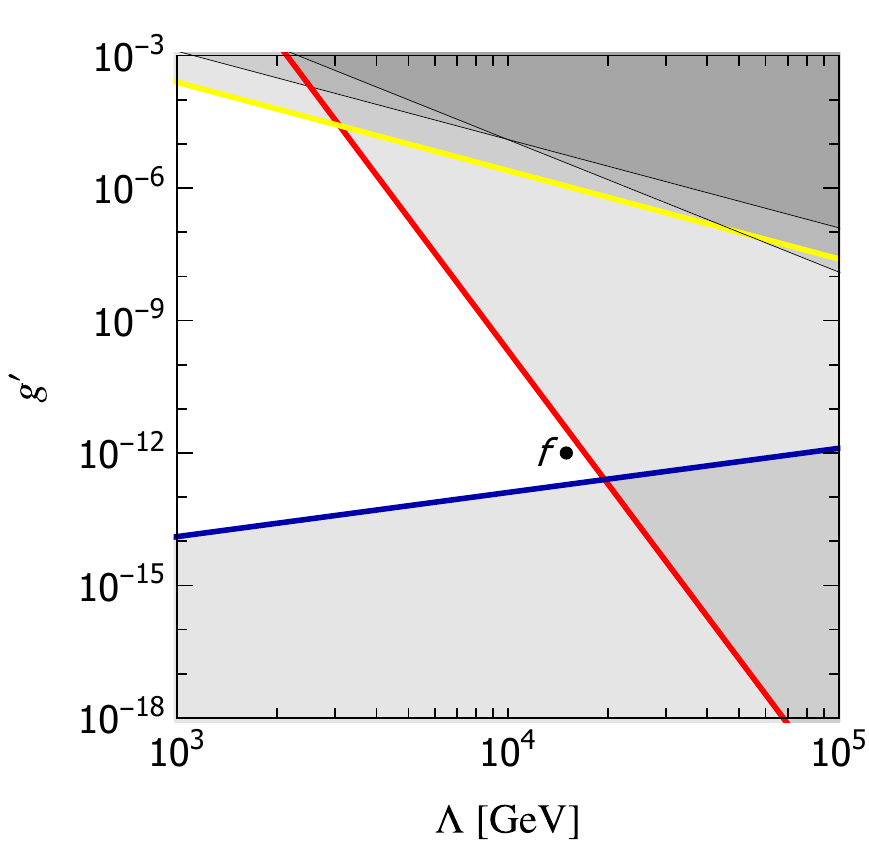}
\includegraphics[width=.45\textwidth]{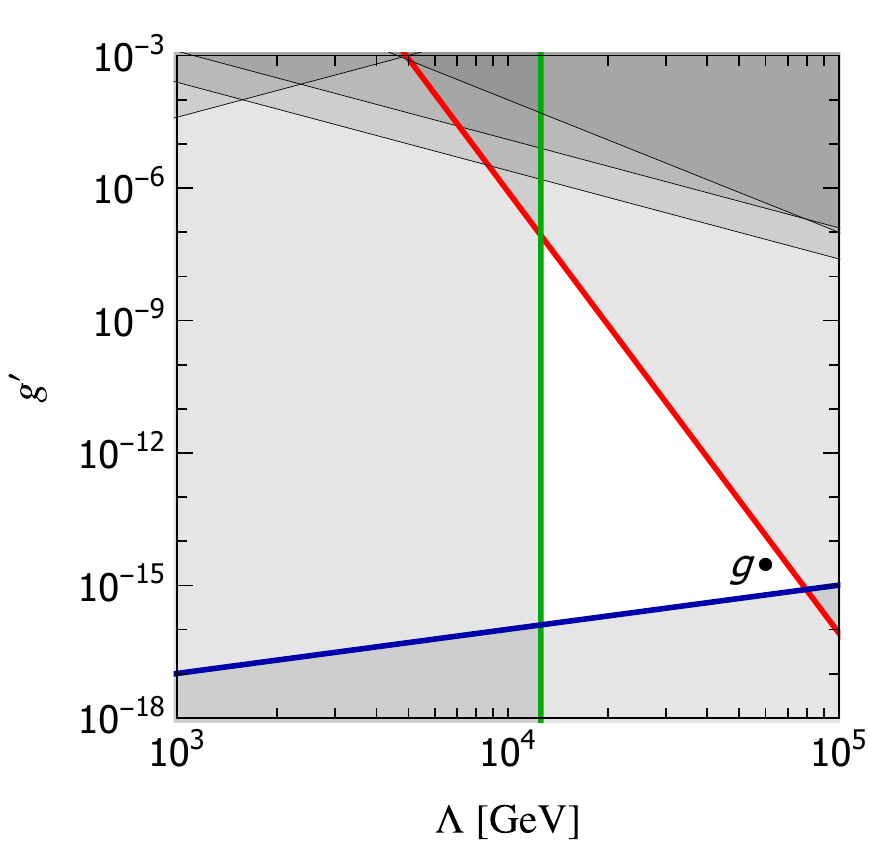}
\begin{minipage}{1.0\textwidth}
\begin{multicols}{1}
\begin{itemize}

\item[{\tikz\fill[rounded corners = .5mm, Red](0,0)rectangle(0.3,0.2);}] Microscopic origin of the barriers Eq.\,(\ref{eq:stability}) and Symmetry breaking pattern Eq.\,(\ref{eq:eftvalidity})
\item[{\tikz\fill[rounded corners = .5mm, Blue](0,0)rectangle(0.3,0.2);}] No slow-roll Eq.\,(\ref{eq:no slow-roll})
\item[{\tikz\fill[rounded corners = .5mm, Yellow](0,0)rectangle(0.3,0.2);}] Velocity larger than $\Lambda_b^2$ and Symmetry breaking pattern Eq.\,(\ref{eq:eftvalidity})
\item[{\tikz\fill[rounded corners = .5mm, Green](0,0)rectangle(0.3,0.2);}]  Slope can be neglected Eq.\,(\ref{eq:slope bound (H=0)})

\end{itemize}
\end{multicols}
\end{minipage}
\caption{\label{fig:gpLambNOInflationApp}
Details of the origin of the constraints delimitating the regions presented in Section \ref{sec:withoutinflation} and Fig.~\ref{fig:gpLambdaNOINFLATION}, for benchmark points $e$, $f$ and $g$, corresponding to the case where relaxation happens through Higgs-dependent barriers, NOT during a stage of inflation,  where the relaxion stops because of {\it fragmentation}.
Top:~$\dot{\phi}_0 = \sqrt{2 g/g'} \,\Lambda^2 $ within $g/g'=1$ (left) and $g/g'= 1/(4 \pi)^2$ (right). Bottom:
$\dot{\phi}_0 = 10^{-2} \Lambda^2$. }
\end{figure}

\begin{figure}[!h]
\centering
\includegraphics[width=.43\textwidth]{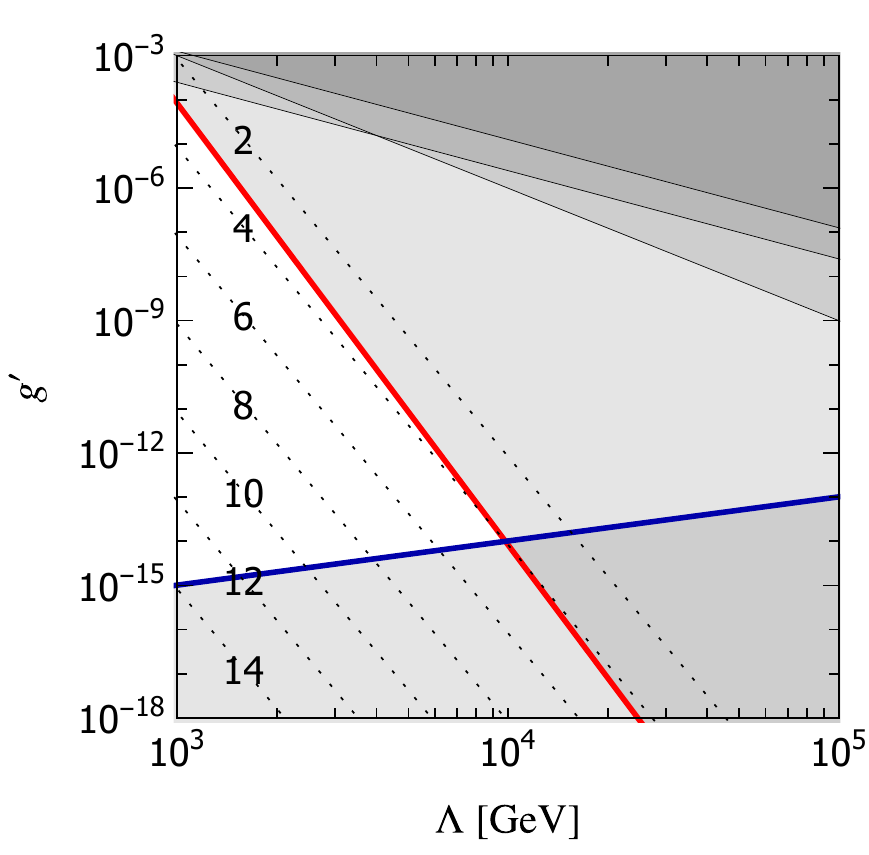}
\includegraphics[width=.43\textwidth]{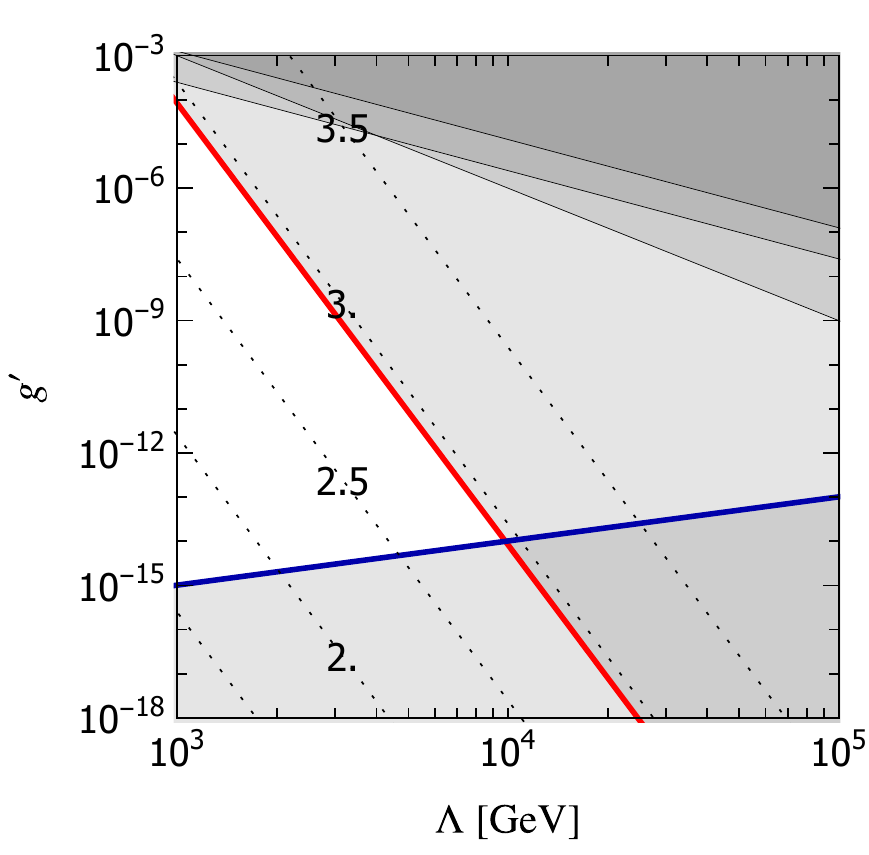}
\includegraphics[width=.43\textwidth]{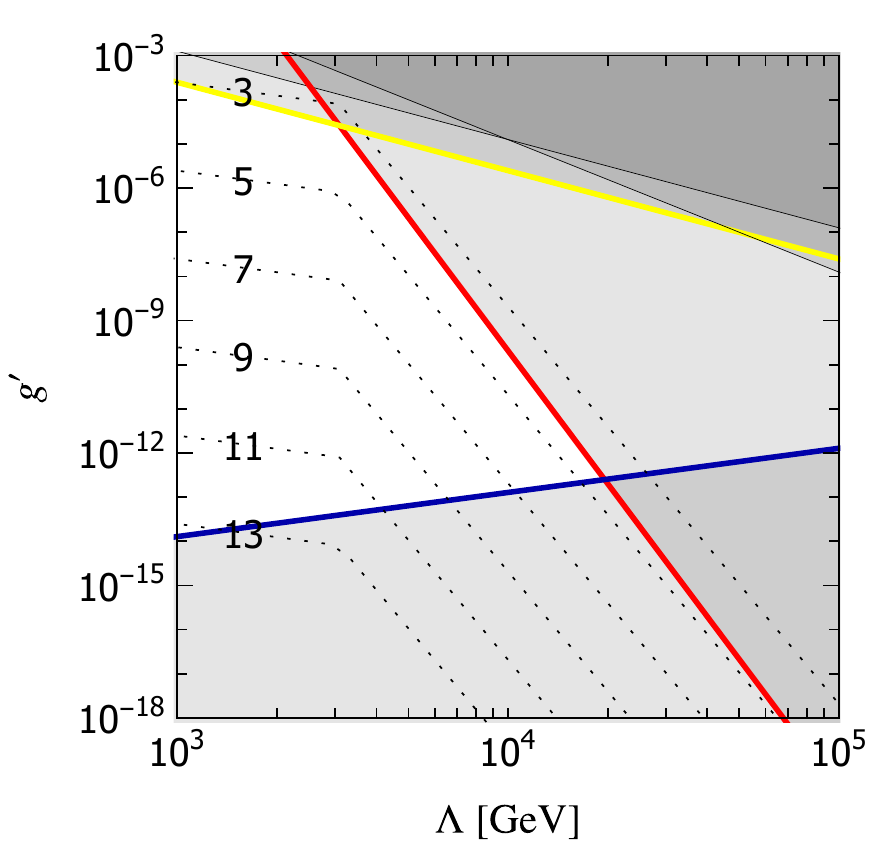}
\includegraphics[width=.43\textwidth]{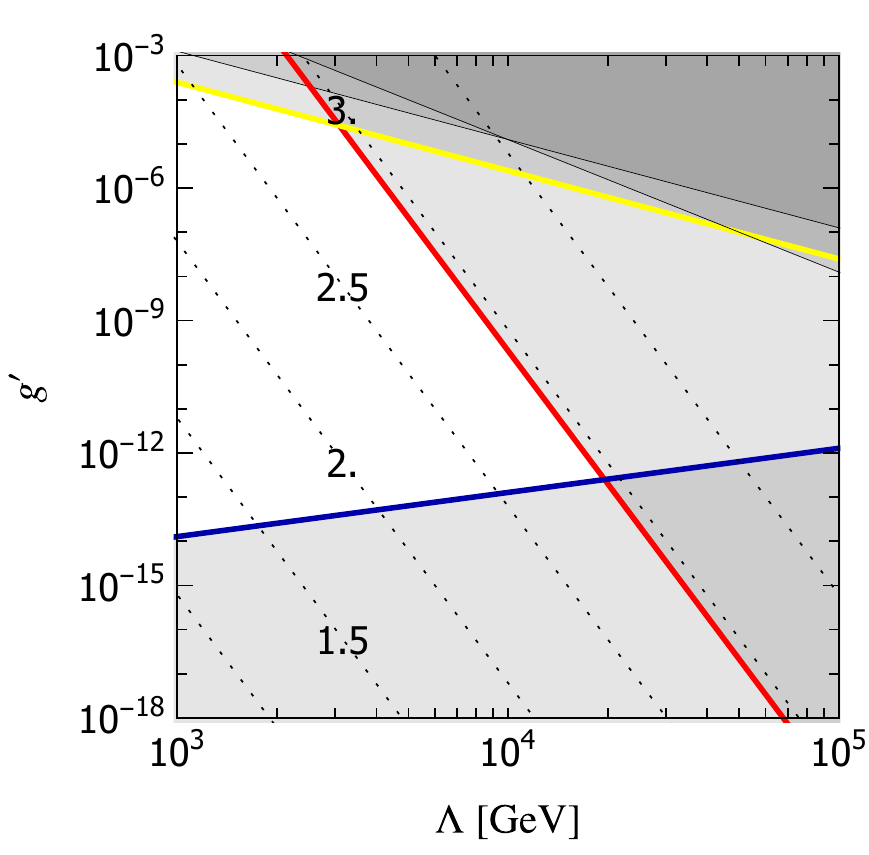}
\includegraphics[width=.43\textwidth]{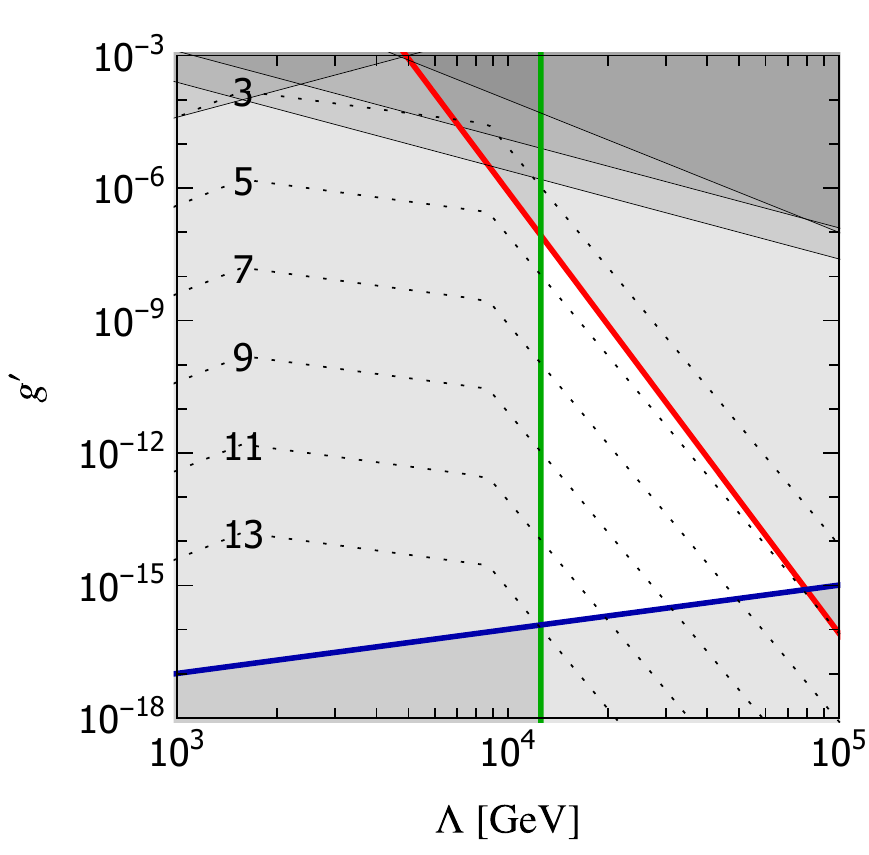}
\includegraphics[width=.43\textwidth]{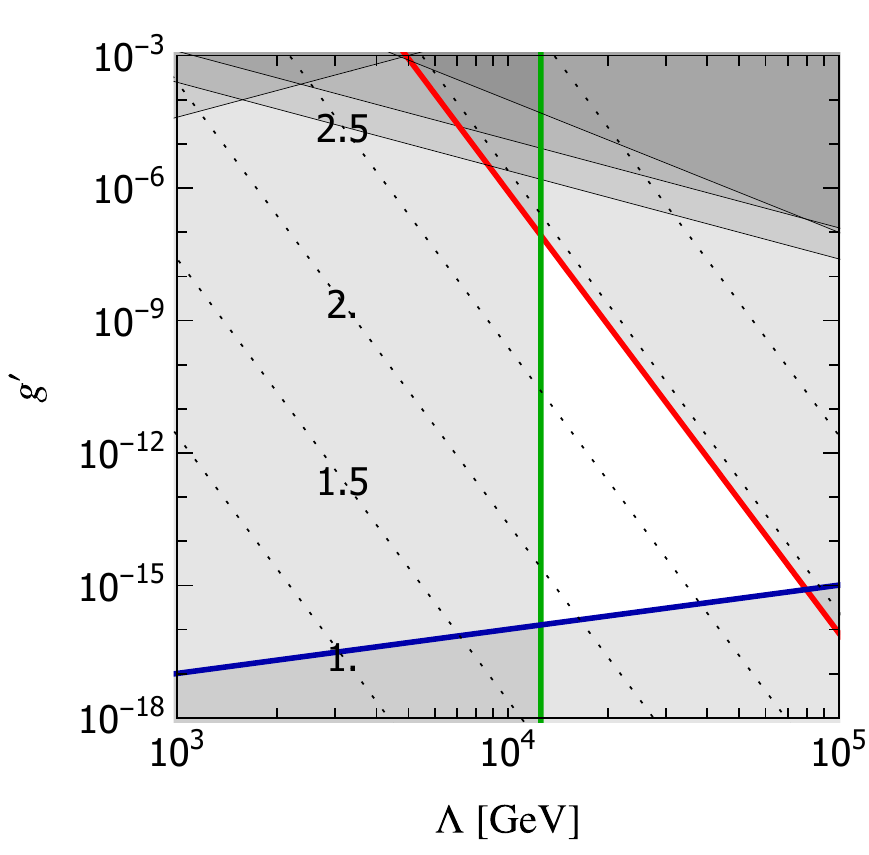}
\begin{minipage}{1.0\textwidth}
\begin{multicols}{1}
\begin{itemize}

\item[{\tikz\fill[rounded corners = .5mm, Red](0,0)rectangle(0.3,0.2);}] Microscopic origin of the barriers Eq.\,(\ref{eq:stability}) and Symmetry breaking pattern Eq.\,(\ref{eq:eftvalidity})
\item[{\tikz\fill[rounded corners = .5mm, Blue](0,0)rectangle(0.3,0.2);}] No slow-roll Eq.\,\ref{eq:no slow-roll}
\item[{\tikz\fill[rounded corners = .5mm, Yellow](0,0)rectangle(0.3,0.2);}] Velocity larger than $\Lambda_b^2$ and Symmetry breaking pattern Eq.\,(\ref{eq:eftvalidity})
\item[{\tikz\fill[rounded corners = .5mm, Green](0,0)rectangle(0.3,0.2);}] Slope can be neglected Eq.\,\ref{eq:slope bound (H=0)}

\end{itemize}
\end{multicols}
\end{minipage}
\caption{\label{fig:gpLambContoursApp} 
Same regions as in Fig.~\ref{fig:gpLambNOInflationApp} (self-stopping relaxion NOT during inflation) where we show the contours  of $\textrm{log}_{10}(f{\textrm{max}})$ (left) and $\textrm{log}_{10}(\Lambda_{b,\textrm{min}})$ (right) in GeV. From top to bottom: $\dot{\phi}_0 = \sqrt{2 g/g'} \,\Lambda^2 $ with $g/g'=1$,  $\dot{\phi}_0 = \sqrt{2 g/g'} \,\Lambda^2 $ with $g/g'= 1/(4 \pi)^2$, and $\dot{\phi}_0 = 10^{-2} \Lambda^2$.  Contours for  $\Lambda_{b, \textrm{max}}$  and $f_{\textrm{min}}$ are trivial as  they  saturate their respective maximal and minimal  allowed values, i.e.,  $\Lambda_{b, \textrm{max}}= \sqrt{4 \pi} \,\vEW$   and $f_{\textrm{min}}= \Lambda$. }
\end{figure}

\begin{figure}[!h]
\centering
\includegraphics[width=.45\textwidth]{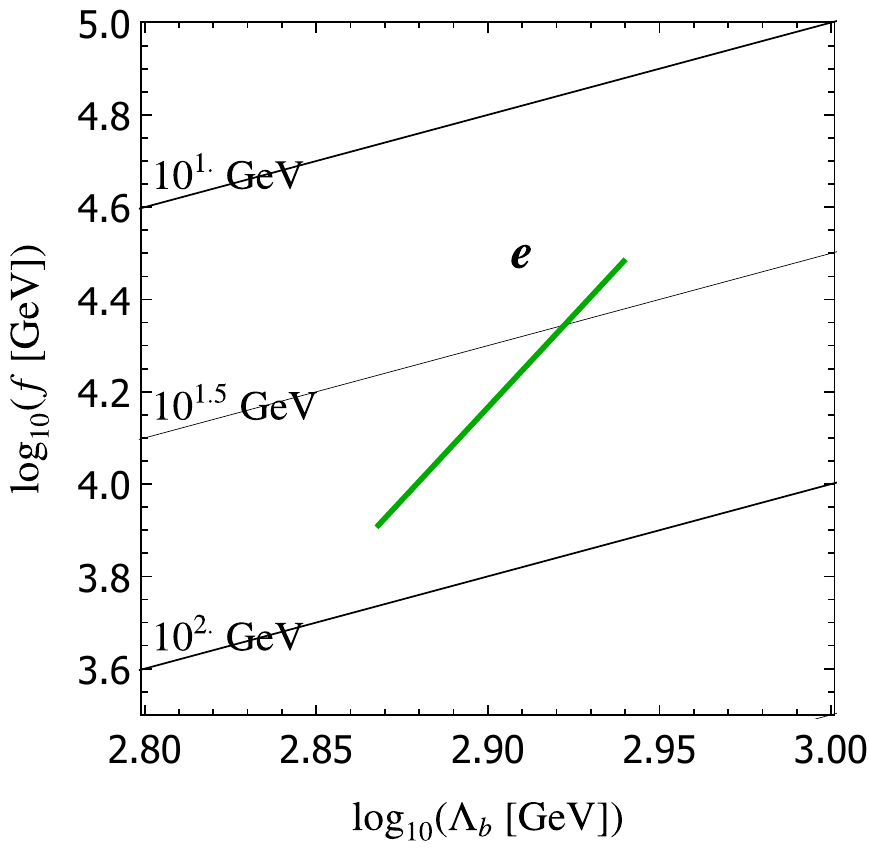}
\includegraphics[width=.45\textwidth]{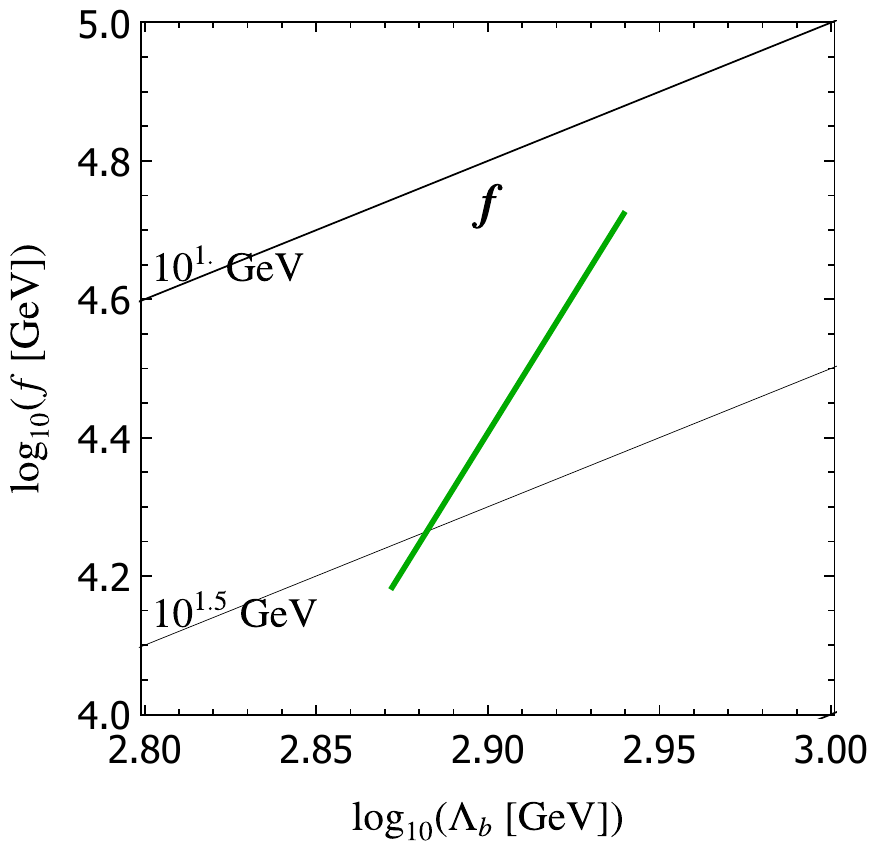}
\includegraphics[width=.45\textwidth]{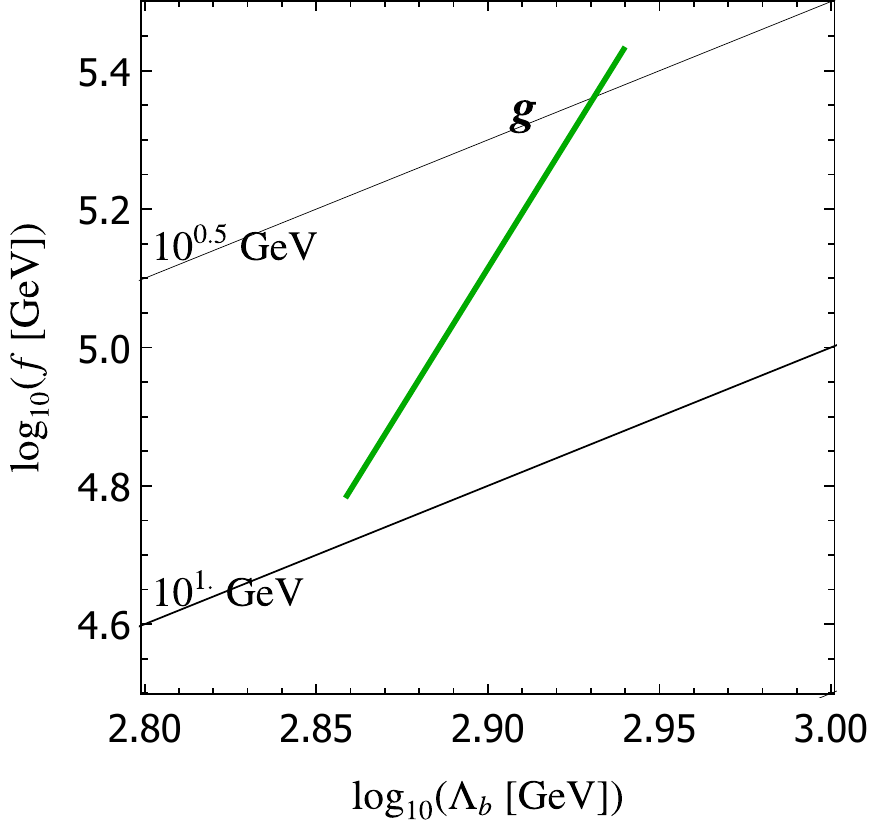}
\caption{\label{fig:contoursMassApp} Contours of $\mphi$, as defined in \Eq{eq:massappr}, in the allowed range of $f$ and $\Lambda_b$ for the three benchmark points $e$, $f$, $g$ in \Fig{fig:gpLambdaNOINFLATION} corresponding to the case of self-stopping relaxion NOT during inflation discussed in Sec.\,\ref{sec:withoutinflation}.}
\end{figure}

\clearpage
\subsection{Self-stopping relaxion triggering a stage of inflation}
\label{subsec:relaxioninflaton}

\begin{figure}[!h]
\centering
\includegraphics[width=.45\textwidth]{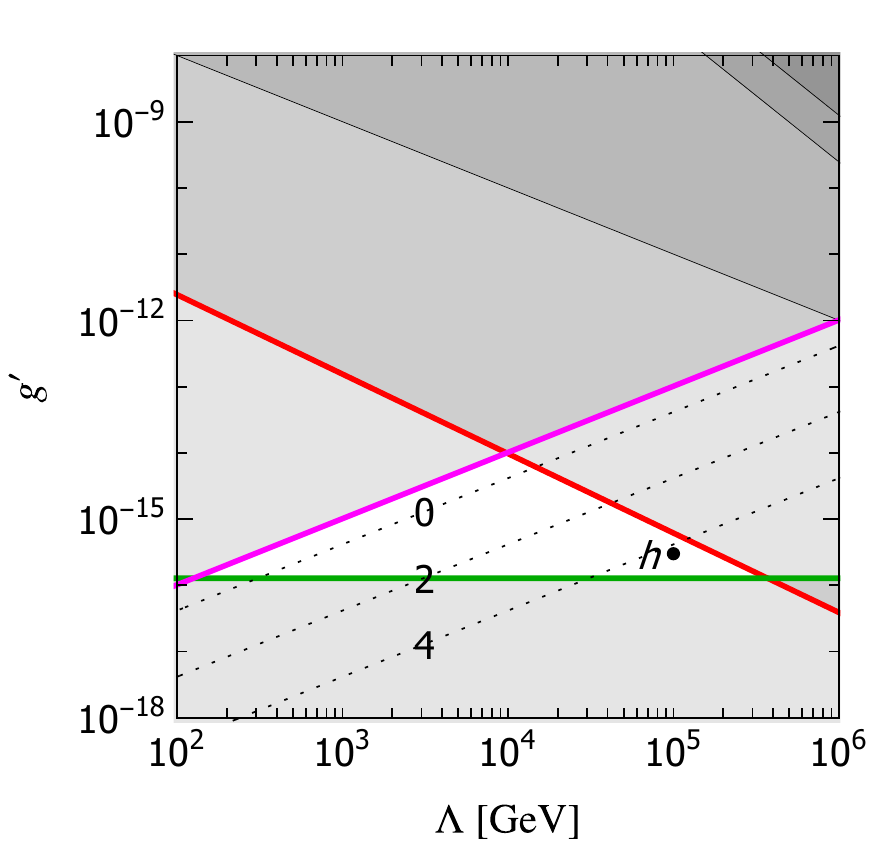}
\begin{minipage}{1.0\textwidth}
\begin{multicols}{1}
\begin{itemize}

\item[{\tikz\fill[rounded corners = .5mm, Red](0,0)rectangle(0.3,0.2);}] Microscopic origin of the barriers Eq.\,(\ref{eq:stability}) and Symmetry breaking pattern Eq.\,(\ref{eq:eftvalidity})
\item[{\tikz\fill[rounded corners = .5mm, Green](0,0)rectangle(0.3,0.2);}] Slope can be neglected Eq.\,(\ref{eq:slope bound (H=0)})
\item[{\tikz\fill[rounded corners = .5mm, Yellow](0,0)rectangle(0.3,0.2);}] Velocity larger than $\Lambda_b^2$ and Symmetry breaking pattern Eq.\,(\ref{eq:eftvalidity})
\item[{\tikz\fill[rounded corners = .5mm, Magenta](0,0)rectangle(0.3,0.2);}] Slow roll velocity smaller than the cutoff \Eq{eq:phisrconsistent}

\end{itemize}
\end{multicols}
\end{minipage}
\caption{\label{fig:gpLambINFrelaxiondominates}
Allowed parameter space in the plane $g', \Lambda$ for the case discussed in Sec.~\ref{sec:relaxioninflating}, of a self-stopping relaxion,   where the relaxion dominates the energy density of the universe during relaxation and drives an inflationary period.  The dotted lines are the contours of number of efolds of inflation, $\textrm{log}_{10}(N_{\textrm{efolds}})$.}
\end{figure}

\begin{figure}[!h]
\centering
\includegraphics[width=.45\textwidth]{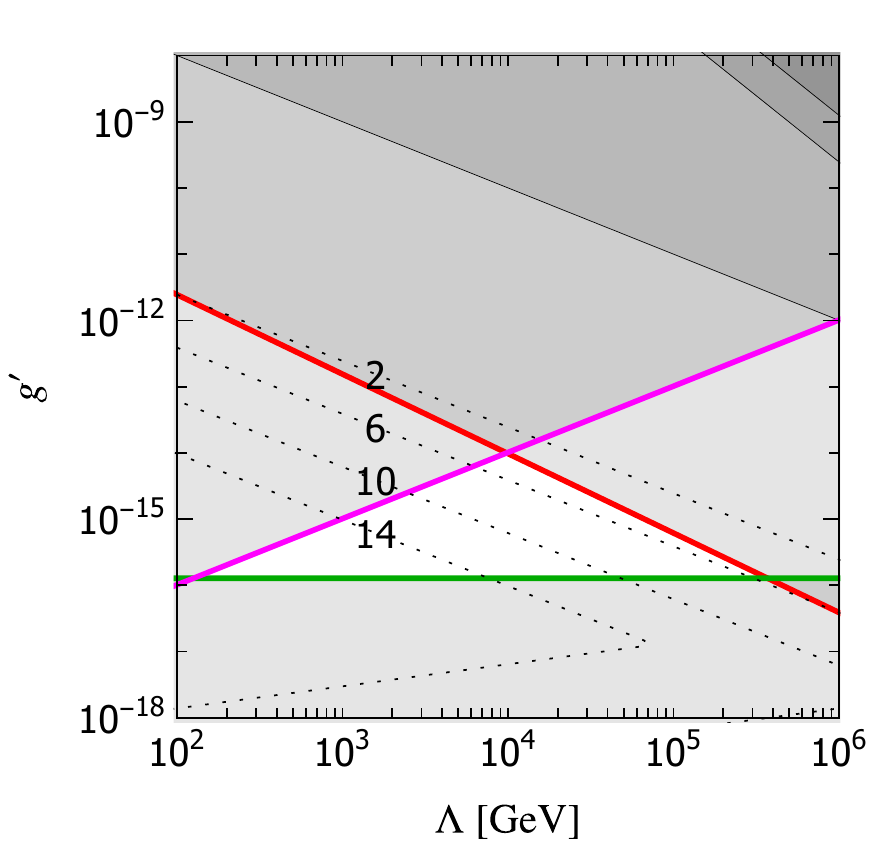}
\includegraphics[width=.45\textwidth]{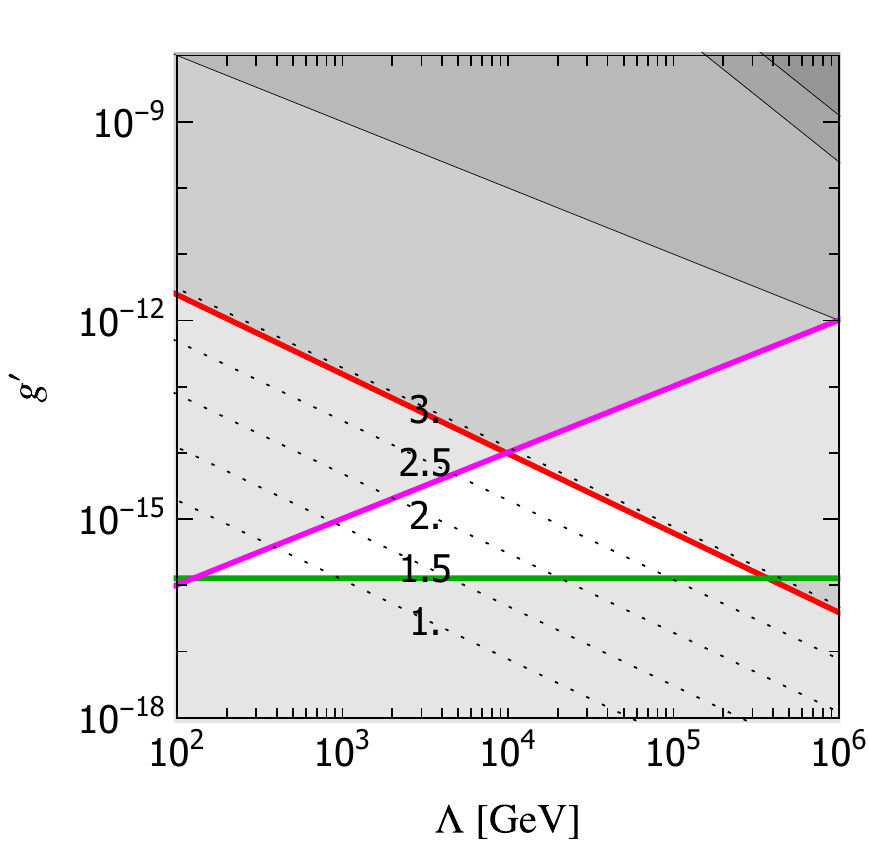}
\begin{minipage}{1.0\textwidth}
\begin{multicols}{1}
\begin{itemize}

\item[{\tikz\fill[rounded corners = .5mm, Red](0,0)rectangle(0.3,0.2);}] Microscopic origin of the barriers Eq.\,(\ref{eq:stability}) and Symmetry breaking pattern Eq.\,(\ref{eq:eftvalidity})
\item[{\tikz\fill[rounded corners = .5mm, Green](0,0)rectangle(0.3,0.2);}] Slope can be neglected \Eq{eq:slope bound (H=0)}
\item[{\tikz\fill[rounded corners = .5mm, Yellow](0,0)rectangle(0.3,0.2);}] Velocity larger than $\Lambda_b^2$ and Symmetry breaking pattern \Eq{eq:eftvalidity}
\item[{\tikz\fill[rounded corners = .5mm, Magenta](0,0)rectangle(0.3,0.2);}] Slow roll velocity smaller than the cutoff \Eq{eq:phisrconsistent}

\end{itemize}
\end{multicols}
\end{minipage}
\caption{\label{fig:gpLambINFrelaxiondominatesCONTOURS}
Same region as Fig.~\ref{fig:gpLambINFrelaxiondominates} where we show the
  contours   of $\textrm{log}_{10}(f_{\textrm{max}})$ (left) and $\textrm{log}_{10}(\Lambda_{b,\textrm{min}})$ (right).  The values of  $\Lambda_{b, \textrm{max}}$  and $f_{\textrm{min}}$   saturate to   $\Lambda_{b, \textrm{max}}= \sqrt{4 \pi} \,\vEW$  and $f_{\textrm{min}}= \Lambda$, respectively.  }
\end{figure}

\begin{figure}
\centering
\includegraphics[width=.45\textwidth]{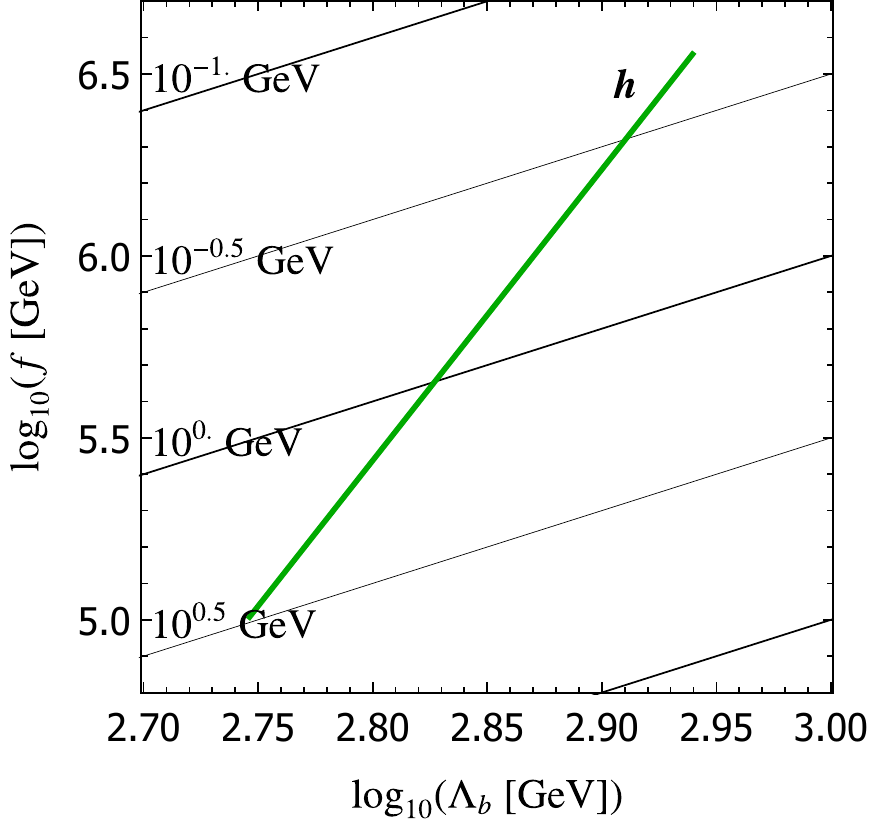}
\caption{\label{fig:gpLambINFrelaxiondominatesCONTOURSmphi}
 Contours of $\mphi$, as defined in \Eq{eq:massappr}, in
  the allowed range of $f$ and $\Lambda_b$ for the benchmark point $h$ in \Fig{fig:gpLamb-relaxion dominates}, discussed in  Sec.~\ref{sec:relaxioninflating}, corresponding to the scenario of a self-stopping relaxion,   where the relaxion dominates the energy density of the universe during relaxation and drives an inflationary period. }
\end{figure}

\clearpage

\subsection{Relaxation through EW gauge boson production during inflation}
\label{sec:AppendixHMTduring}

\begin{figure}[h!]
\centering
\begin{minipage}{.49\textwidth}
\includegraphics[width=\textwidth]{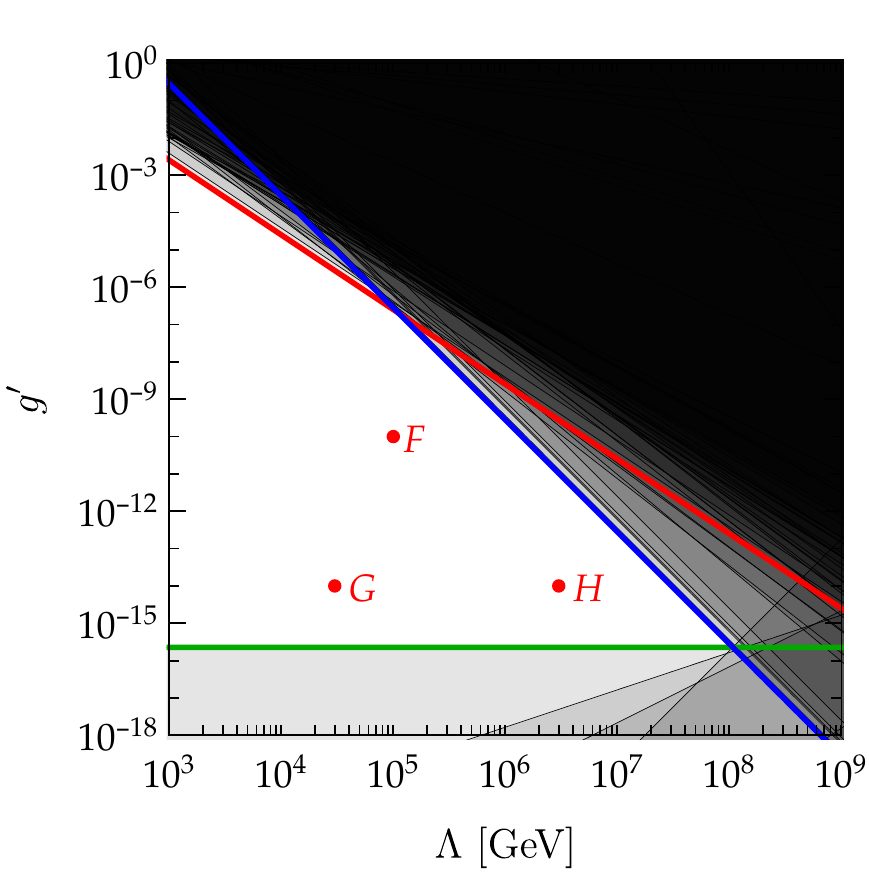}
\end{minipage}
\begin{minipage}{.49\textwidth}
\begin{itemize}
\item[{\tikz\fill[rounded corners = .5mm, blue](0,0)rectangle(0.3,0.2);}] Small variation of the Higgs mass \Eq{eq:f42}, consistency of the EFT \Eq{eq:HMT F>Lambda} and large velocity \Eq{eq:HMT large velocity}
\item[{\tikz\fill[rounded corners = .5mm, red](0,0)rectangle(0.3,0.2);}] Symmetry breaking pattern \Eq{eq:HMT eftvalidity} and precision of the Higgs mass scanning \Eq{eq:HMT scanning precision}
\item[{\tikz\fill[rounded corners = .5mm, Green](0,0)rectangle(0.3,0.2);}] Relaxion subdominant with respect to inflaton \Eq{eq:HMT relaxion subdominant} and \Eq{eq:HMT F>Lambda}
\end{itemize}
\end{minipage}
\caption{\label{fig:HMT inflation contours gp Lambda} Details of the origin of the constraints delimitating the allowed regions presented in Section \ref{subsectionHMTduring} and Fig.~\ref{fig:HMT inflation}, corresponding to the case where relaxation happens through Higgs-INdependent barriers, during a stage of inflation, where the relaxion is a subdominant component of the energy density of the universe and where the relaxion stops because of {\it EW gauge boson production}.}
\end{figure}

\begin{figure}[h!]
\centering
\includegraphics[width=.32\textwidth]{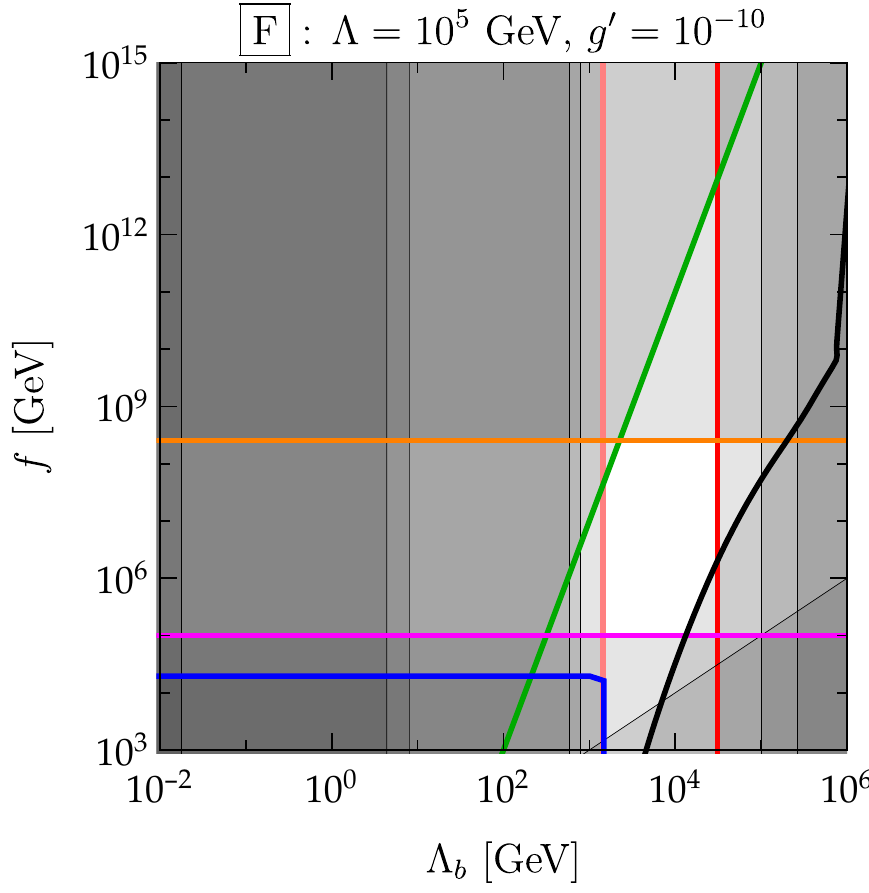}
\includegraphics[width=.32\textwidth]{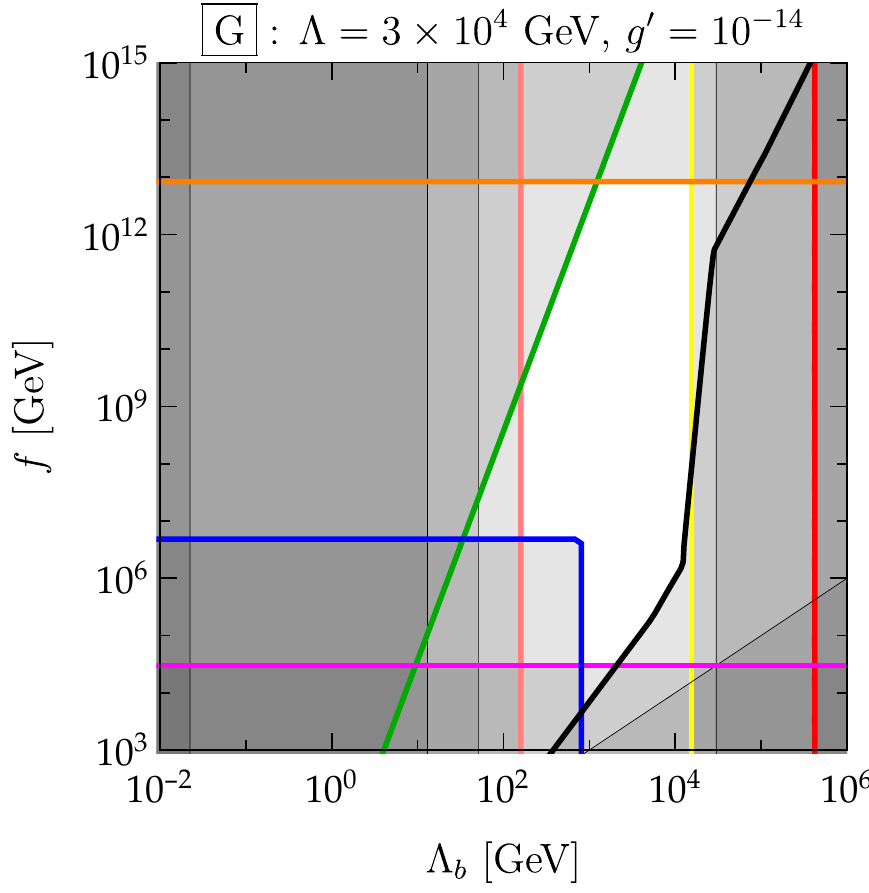}
\includegraphics[width=.32\textwidth]{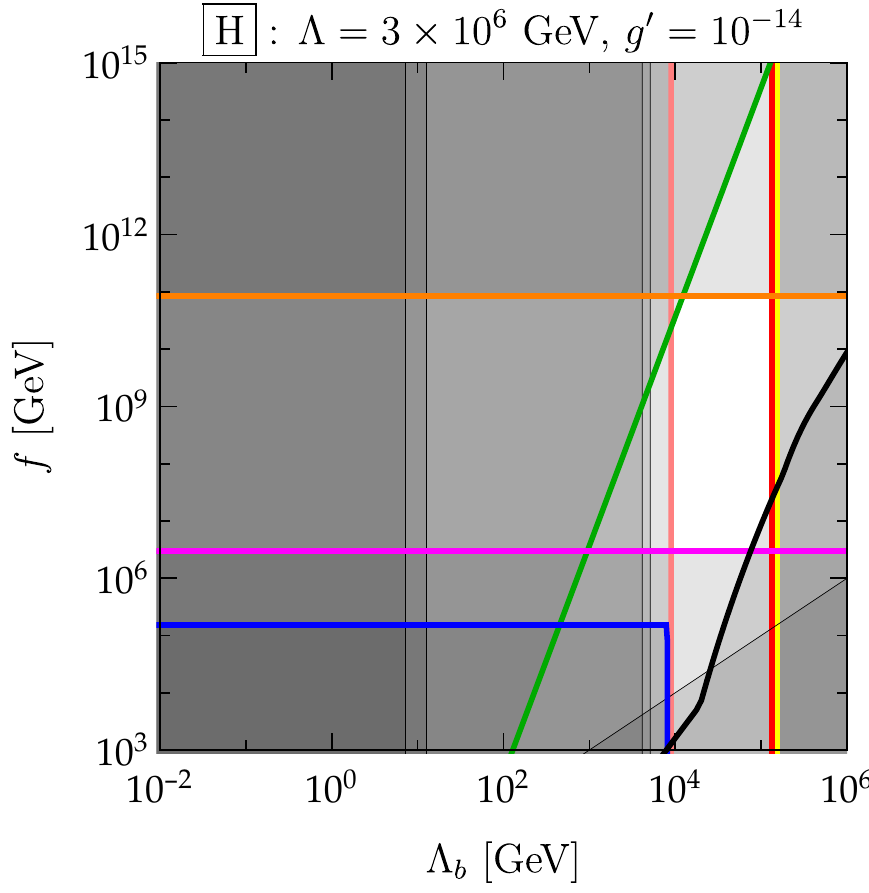}
%
%\begin{minipage}{1.0\textwidth}
\begin{multicols}{1}
\begin{itemize}
\item[{\tikz\fill[rounded corners = .5mm, pink](0,0)rectangle(0.3,0.2);}] Efficient energy dissipation \Eq{eq:condition energy dissipation} and consistency of the EFT \Eq{eq:HMT F>Lambda}
\item[{\tikz\fill[rounded corners = .5mm, red](0,0)rectangle(0.3,0.2);}] Small variation of the Higgs mass \Eq{eq:f42} and large velocity \Eq{eq:HMT large velocity}
\item[{\tikz\fill[rounded corners = .5mm, Yellow](0,0)rectangle(0.3,0.2);}] Relaxion subdominant with respect to inflaton \Eq{eq:HMT relaxion subdominant} and \Eq{eq:HMT large velocity}
\item[{\tikz\fill[rounded corners = .5mm, Green](0,0)rectangle(0.3,0.2);}] Large barriers \Eq{eq:HMT large barriers}
\item[{\tikz\fill[rounded corners = .5mm, orange](0,0)rectangle(0.3,0.2);}] Precision of the Higgs mass scanning \Eq{eq:HMT scanning precision}
\item[{\tikz\fill[rounded corners = .5mm, magenta](0,0)rectangle(0.3,0.2);}] Symmetry breaking pattern \Eq{eq:HMT eftvalidity}
\item[{\tikz\fill[rounded corners = .5mm, blue](0,0)rectangle(0.3,0.2);}] No restoration of the shift symmetry \Eq{eq:H12} and \Eq{eq:HMT F>Lambda}
\item[{\tikz\fill[rounded corners = .5mm, black](0,0)rectangle(0.3,0.2);}] Suppressed coupling to photons \Eq{eq:photon dilution}
\end{itemize}
\end{multicols}
%\end{minipage}
%
\caption{\label{fig:HMT inflation contours f Lambdab}
Same as Fig.~\ref{fig:HMT inflation contours gp Lambda} (EW gauge boson production during inflation) but in the ($\Lambda_b$, $f$) plane, for each benchmark point $F$, $G$, $H$. Here we only show the contours that do not depend on fragmentation. The new exclusion lines due to relaxion fragmentation are shown in Fig.~\ref{fig:HMT inflation contours f Lambdab frag}.}
\end{figure}

\begin{figure}[h!]
\centering
\includegraphics[width=.32\textwidth]{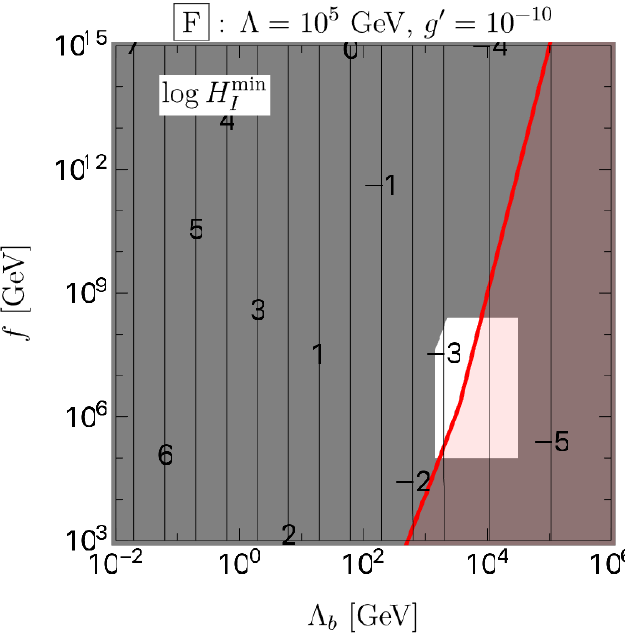}
\includegraphics[width=.32\textwidth]{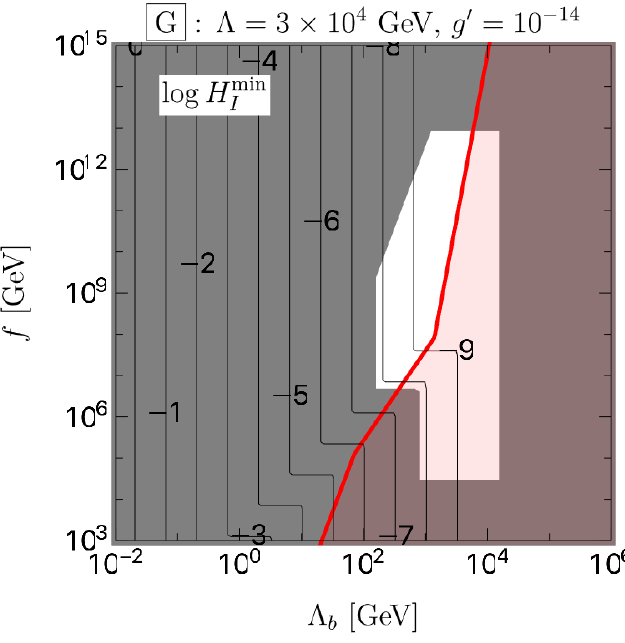}
\includegraphics[width=.32\textwidth]{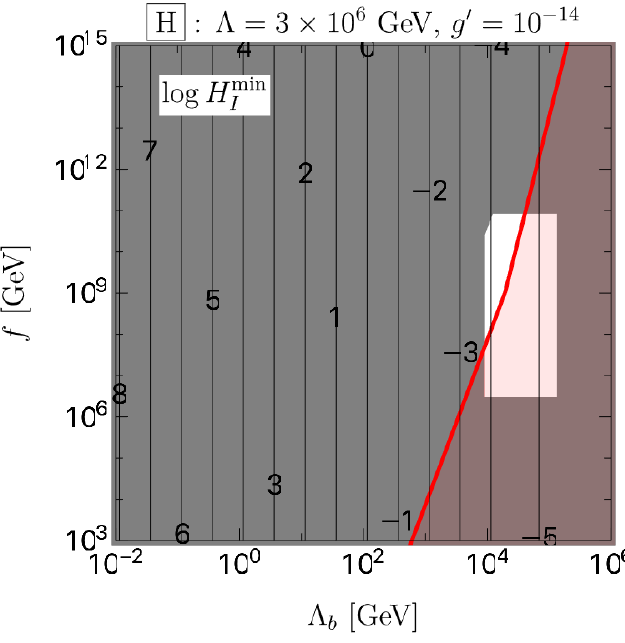}
\includegraphics[width=.32\textwidth]{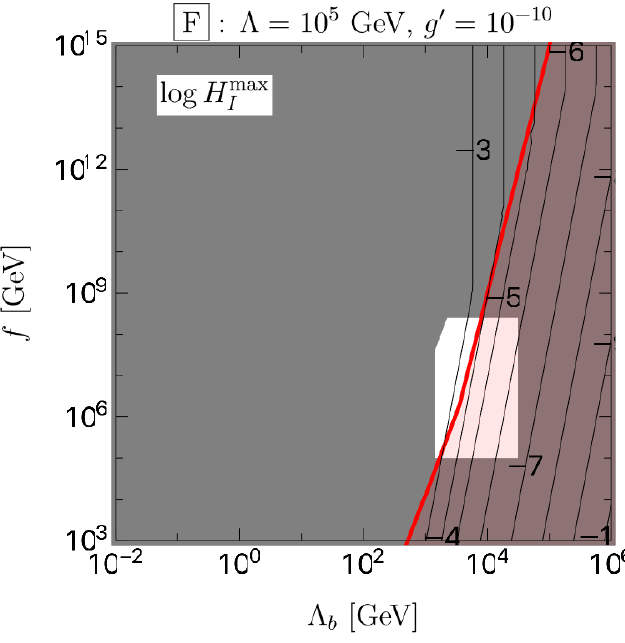}
\includegraphics[width=.32\textwidth]{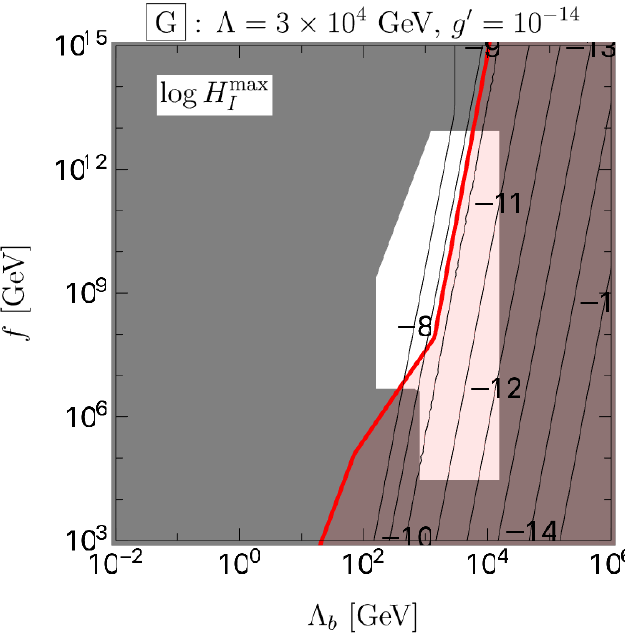}
\includegraphics[width=.32\textwidth]{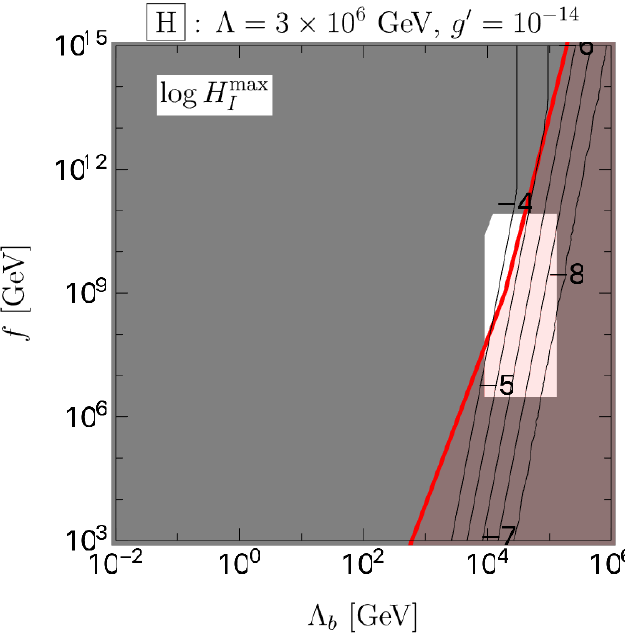}
\includegraphics[width=.32\textwidth]{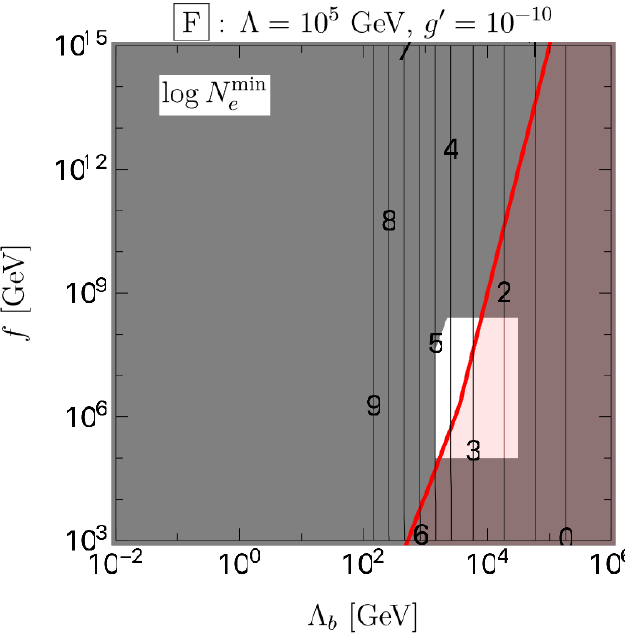}
\includegraphics[width=.32\textwidth]{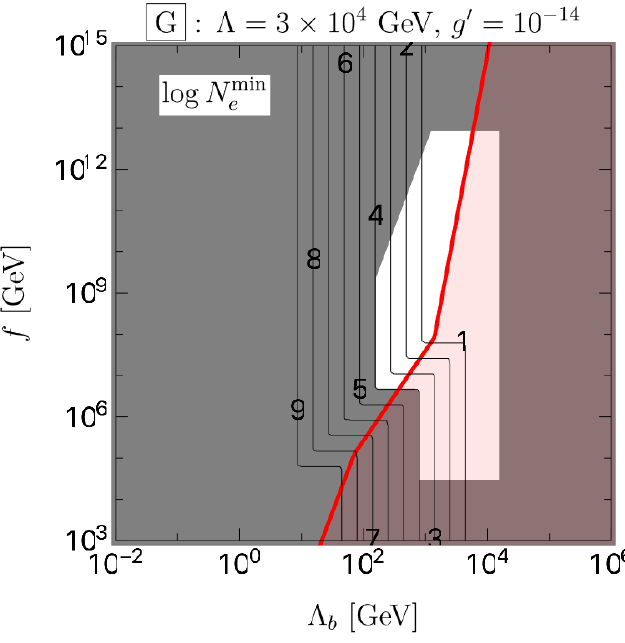}
\includegraphics[width=.32\textwidth]{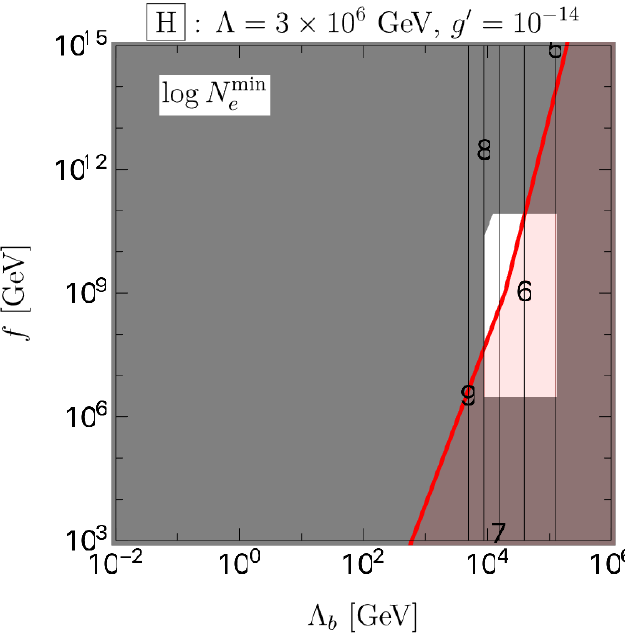}
\includegraphics[width=.32\textwidth]{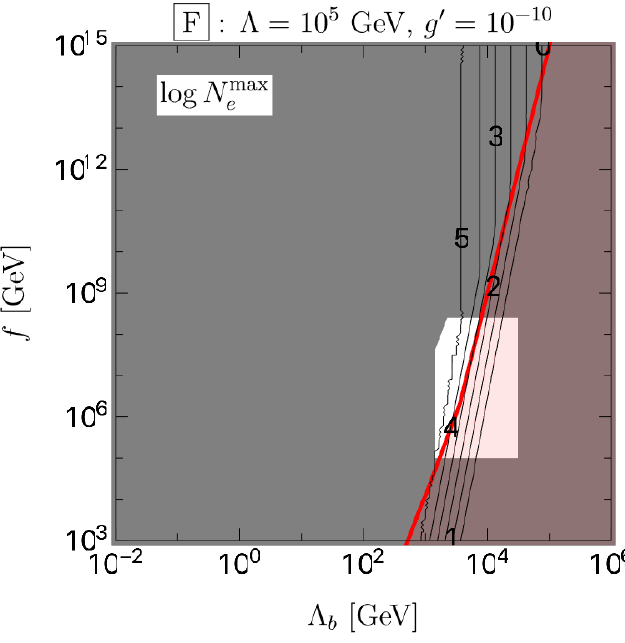}
\includegraphics[width=.32\textwidth]{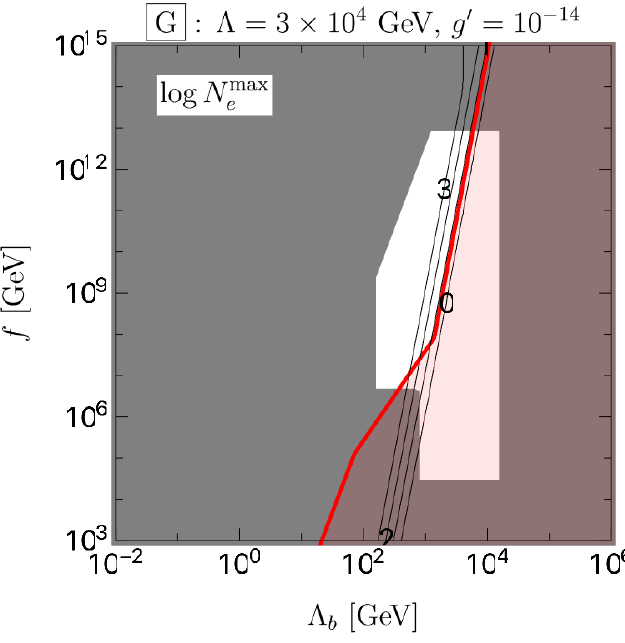}
\includegraphics[width=.32\textwidth]{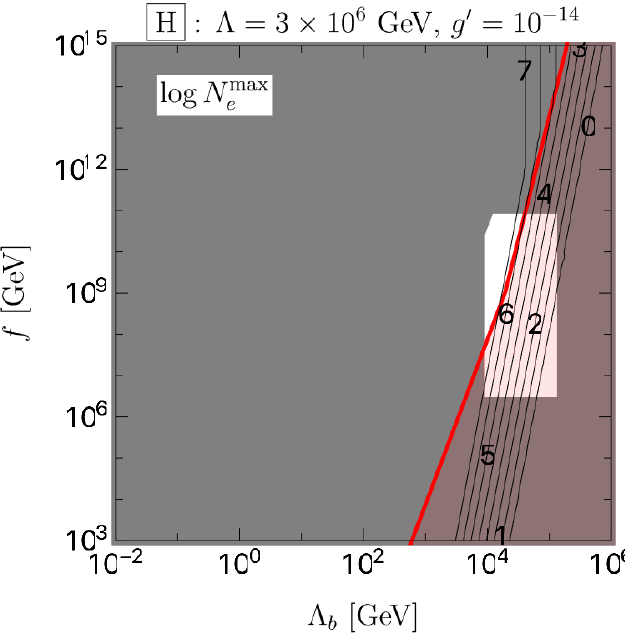}
\caption{\label{fig:HMT inflation contours HI Ne f Lambdab}
Same as Fig.~\ref{fig:HMT inflation contours f Lambdab} (EW gauge boson production during inflation), highlighting the contours for the minimal and maximal value of the Hubble rate $H_I$ (in GeV) and of the number of efolds of inflation that relaxation takes to complete, for each benchmark point $F$, $G$, $H$.}
\end{figure}

\begin{figure}[h!]
\centering
\includegraphics[width=.32\textwidth]{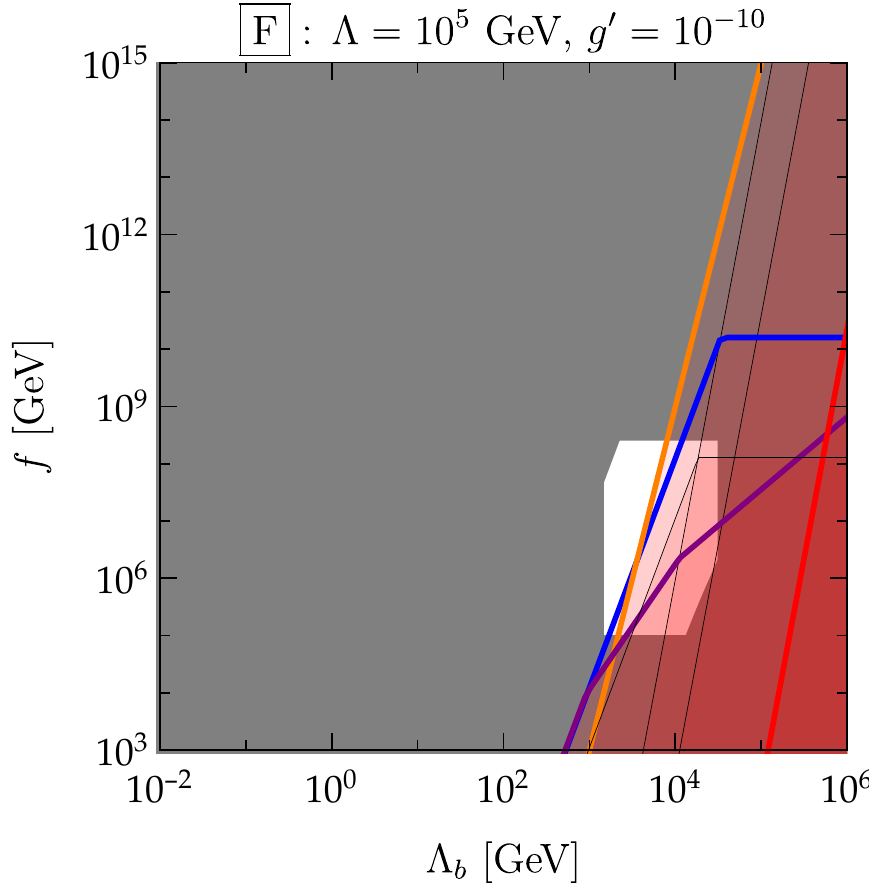}
\includegraphics[width=.32\textwidth]{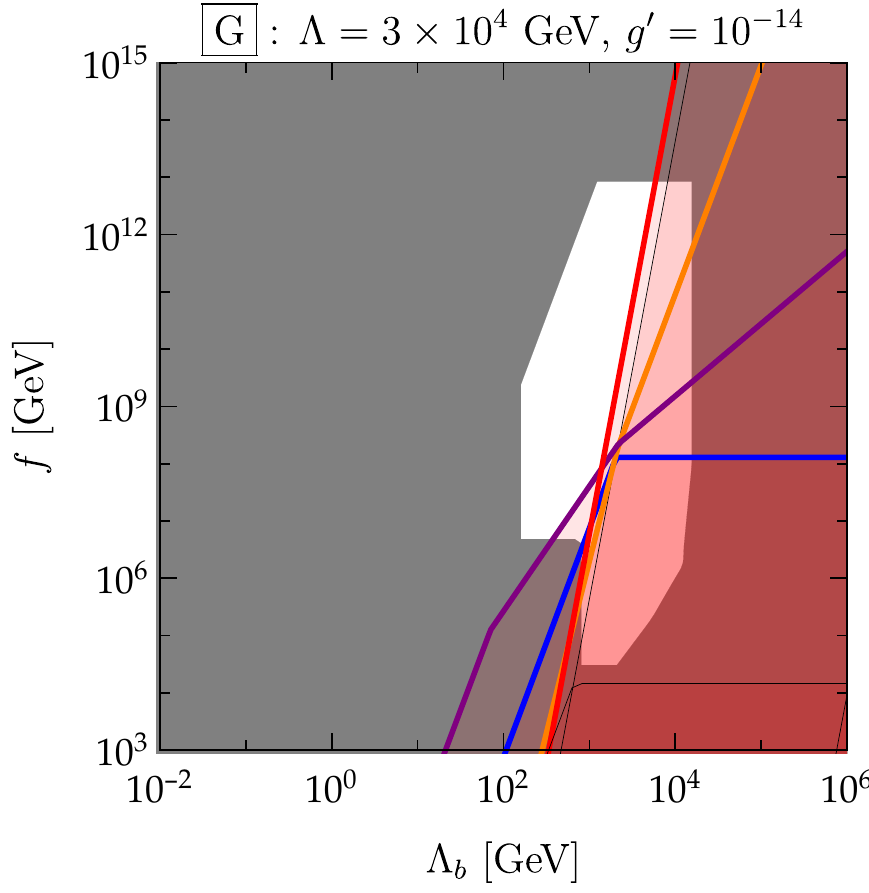}
\includegraphics[width=.32\textwidth]{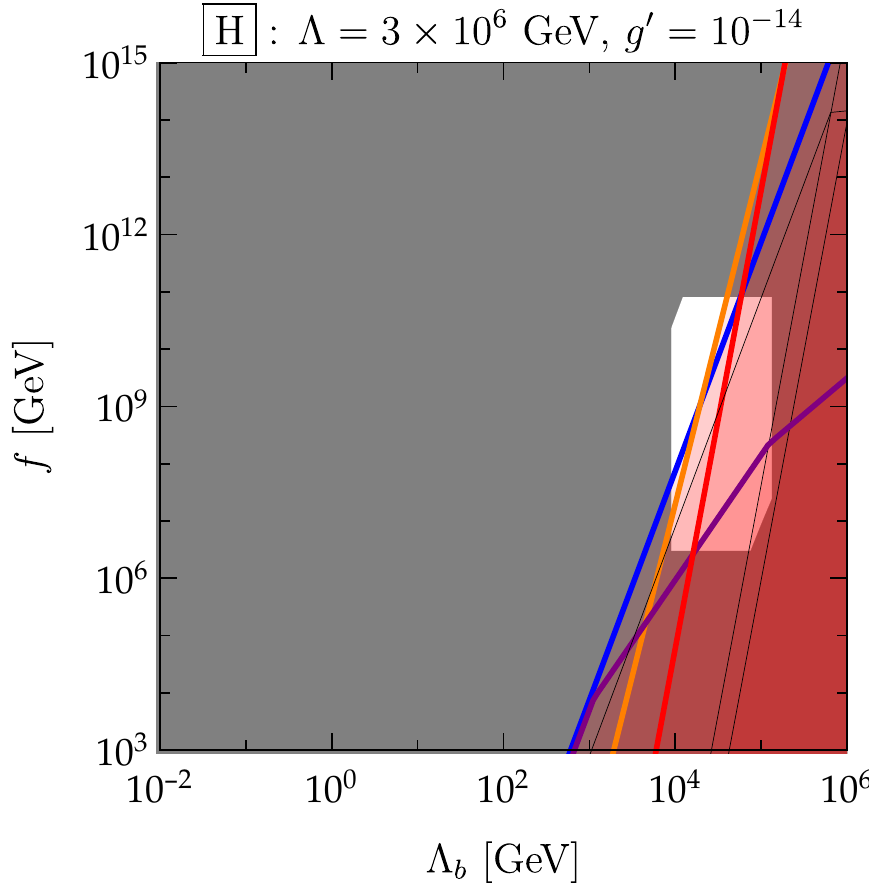}
\begin{multicols}{1}
\begin{itemize}
\item[{\tikz\fill[rounded corners = .5mm, blue](0,0)rectangle(0.3,0.2);}] Efficient energy dissipation \Eq{eq:condition energy dissipation} and inefficient fragmentation \Eq{eq:HMT slope fragmentation}
\item[{\tikz\fill[rounded corners = .5mm, orange](0,0)rectangle(0.3,0.2);}] Small variation of the Higgs mass \Eq{eq:f42} and \Eq{eq:HMT slope fragmentation}
\item[{\tikz\fill[rounded corners = .5mm, Purple](0,0)rectangle(0.3,0.2);}] No restoration of the shift symmetry \Eq{eq:H12} and \Eq{eq:HMT slope fragmentation}
\item[{\tikz\fill[rounded corners = .5mm, red](0,0)rectangle(0.3,0.2);}] Relaxion subdominant with respect to inflaton \Eq{eq:HMT relaxion subdominant} and \Eq{eq:HMT slope fragmentation}
\end{itemize}
\end{multicols}
\caption{\label{fig:HMT inflation contours f Lambdab frag}
Same as Fig.~\ref{fig:HMT inflation contours f Lambdab} (EW gauge boson production during inflation). The gray region is the envelope of the contours shown in Fig.~\ref{fig:HMT inflation contours f Lambdab frag}. The coloured contours are obtained by imposing that fragmentation is not efficient (see Sec.~\ref{sec:HMT} in the main text for further explanation)}
\end{figure}

\clearpage

%%%%%%%%%%%%%%%%%%%%%%%%%%%%%%%%%%%%%%%%%%%%%%%%%%%%%%%%%%%%%%%%%%%%%%%%%%%%%%%%%%
%%%%%%%%%%%%%%%%%%%%%%%%%%%%%%%%%%%%%%%%%%%%%%%%%%%%%%%%%%%%%%%%%%%%%%%%%%%%%%%%%%
%%%%%%%%%%%%%%%%%%%%%%%%%%%%%%%%%%%%%%%%%%%%%%%%%%%%%%%%%%%%%%%%%%%%%%%%%%%%%%%%%%
\newpage

\section{Cosmological histories}
\label{sec:cosmologicalhistory}

In this appendix, we illustrate and comment on the possible cosmological scenarios that could arise in the various cases we have discussed.

\begin{figure}[b!]
 %\label{fig:sketchinflation}
   \centering
\includegraphics[scale=1.1]{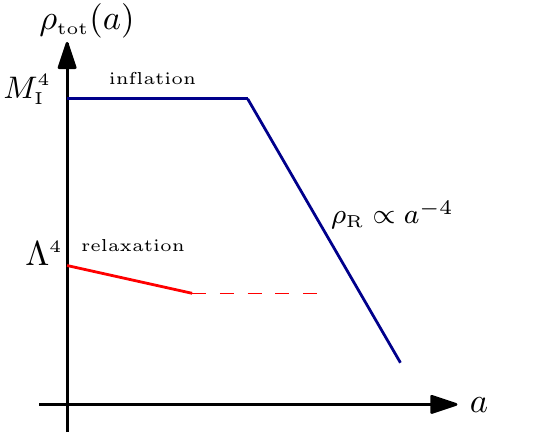}
\caption{\label{fig:sketchinflation} Sketch of the energy density of the universe as a function of the scale factor  in the scenario discussed in Sec.\,\ref{sec:during inflation} where relaxation happens when the energy universe is dominated by the inflaton potential. }
\end{figure}

First, in the case of relaxation with Higgs-dependent barriers which happens during inflation (Section  \ref{sec:during inflation}), we assume the energy density of the universe is dominated by the inflaton, and the relaxion is a subdominant component (see Fig.~\ref{fig:sketchinflation}). The universe is eventually reheated from the inflaton energy density and cosmological perturbations are inherited from the inflaton. Generally, the energy density stored in relaxion oscillations is subdominant, see e.g. \cite{Espinosa:2015eda}. One has nevertheless to make sure that the relaxion vacuum energy density does not eventually take over, so it should decay  (for instance by introducing a new coupling to gauge bosons) or the corresponding cosmological constant should  be cancelled. 
In the new scenario that we have discussed where the relaxion stops because of fragmentation (section 3.1.3), most of the relaxion kinetic energy goes into relaxion particles which behave as hidden radiation that gets diluted away by inflation.
Note also that in this case, the number of efolds and the inflation scale can be small (see Fig.~\ref{fig:comparisonINFL} and Fig.~\ref{fig:efoldsduringinflation}), this means even a short late stage of inflation is enough for relaxation of the EW scale, and we do not have to impose necessarily that this stage of inflation is responsible for cosmological perturbations.

\begin{figure}[!h]
   \centering
\includegraphics[scale=1.1]{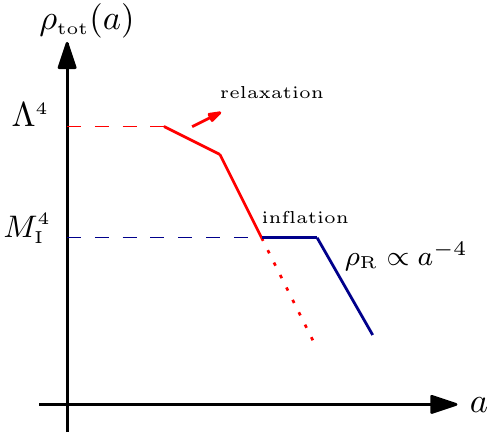}
\caption{ \label{fig:beforeinflation relaxion dom} Sketch of the energy density of the universe as a function of the scale factor when relaxation happens \emph{before} inflation and ends because of axion fragmentation.}
\end{figure}

In the case of relaxation before inflation with axion fragmentation (section  \ref{sec:withoutinflation}), there is no concern and standard big bang cosmology can proceed. This case is shown in Fig.~\ref{fig:beforeinflation relaxion dom}.

\begin{figure}[!h]
   \centering
\includegraphics[scale=1.1]{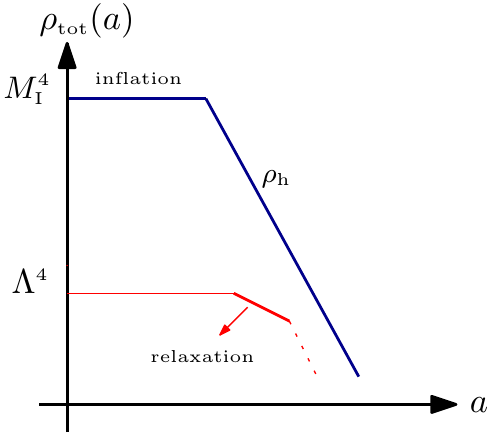}
\caption{ \label{fig:afterinflation relaxion NOT dom}    Sketch of the energy density of the universe as a function of the scale factor in the case where relaxation takes place \emph{after inflation} and when the energy density of the universe is dominated by a hidden sector that later decays into the SM. }
\end{figure}

\begin{figure}[!h]
    \centering
    \subfloat[]{{\includegraphics[scale=1.1]{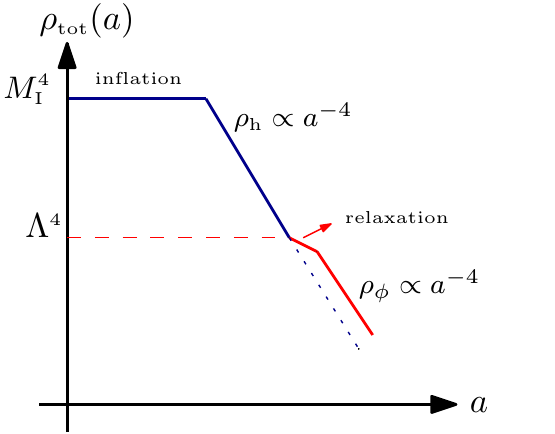}  } \label{fig:a} }
   \qquad \qquad
       \subfloat[]{{\includegraphics[scale=1.1]{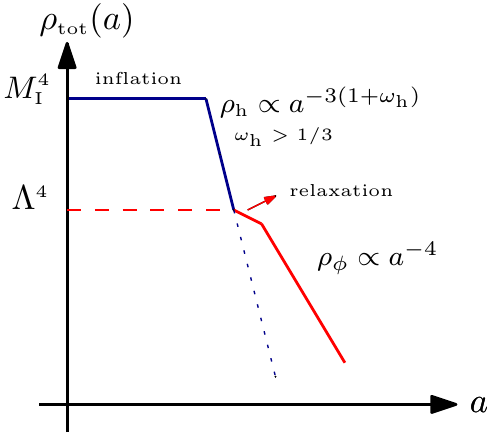} }  \label{fig:b} }
          \qquad \qquad
          \subfloat[]{{\includegraphics[scale=1.1]{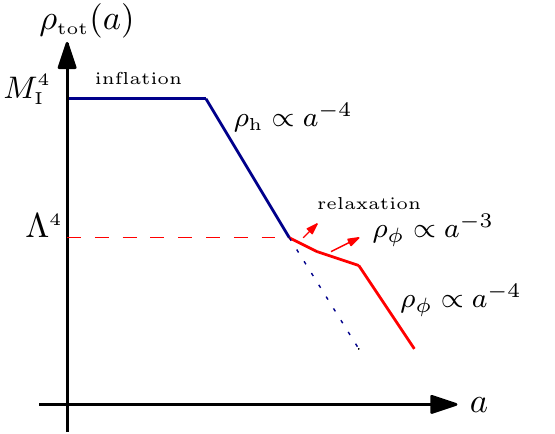}  \label{fig:c} }}
            \qquad \qquad
          \subfloat[]{{\includegraphics[scale=1.1]{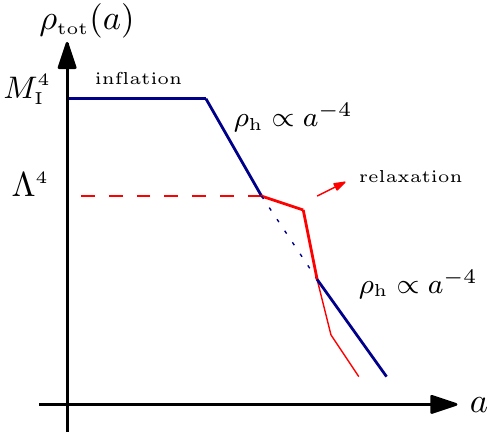}  \label{fig:d} }}
       \caption{Sketch  of the energy density of the universe as a function of the scale factor for cases where relaxation happens \emph{after inflation} when the energy density is  dominated by the relaxion field, i.e. $H\sim \Lambda^2/(\sqrt{3}\MPl)$. (a): Hidden sector red-shifts as radiation, which makes this scenario very constrained by dark radiation bounds;   (b): Hidden sector red-shifts faster than radiation $\omega_{\textrm{h}}> 1/3$ (as a kination-like period); (c): There is a period of matter domination after relaxation. (d): Hidden sector red-shifts as radiation and at the end of relaxation $\omega_\phi> 1/3$ (kination-like). At late times the hidden sector decays  into the Standard Model particles.  \label{fig:afterinflation relaxion dom}}
\end{figure}

\begin{figure}[!h]
\centering
\includegraphics[width=9cm]{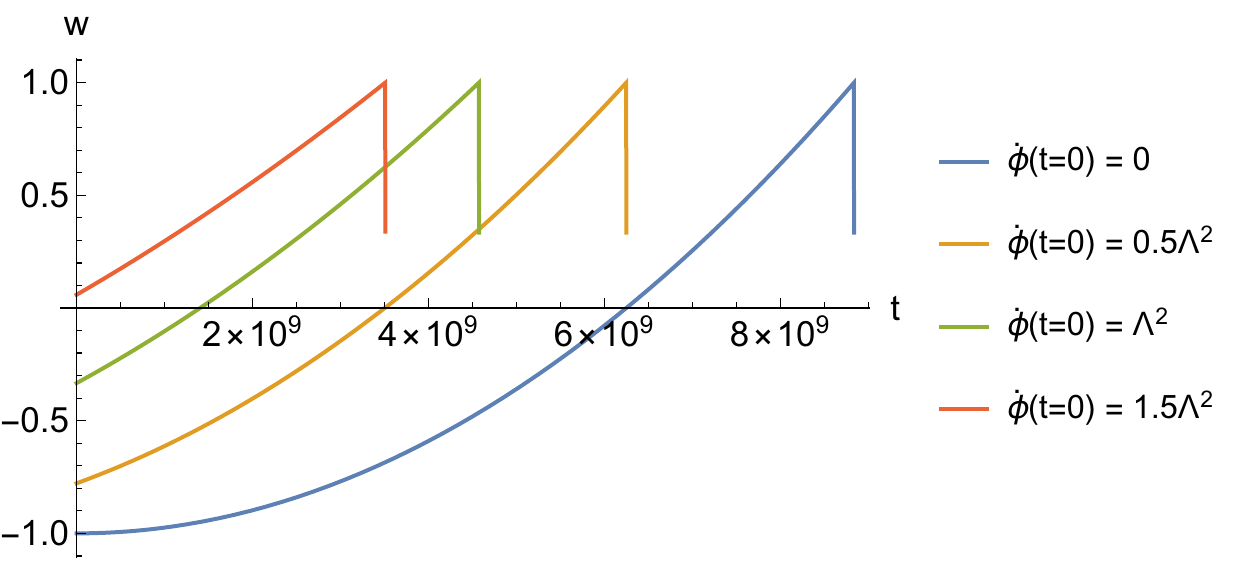}
\caption{\label{fig:equationofstate} Equation of state of the universe in model with self-stopping relaxion,  for $\Lambda_b = 800$ GeV, $\Lambda = 8$ TeV, $g = 2 \times 10^{-14}$, and $f$ is determined by the stopping condition. This corresponds to the cases in \Fig{fig:afterinflation relaxion dom} where the relaxion dominates the energy density.}
\end{figure}

We now discuss the case where  relaxation takes place after inflation, while the universe has been reheated into some hidden radiation,  and ends because of axion fragmentation (Sec.\,\ref{sec:withoutinflation}).
Fig.~\ref{fig:afterinflation relaxion NOT dom} shows the situation where the relaxion is a subdominant component of the energy density during relaxation, for which there is no need to worry about overclosure of the universe by the relaxion.  The underlying assumption is that the relaxion potential emerges from a sector independent from the one dominating the energy density.
In contrast, Fig.~\ref{fig:afterinflation relaxion dom} corresponds to the case where the relaxion potential emerges due to coupling to the hidden radiation and therefore relaxation starts when the relaxion energy density dominates.  In this case, we need to introduce a coupling to photons as discussed in Sec.~\ref{subsub:cosmohisto} to avoid that the relaxion energy density takes over eventually and overclose the universe. 
The evolution of the equation of state of the universe until fragmentation starts is shown in Fig.~\ref{fig:equationofstate}.
 In the cases (a-b-c) of Fig.~\ref{fig:afterinflation relaxion dom}, we need to assume that the relaxion eventually decays into photons to recover a standard radiation era. 
In case (d), we  assume a stage of kination domination may enable hidden radiation to dominate after relaxation.

\begin{figure}[!h]
   \centering
\includegraphics[scale=1.1]{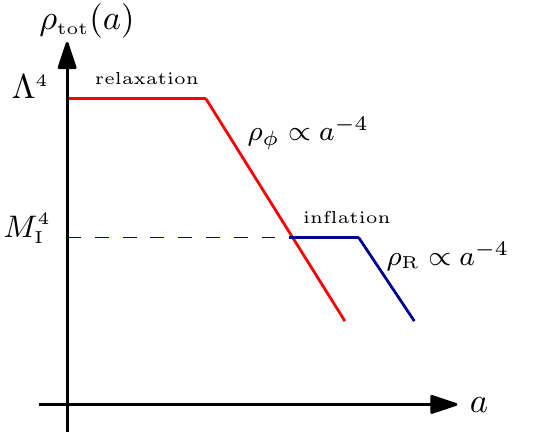}
\caption{ \label{fig:duringinflation rel dom} Sketch of the energy density of the universe as a function of the scale factor in the scenario where the relaxion drives an inflationary period.}
\end{figure}

Finally, there is the case where the relaxion drives a stage of inflation as discussed in Sec.~\ref{sec:relaxioninflating}. This is illustrated in Fig.~\ref{fig:duringinflation rel dom}.
At the end of relaxation, the energy density of the universe is in relaxion radiation. This should be followed by a stage of standard inflation and then  by reheating (we know that the relaxion cannot lead to the correct size of perturbations \cite{Tangarife:2017rgl}).
Alternatively, this period can follow the standard inflationary epoch in which curvature perturbations are generated, provided that it lasts for less than $\mathcal{O}(10)$ efolds.

\clearpage
\bibliography{ref}
\bibliographystyle{JHEP}

\end{document}